\documentclass{aa}
\usepackage{graphicx}
\usepackage{txfonts}
\usepackage{natbib}
\usepackage{color}
\definecolor{darkblue}{rgb}{0.00,0.00,0.50}
\definecolor{darkdarkblue}{rgb}{0.00,0.00,0.3}
\usepackage[colorlinks=true,citecolor=darkdarkblue,linkcolor=darkblue]{hyperref}
\usepackage{etoolbox}
\makeatletter
\patchcmd\@combinedblfloats{\box\@outputbox}{\unvbox\@outputbox}{}{%
  \errmessage{\noexpand\@combinedblfloats could not be patched}%
}%
\makeatother

\usepackage{xspace}

\newcommand{\eg}{e.\,g.\xspace}
\newcommand{\ie}{i.\,e.\xspace}

\newcommand{\fluxspx}{\,erg\,s$^{-1}$\,cm$^{-2}$\,spaxel$^{-1}$\xspace}
\newcommand{\fluxa}{\,erg\,s$^{-1}$\,cm$^{-2}$\,arcsec$^{-2}$\xspace}

\newcommand{\flux}{\,erg\,s$^{-1}$\,cm$^{-2}$\xspace}
\newcommand{\lum}{\,erg\,s$^{-1}$\xspace}
\newcommand{\nexp}[1]{$\times10^{-#1}$\xspace}
\newcommand{\pexp}[1]{$\times10^{#1}$\xspace}

\newcommand{\ngc}{NGC\,40} % for Antennae NGC numbers
\newcommand{\kms}{\,km\,s$^{-1}$\xspace}
\newcommand{\hiireg}{\textsc{H\,ii}~region\xspace}
\newcommand{\hiiregs}{\textsc{H\,ii}~regions\xspace}
\newcommand{\ha}{H$\alpha$\xspace}
\newcommand{\hb}{H$\beta$\xspace}

\newcommand{\hi}{\ion{H}{i}\xspace}

\newcommand{\hamath}{\ensuremath {\mathrm{H}\alpha}}
\newcommand{\hbmath}{\ensuremath {\mathrm{H}\beta}}
\newcommand{\fesc}{$f_\mathrm{esc}$\xspace}

\newcommand{\oi}{[\ion{O}{i}]\xspace}
\newcommand{\oii}{[\ion{O}{ii}]\xspace}
\newcommand{\oiii}{[\ion{O}{iii}]\xspace}
\newcommand{\nii}{[\ion{N}{ii}]\xspace}
\newcommand{\sii}{[\ion{S}{ii}]\xspace}
\newcommand{\siii}{[\ion{S}{iii}]\xspace}
\definecolor{pmwgrey}{rgb}{0.4,0.4,0.4}
\definecolor{darkgreen}{rgb}{0.0,0.3,0.0}
\definecolor{darkviolet}{rgb}{0.4,0.0,0.4}
\definecolor{darkred}{rgb}{0.4,0.0,0.0}

\begin{document}
\title{On the Origin of Diffuse Ionized Gas in the Antennae Galaxy}
\author{Peter M.\ Weilbacher\inst{1}
        \and
        Ana Monreal-Ibero\inst{2,3}
        \and
        Anne Verhamme\inst{4,5}
        \and
        Christer Sandin\inst{1}
        \and
        Matthias Steinmetz\inst{1}
        \and
        Wolfram Kollatschny\inst{6}
        \and
        Davor Krajnovi\'{c}\inst{1}
        \and
        Sebastian Kamann\inst{6}
        \and
        Martin M.\ Roth\inst{1}
        \and
        Santiago Erroz-Ferrer\inst{7}
        \and
        Raffaella Anna Marino\inst{7}
        \and
        Michael V.\ Maseda\inst{8}
        \and
        Martin Wendt\inst{1,9}
        \and
        Roland Bacon\inst{4}
        \and
        Stefan Dreizler\inst{6}
        \and
        Johan Richard\inst{4}
        \and
        Lutz Wisotzki\inst{1}
  }
  \institute{Leibniz-Institut f\"ur Astrophysik Potsdam (AIP),              
             An der Sternwarte 16, D-14482 Potsdam, Germany\\
             \email{pweilbacher@aip.de}
        \and
             Instituto de Astrof\'{\i}sica de Canarias (IAC),               
             E-38205 La Laguna, Tenerife, Spain
        \and
             Universidad de La Laguna, Dpto.\ Astrof\'{\i}sica,             
             E-38206 La Laguna, Tenerife, Spain
        \and
             Univ Lyon, Univ Lyon1, Ens de Lyon, CNRS, Centre de Recherche  
             Astrophysique de Lyon UMR5574, F-69230, Saint-Genis-Laval, France
        \and
             Observatoire de Gen\`eve, Universit\'e de Gen\`eve, 51 Ch. des 
             Maillettes, 1290 Versoix, Switzerland
        \and
             Institut f\"ur Astrophysik, Friedrich-Hund-Platz 1,            
             D-37077 G\"ottingen, Germany
        \and
             Department of Physics, ETH Z\"urich, Wolfgang-Pauli-Strasse 27,
             CH-8093 Z\"urich, Switzerland
        \and
             Leiden Observatory, Leiden University, P.\,O.\ Box 9513,       
             2300 RA, Leiden, The Netherlands
        \and
             Institut f\"ur Physik und Astronomie, Universit\"at Potsdam,   
             Karl-Liebknecht-Str. 24/25,  14476 Golm, Germany
   }
   \date{Received 28 July 2017 / accepted 13 December 2017}
\authorrunning{Weilbacher et al.}
\abstract
  {The ``Antennae Galaxy'' (NGC 4038/39) is the closest major interacting
   galaxy system and therefore often taken as merger prototype.
   We present the first comprehensive integral field spectroscopic dataset of
   this system, observed with the MUSE instrument at the ESO VLT. We cover the
   two regions in this system which exhibit recent star-formation: the central
   galaxy interaction and a region near the tip of the southern tidal tail.
   In these fields, we detect \hiiregs and diffuse ionized gas to unprecedented
   depth. About 15\% of the ionized gas was undetected by previous observing
   campaigns. This newly detected faint ionized gas is visible everywhere around
   the central merger, and shows filamentary structure. We estimate diffuse gas
   fractions of about 60\% in the central field and 10\% in the southern region.
   We are able to show that the southern region contains a significantly
   different population of \hiiregs, showing fainter luminosities.
   By comparing \hiireg luminosities with the HST catalog of young star
   clusters in the central field, we estimate that there is enough
   Lyman-continuum leakage in the merger to explain the amount of diffuse
   ionized gas that we detect.  We compare the Lyman-continuum escape fraction
   of each \hiireg against ionization-parameter sensitive emission line ratios.
   While we find no systematic trend between these properties, the most extreme
   line ratios seem to be strong indicators of density bounded ionization.
   Extrapolating the Lyman-continuum escape fractions to the southern
   region, we conclude that just from the comparison of the young stellar
   populations to the ionized gas there is no need to invoke other ionization
   mechanisms than Lyman-continuum leaking \hiiregs for the diffuse ionized
   gas in the Antennae.
}
\maketitle

\section{Introduction}
In the hierarchical paradigm of galaxy formation, interactions and merging are
major events in galaxy evolution \citep{WR78,1993MNRAS.262..627L}. They create some
of the strongest starburst galaxies that we know \citep{SM96} and shape the
galaxies we see today in the nearby universe \citep{2002NewA....7..155S,2003ApJS..147....1C}.
From the theoretical point of view, mergers occur by the thousands in cosmological
simulations \citep{2015MNRAS.446..521S,2014MNRAS.444.1518V}.
From the observational point of view, major mergers have been
identified at intermediate to high-redshift \citep[\eg][]{Tacconi08,Ivison12,Ventou17}.
They are also being studied in the nearby universe, often in galaxies classified
as infrared-bright \citep[LIRG or ULIRG,][]{2010A&A...522A...7A,RKD11},
highlighting aspects as diverse as central shocks \citep{mon06} and Tidal
Dwarf Galaxies \citep{WDF03}.
One has to study the most nearby mergers in detail, at high spatial
resolution, to be able to disentangle and characterize the different elements
playing a role in their evolution, in order to be able to properly interpret
the high-redshift cases.

In that sense, \object{the Antennae} (\object{NGC\,4038/39}, \object{Arp\,244}),
one of the most spectacular examples of gas-rich major mergers, and at a
distance of $22\pm3$\,Mpc \citep{2008AJ....136.1482S}
the closest one, constitutes a desirable laboratory to study the
interplay of gas and strong recent star formation during the merger evolution.
The system has a solar or slightly super-solar metallicity
\citep{BTK+09,2015ApJ...812..160L}
and not only exhibits one of the most violent
star-forming events in the nearby universe, forming a multitude of young,
massive stellar clusters in the central merger
\citep{WGL+05,WCS+10,BEK+06},
but also shows a more quiescent star-formation mode at the end of the southern
tidal tail \citep{Hibbard05}.
Being such a paradigmatic object, it has a long history of being the subject of
tailored simulations \citep[\eg][]{TT72,KNJ+10,2015MNRAS.446.2038R}
and being extensively observed at all wavelengths, from radio to X-rays (\eg radio:
\citealt{Bigiel15}, \citealt{Whitmore14}; far- and mid- infrared:
\citealt{BSdB+09}, \citealt{Schirm14}; near-infrared: \citealt{Brandl05},
\citealt{2005A&A...443...41M}; optical: \citealt{WCS+10}; ultraviolet:
\citealt{Hibbard05}).

\begin{figure}
\centering
\includegraphics[width=\linewidth]{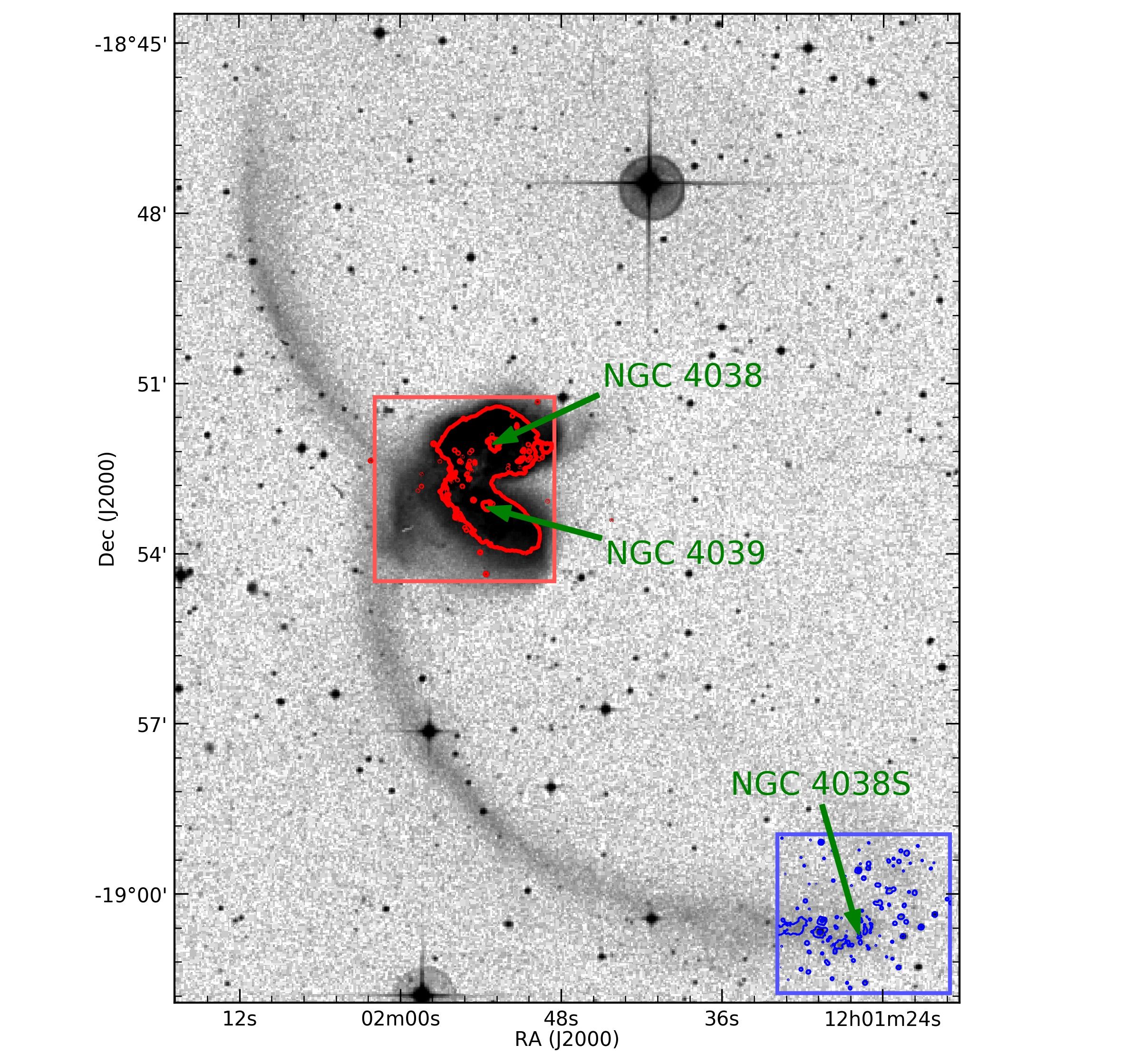}
\caption{The Antennae as seen on the blue Digital Sky Survey, version 2.
         The light red and light blue boxes show the outer edges of the MUSE
         coverage as used in all further plots.
         The red contours mark arbitrary continuum levels derived from a
         smoothed HST ACS image in the F814W filter in the central interacting
         galaxy. The two galaxy nuclei can be used to relate their location in
         other figures in this paper.
         The blue contours are similarly derived from HST data in the region
         near the tip of the southern tidal tail, they mark mostly foreground
         stars and background objects. The two bright stars in this region can
         be used to relate the location of this field to the other figures in
         this paper. This region is sometimes called \ngc38\,S in the literature;
         here, we use the term ``South'' or ``southern'' to describe it.}
\label{fig:ant}
\end{figure}

An image of the Antennae system is shown in Fig.~\ref{fig:ant}. In this
publication, we deal with data of two fields, the central field and a southern
one, both of which are marked with continuum contours taken in the HST ACS image; the
same continuum contours are used throughout this paper.

An important missing piece of information would be a spectroscopic optical
mapping of the whole system, ideally at high spatial and spectral resolution,
and covering as much of the optical spectral range as possible.
Data taken with the GH$\alpha$FaS instrument presented by
\citet{2014MNRAS.445.1412Z} were a first step in this direction.
They cover the main body of both galaxies at very good spectral resolution (8\kms).
However, these data were gathered with a Fabry-Perot instrument, covering only
a narrow spectral range around \ha.
On the other side, long-slit observations can address these points, but only at
very specific locations in the system \citep[like individual star clusters, \eg][]{WGL+05,BTK+09}.
Data from an Integral Field Spectrograph (IFS) would provide both, large
spectral coverage and spatial mapping.
\citet[][]{BEK+06} nicely illustrate the potential of this technique:
with only one set of observations, both the ionized gas and the stellar
populations can be mapped and characterized, although they still map only a
small portion of the system.

The advent of the \emph{Multi Unit Spectroscopic Explorer}
\citep[MUSE][]{Bacon+10} at the 8\,m Very Large Telescope (VLT), with a
large field of view ($1\arcmin\times1\arcmin$) and a wide spectral coverage ($\sim4600\dots9350$\,\AA\ at $R\sim3000$), is the next step to overcome the shortcomings
of previous observations. This motivated us to perform a thorough optical spectrocopic
mapping of the system. In this first paper of a series, we will present the
survey and, in particular, address one specific question: Can the photons leaking from the
\hiiregs in the system explain the detected diffuse component of the ionized
gas?

The diffuse ionized gas (DIG, also called \textsl{warm ionized medium}, WIM)
has proven to be ubiquitous in star-forming galaxies, including ours
(see \citealt{2000ApJ...544..347M};
\citealt{2009RvMP...81..969H} for a review).
Strong emission line ratios in DIGs differ from those found in \hiiregs.
Typically, they present higher forbidden-to-Balmer line ratios
\citep[\eg \sii{6717,6731}/\ha, \nii{6584}/\ha, \oi{6300}/\ha;][]{2003ApJ...586..902H,Madsen06,2006ApJ...644L..29V} than \hiiregs.

Several mechanisms have been proposed to explain the detection of DIG in other
galaxies and the unusual line ratios of which the most prominent are:
(a) Lyman-continuum (LyC; $\lambda<912$\,\AA) photons leaking from \hiiregs
\citep{2002A&A...386..801Z} into the interstellar medium. Part of the UV
continuum is absorbed by the gas, leading to a harder ionizing spectrum
\citep{2003ApJ...586..902H} which can at least partly explain the observed
properties.
(b) Shocks are another mechanism to change the line ratios
\citep{dop95}, and these are also frequently observed in merging galaxies
\citep{MonrealIbero10,Soto12}. However, they are not always observed in the same parts of
the galaxy as the DIG or do not explain its observed properties
\citep[\eg, all line ratios at the same time;][]{2016A&A...585A..79F}.
Finally, (c) evolved stars have a hard UV spectrum that could explain the DIG
\citep{2017MNRAS.466.3217Z}. This is most relevant for early-type galaxies
\citep{Kehrig12,Papaderos13}. However, in starburst galaxies their
contribution to the Lyman-continuum is likely negligible compared to hot stars
in star-forming regions,
which produce orders of magnitude more ionizing photons \citep{LSG+99}.

Since this is the first publication of the MUSE data of the Antennae, we
explain the data reduction and properties in some detail in
Sect.~\ref{sec:data}.  In Sect.~\ref{sec:strct} we then present the morphology
of the ionized gas in the system, and specifically discuss the structure of the
diffuse ionized gas in Sect.~\ref{sec:dig}.  In Sect.~\ref{sec:hiiregs} we
present a brief analysis of the \hiiregs and investigate to what extent the amount of
diffuse gas can be explained by ionizing photons originating from the
star-forming regions. We summarize our results in Sect.~\ref{sec:concl}.
In this publication we restrict ourselves to this narrow topic, but
would like to emphasize that topics like detailed stellar population modeling
and kinematics, among others, are to be analyzed in forthcoming papers.

\section{Data description}\label{sec:data}
\subsection{MUSE Observations}
The Antennae were observed during multiple nights in April and May 2015, and
February and May 2016, with the MUSE instrument \citep[][and in
prep.]{BAA+12}. We employ the wide-field mode. This samples the sky at 0\farcs2
and covers a field of view of about $1\sq\arcmin$. The extended wavelength
configuration (WFM-NOAO-E) was set up to attain a contiguous wavelength coverage
from 4600 to 9350\,\AA. This mode incurs a faint and broad second-order overlap
beyond $\sim8100$\,\AA\ \citep[see][]{WeilbacherM42,2015Msngr.162...37W} that is,
however, not affecting our analysis.

\begin{figure*}
\includegraphics[width=0.486\linewidth]{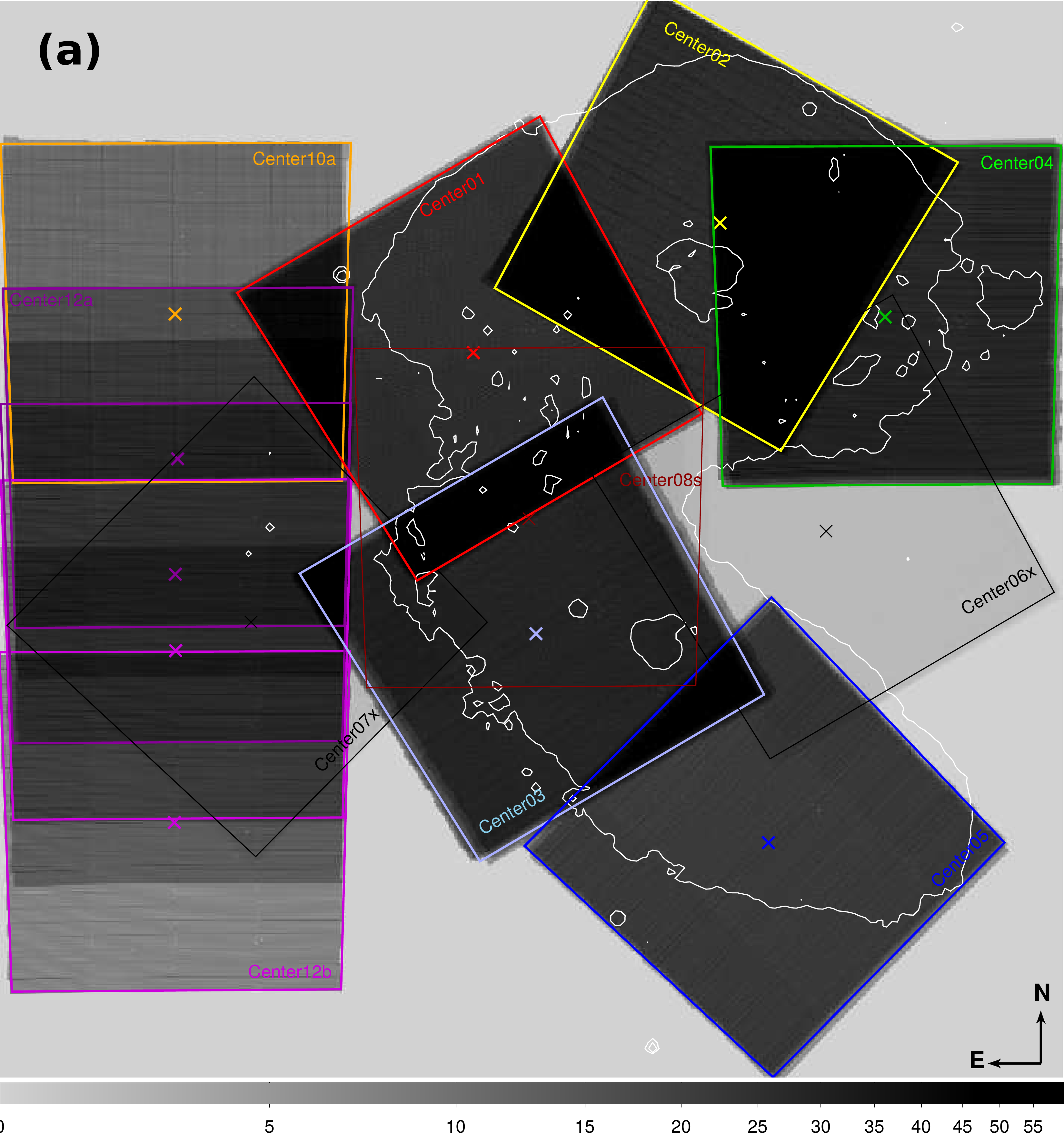}
\includegraphics[width=0.512\linewidth]{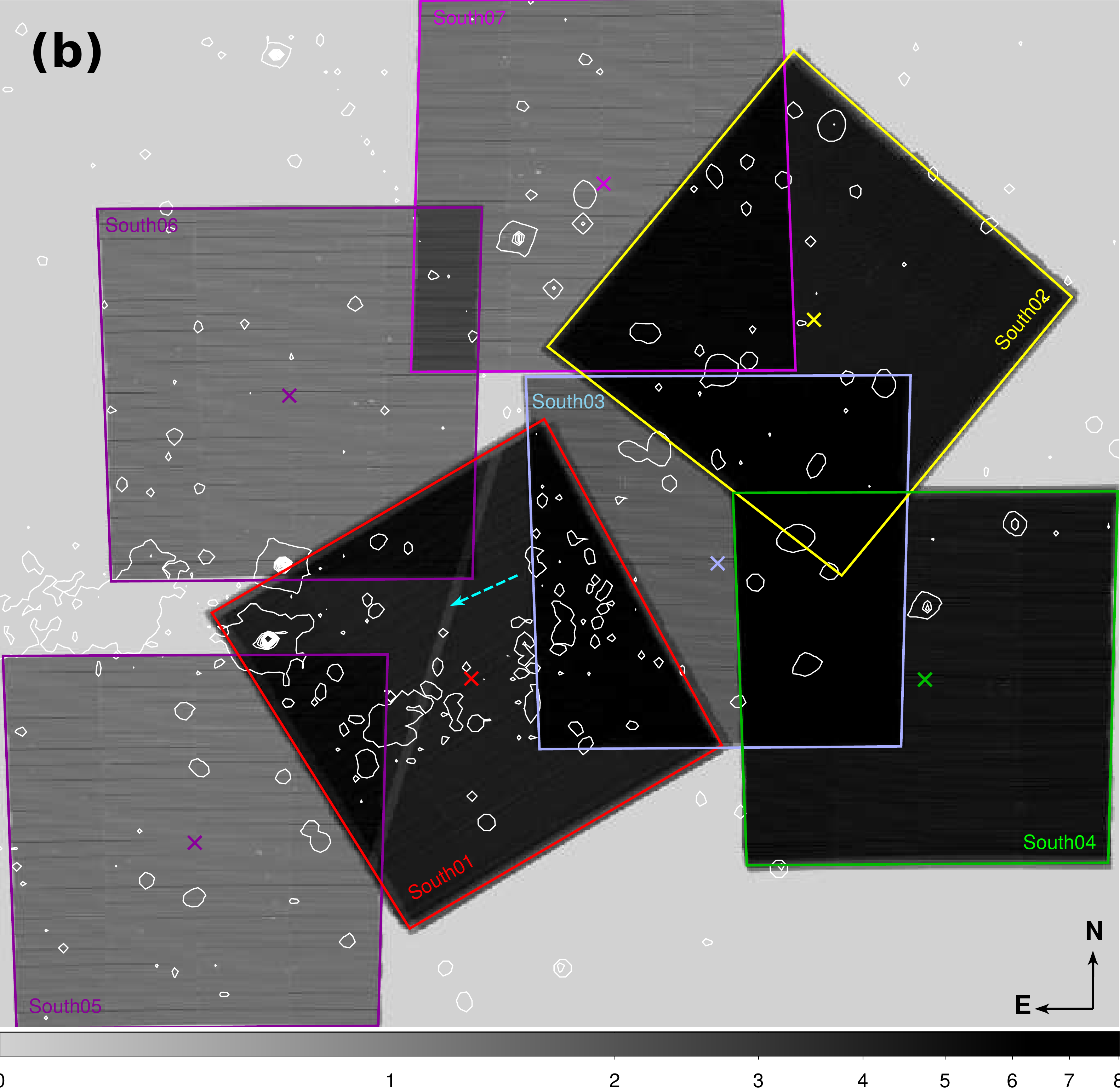}
\caption{Inverse grayscale map of the relative weights of the data of
         the MUSE data of the Antennae. In this representation, the deepest
         regions appear black while those parts of the data only covered by
         exposures taken in poor conditions appear light gray.
         In color, boxes representing the MUSE fields (each approx.\
         $1\arcmin\times1\arcmin$ in size) are shown, with the annotated field
         designation and a cross marking the field center.
         The white contours are similar to the continuum levels shown in
         Fig.~\ref{fig:ant}.
         In panel {\bf (a)} we show the pointings of the central Antennae
         while in {\bf (b)} the pointings around the tip of the southern tidal
         tail are presented. In panel (b) a removed satellite trail that
         decreased the effective exposure time is marked with a dashed cyan
         arrow.}
\label{fig:fields}
\end{figure*}

\begin{table}
\caption{Layout of the observations}
\label{tab:obs}

\begin{tabular}{l ll rr}
Night      & Field     & Depth         & \multicolumn{2}{c}{Seeing}   \\
(UT)$^a$   & $^b$      & [s]           & AG$^c$    & MUSE$^d$         \\
\hline\hline
2015-04-22 & Center06x & 2$\times$1350 & 1\farcs00 & $\sim$1\farcs00 \\
2015-04-22 & Center01  & 4$\times$1350 & 0\farcs77 & 0\farcs61 \\
2015-04-23 & South01   & 1$\times$1350 & 0\farcs95 & 0\farcs93 \\
2015-04-25 & South01   & 3$\times$1350 & 0\farcs83 & 0\farcs79 \\
\hline
2015-05-10 & Center02  & 4$\times$1350 & 0\farcs73 & 0\farcs57 \\
2015-05-10 & Center03  & 2$\times$1350 & 0\farcs80 & 0\farcs64 \\
2015-05-11 & South02   & 2$\times$1350 & 0\farcs82 & 0\farcs64 \\
2015-05-12 & Center04  & 2$\times$1350 & 0\farcs64 & 0\farcs57 \\
2015-05-13 & Center03  & 2$\times$1350 & 0\farcs58 & 0\farcs45 \\
2015-05-19 & South02   & 2$\times$1350 & 0\farcs76 & 0\farcs61 \\
2015-05-20 & Center04  & 2$\times$1350 & 0\farcs74 & 0\farcs65 \\
2015-05-20 & Center05  & 2$\times$1350 & 0\farcs79 & 0\farcs62 \\
2015-05-21 & Center05  & 2$\times$1350 & 1\farcs01 & 0\farcs85 \\
2015-05-21 & South03   & 2$\times$1350 & 1\farcs02 & 0\farcs93 \\
2015-05-22 & Center07x & 2$\times$1350 & 0\farcs88 & 0\farcs76 \\
\hline
2016-01-31 & Center08s & 2$\times$\ \,100& 0\farcs79 & 0\farcs63 \\
2016-01-31 & South04   & 2$\times$1350 & 0\farcs78 & 0\farcs58 \\
2016-01-31 & Center12a & 3$\times$1350 & 0\farcs81 & 0\farcs70 \\
2016-02-01 & South05   & 1$\times$1350 & 0\farcs69 & 0\farcs53 \\
2016-02-01 & South06   & 1$\times$1350 & 0\farcs71 & 0\farcs60 \\
2016-02-01 & South07   & 1$\times$1350 & 0\farcs70 & 0\farcs56 \\
2016-02-01 & South04   & 2$\times$1350 & 0\farcs73 & 0\farcs58 \\
2016-02-02 & Center12b & 1$\times$1350 & 0\farcs72 & 0\farcs56 \\
2016-02-03 & Center12b & 2$\times$1350 & 0\farcs72 & 0\farcs67 \\
\hline
2016-05-11 & Center10a & 2$\times$1350 & 0\farcs83 & 0\farcs72 \\
\hline
\end{tabular}\\
$^a$ This column gives the UTC date of the start of the night.\\
$^b$ The postfix characters are:
     {\it x}: extra field taken in non-photometric conditions;
     {\it s}: short exposures to avoid line saturation;
     {\it a} and {\it b}: fields with large offsets.\\
$^c$ Measured using Gaussian fits by the VLT autoguiding system, averaged over
     each exposure.\\
$^d$ Measured using Moffat fits around $7000\pm100$\,\AA\ in the MUSE cubes.
     The field Center06x does not contain any bright enough point sources to
     reliably determine the seeing in the cube.\\
\end{table}

The layout of the observations is indicated in Fig.~\ref{fig:fields}.
These maps display the relative weights used for the creation of the datacube
(see below) which can be used to judge the relative depth of the data and is
annotated with the MUSE field number of the pointings.
Most pointings were taken with a spatial dither pattern at fixed position
angle, with 1350\,s per exposure.
The shallow extra pointings (featuring an {\it x} in the name) were observed at two
angles separated by 90$^\circ$, Center02 was observed at position angles of
150, 240, and 2$\times$330$^\circ$. All observations in 2015 were interleaved
with 200\,s exposures of a blank sky field. This was skipped for the {\it
South} pointings taken in 2016, since it was realized that those fields offer
enough blank sky already.
Except for one pointing that was observed through moderate clouds (Center06x),
all exposures were done in clear or photometric conditions.  The seeing as
measured by the autoguider probe of the VLT varied during the observations
between about 0\farcs5 and 1\farcs2 but was generally in the sub-arcsec regime.
A detailed timeline of the observations of the different fields is given in
Table~\ref{tab:obs}.

\subsection{Data processing}
All data were consistently reduced using the public MUSE pipeline \citep[][Weilbacher
et al.\ in prep.]{WSU+12,2014ASPC..485..451W}\footnote{Another operational
  update of the pipeline was discussed in \citet{2015scop.confE..53W}.}
in v1.6.

Basic data reduction followed standard steps for MUSE data. Master calibrations
were created from biases, lamp flat-fields, and arc exposures, and resulted in
master biases, master flats, trace tables, and wavelength calibration tables.
One set of arcs of each run was used to determine the line spread function
(LSF) of each slice in the MUSE field of view. The twilight sky flats in
extended mode were converted into three-dimensional corrections of the
instrument illumination. These calibrations and extra illumination-flats were
then applied onto the on-sky exposures (standard stars, sky fields, and object
exposures), using the calibration closest in time. The instrument geometry was
taken from the master calibration created by the MUSE team for each
corresponding run of guaranteed time observations (GTO).

We treated each extended-mode standard star exposure in the same way as
\citet{WeilbacherM42}: two reductions were run, using circular flux integration
and Moffat fits to extract the flux. The separate response functions were then
merged at 4600 and 8300\,\AA, with the circular measurements for the central part
of the wavelength range, equalizing the response level at the merging positions
to the central part. This procedure reduces the effect of the 2nd order contamination in the
very red and allows flux calibration even in the partially incomplete planes
below 4600\,\AA.
Next, all offset sky fields were processed, using response curve and telluric
correction derived from the standard star closest in time. This produced an
adapted list of sky line strengths and a sky continuum. If these sky properties
needed to be applied to a science exposure taken in between two sky fields,
both tables were averaged. For both sky and science fields, we used a modified
initial sky line list, where all lines below 5197\,\AA\ were removed. This
reduced broad artifacts near \hb and in other regions in the blue spectral
range where very faint OH bands but no strong telluric emission lines were
present.

The science post-processing then used the closest-in-time or averaged sky
properties, the LSF of the run, the response and telluric correction as well as
the astrometric solution that matches the geometry table for the respective
run. To help the pipeline with the creation of an optimal sky spectrum, we used
the approximate sky fraction as processing parameter. This was 10\% for the
central fields where the sky spectrum of the science exposure was just used
to adapt the sky line strengths, and 50\% for the southern fields where at
least half the field consisted of sky and hence both line adaptation and
continuum computation were done using the science field itself.
One exposure of the field South01 was affected by a satellite trail. This was
removed by masking the data within $\pm5.5$ pixels from the center of the
trail. (This is visible as a lighter linear feature in
Fig.~\ref{fig:fields}b and pointed out by an arrow.)

To be able to combine all exposures, the spatial shifts of all individual MUSE cubes
were computed against a corresponding HST exposure. For this, the MUSE data was
integrated using the filter-function of the HST ACS F814W filter to create an
image. The HST image was smoothed to approximately 0\farcs6 FWHM.
The centroid of the brightest compact object in each MUSE field was then
measured in sky coordinates using IRAF {\sc imexam} and compared to the HST ACS
F814W images \citep[HST proposal ID 10188,][]{WCS+10} to derive the effective
offset. These were given to the pipeline for the final combination of all
exposures.
The positions of the few bright foreground stars in our final cube agree
with positions given in the 2MASS catalog \citep{2MASS} to better than 1\farcs0
and to the positions in the Gaia DR1 \citep{GaiaDR1} catalog to within
0\farcs3. As a result, our data have a good relative and absolute
astrometric accuracy.
The wavelengths of all exposures were shifted to the solar system barycenter.

We created separate cubes for the central and southern regions, each
encompassing all relevant exposures.  To optimize the spatial resolution of the
cubes, we applied FWHM-based weighting offered by the pipeline. This uses the
average seeing measured by the VLT autoguiding system during the exposure to
create a weight inversely proportional to the FWHM of an exposure
\citep[similar to the procedure of][]{HAG+03}. For the extra exposures
(Center06x and Center07x) we reset the seeing in the FITS headers so that these
were weighted approximately five times less than the other exposures.
This ensures that these exposures only contribute significantly where other
better data is absent. The short exposures (Center08s) were taken to cover part of
the field where \ha was saturated; they contribute significantly only in the
regions around the saturation which were masked by hand in the longer
exposures.
We used the standard (linear) sampling of
$0\farcs2\times0\farcs2\times1.25\,\text{\AA}\,\text{voxel}^{-1}$ but also created
cubes in log-sampled wavelengths in the same spatial grid, where the sampling
ranged from 0.82\,\AA\,pixel$^{-1}$ in the blue to 1.72\,\AA\,pixel$^{-1}$ at
the red end; this corresponds to a velocity scale of 53.5070\,km\,pixel$^{-1}$.
The pipeline also created weight cubes to show the relative contribution at
each position. The wavelength-averaged version of these for the linear sampling
is displayed in Fig.~\ref{fig:fields}.

The effective seeing in the final cube is difficult to assess since many of the
compact objects were not point sources and spatial variations remain.
From the few foreground stars, we estimate a seeing of at a
wavelength of 7000\,\AA\ and find $0\farcs59\pm0\farcs05$ in the main central
part (covered by Center01 to Center05), $0\farcs76\pm0\farcs15$ over the full
central field, and $0\farcs57\pm0\farcs07$ in the southern field.

\section{Structure of the ionized gas}\label{sec:strct}
\begin{figure*}
\includegraphics[trim={0 0 71 0},clip,width=0.486\linewidth]{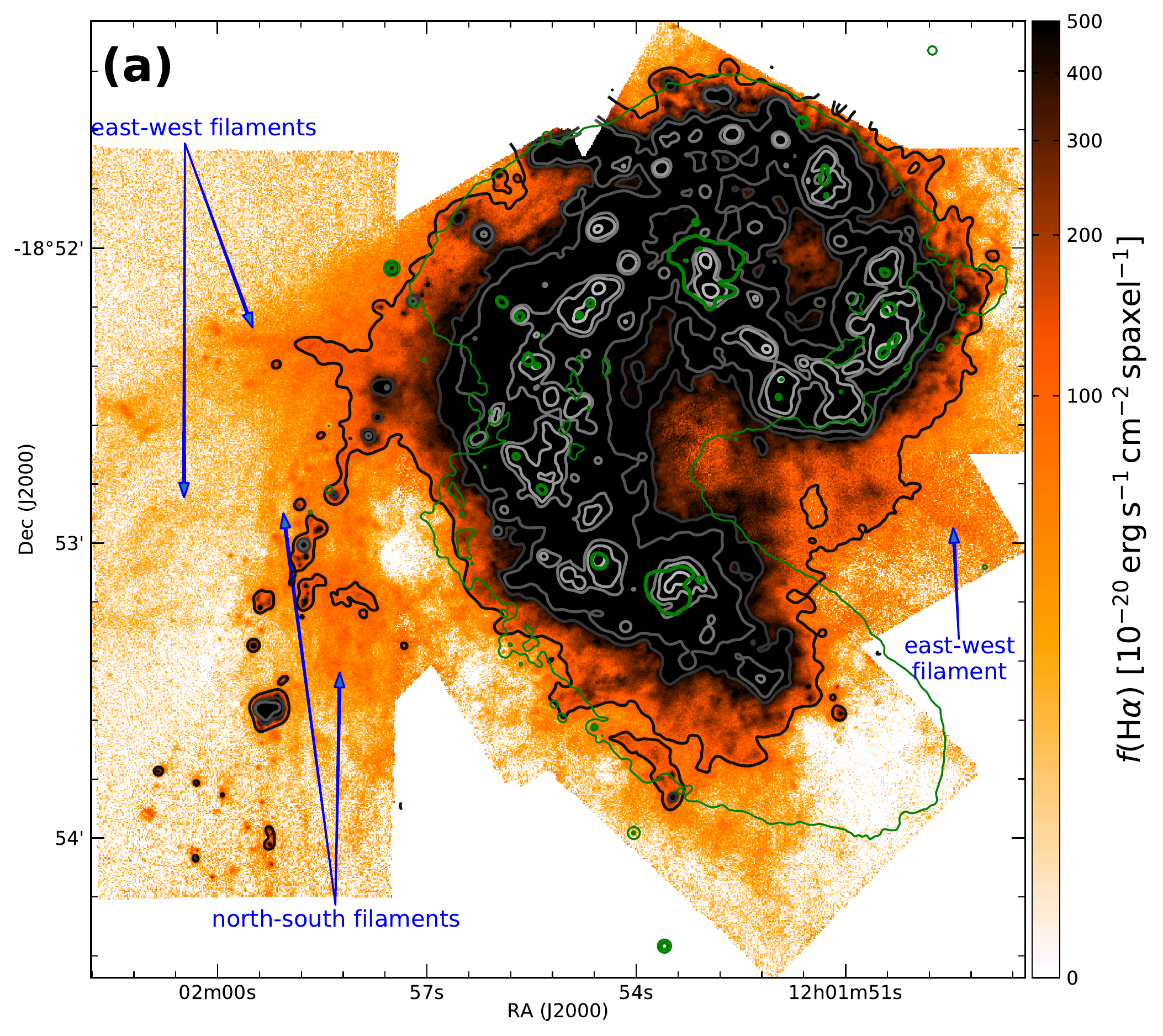}
\includegraphics[trim={120 41 0 0},clip,width=0.512\linewidth]{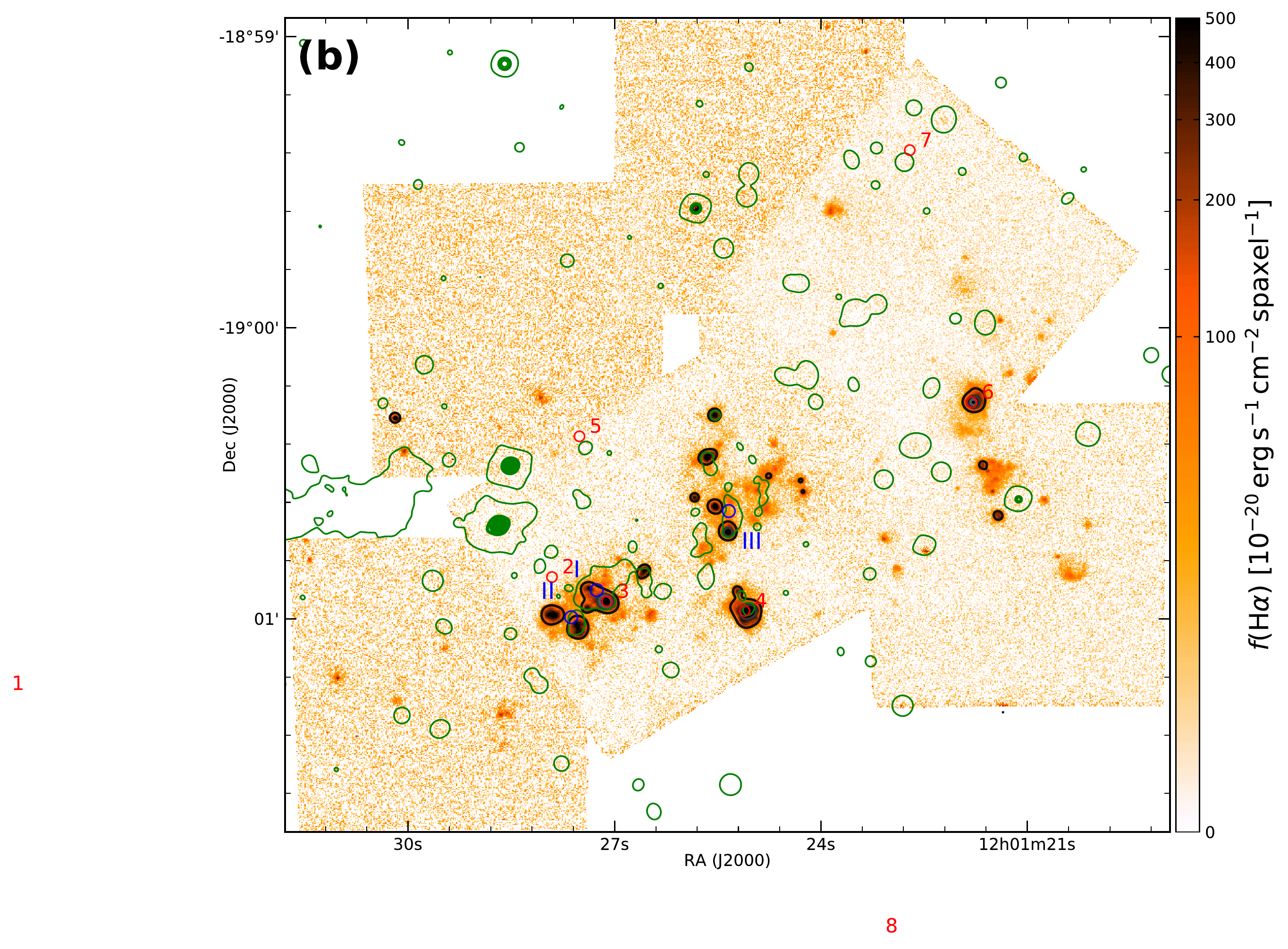}
\caption{Continuum-subtracted \ha flux maps for {\bf (a)} the central and
         {\bf (b)} the southern region, as produced by the narrow-band technique
         (see text).
         The common color scaling (color bar on the right) was chosen to
         highlight faint features. The black-to-gray contours denote the \ha flux levels,
         smoothed by a 3-pixel Gaussian, of
         $2.50\times10^{-17}$, $1.25\times10^{-16}$, $2.50\times10^{-16}$,
         $1.25\times10^{-15}$, $2.50\times10^{-15}$, and
         $1.25\times10^{-14}$\fluxa; since the emission in the southern region
         is fainter, not all contours are visible in panel (b).
         The green contours highlight continuum features and are
         identical to the ones shown in Fig.~\ref{fig:ant}.
         In panel (b), the blue circles and corresponding roman
         numerals denote the detections by \citet{MDL92} with coordinates from
         \citet{HvdH+01}, the red circles and arabic numerals are detections by
         \citet{BDA+04}.}
\label{fig:ha}
\end{figure*}

We created first, simple
continuum-subtracted \ha flux maps from the original cubes employing two
methods:
1.\ We summed the flux in the cubes between 6595 and 6604\,\AA\ and
subtracted the continuum flux averaged over the two wavelength ranges
6552\dots6560\,\AA\ and 6641\dots6649\,\AA. These ranges were tailored to integrate
as much flux as possible of the \ha line at the velocity of the Antennae
without being influenced by the \nii lines. This approach is similar to using
very narrow filters matched to the redshift of the Antennae.
2.\ We ran a single-Gaussian fit over the whole field in the spectral region
around \ha. As a first-guess for the Gaussian profiles we used the systemic
velocity of 1705\,km\,s$^{-1}$ (as listed in NED\footnote{The NASA/IPAC
  Extragalactic Database (NED) is operated by the Jet Propulsion Laboratory,
  California Institute of Technology, under contract with the National
  Aeronautics and Space Administration.}
with reference to the HIPASS survey) and the approximate instrumental FWHM of
2.5\,\AA, the fits used the propagated variance in the datacube.
The latter approach has the advantage of integrating most of the flux of the
\ha line for all spatial positions, independent of the actual gas velocity.
However, it tends to overestimate the flux in regions of very low $S/N$. The
narrow-band approach on the other hand integrates the \ha flux even in places
where the emission line has broad wings or multiple components; it is also
insensitive to the correlated noise in the MUSE datacubes that causes artifacts
in some parts of the field. We present the results of the narrow-band technique
in Fig.~\ref{fig:ha}, while the Gaussian fit result for the central field is
shown in Fig.~\ref{fig:ha_radial}. When accounting for the relative
characteristics, the features visible in both types of maps are the same.

The line detection sensitivity of these simple approaches can be estimated using
the noise near \ha. We did that by measuring the standard deviation across two
empty spectral regions of 6400\dots6570\,\AA\ and 6630\dots6705\,\AA. The
resulting noise is overestimated in places where significant continuum exists
(and hence contains stellar absorption), and it shows a pattern of correlated
noise in several regions (which is normal in MUSE data), especially where
single pointings or dither without rotation dominate the signal.  The typical
$1\sigma$ noise level in regions with 4 overlapping exposures is
$3\times10^{-20}$\fluxspx (variations from 2.4 to 3.6), equivalent to about
$7.5\times10^{-19}$\fluxa. In regions with low-quality data or shorter, single
exposures, the noise can reach $1.6\times10^{-18}$\fluxa.
This can be compared to the $1\sigma$ limit of $4-5\times10^{-18}$\fluxa of
\citet[][their Sect.~4]{2016ApJ...817..177L}, one of the deepest \ha studies of
nearby galaxies to date, using a tunable filter adapted to each object.
Even in the worst case, the MUSE data are still at least $2.5\times$ more
sensitive to \ha emission than the data of Lee et al., and $5\times$ on average.

The resulting flux maps (Fig.~\ref{fig:ha}) show strong \ha emission in the
disks of the merging galaxies. This is highlighted in panel (a) by the light
gray contours and corresponds to the well-known structures detected in previous
narrow-band \citep{WZL99,2005A&A...443...41M} and Fabry-Perot
\citep{AMB+92,2014MNRAS.445.1412Z} observations of the Antennae.  However, in
the much deeper MUSE data, we also detect faint \ha emission around the central
merger, out to the edge of the field covered by the MUSE data. This warm gas is
well visible by eye in the \ha flux map in panel (a) of Fig.~\ref{fig:ha},
beyond the outermost contour.
Simply summing up the detected flux in the complete \ha flux map and within the
$5\times10^{-16}$\fluxa contour -- this corresponds approximately to the
sensitivity limit of the Fabry-Perot data of \citet{2014MNRAS.445.1412Z} --,
suggests that up to 14\% of the \ha flux of the central Antennae that forms the
faint diffuse component has not been detected in previous studies.

The central \ha map shows several noteworthy characteristics in the faint
component.
Everywhere around the \ha emission of high surface brightness (beyond the
$2.5\times10^{-16}$\fluxa contour), one can see filaments
reminiscent of ionized structures visible in edge-on galaxies
\citep[\eg][]{2004AJ....128..674R} and starbursts with outflows
\citep{HDL+95,1996ApJ...462..651L}. Such structures are frequently attributed
to starburst events in the centers that through stellar winds and supernova
explosions give rise to superbubbles and chimneys
\citep{1996AJ....112.2567F,2003A&A...406..493R} and might provide pathways of
low density to allow Lyman-continuum photons to travel into the surroundings
\citep[][\eg in UGC\,5456]{2016ApJ...817..177L}.
Further away, filaments in north-south direction dominate
the MUSE data. These are oriented in the same way as the ridgeline\footnote{
  By ``ridge'' we mean the center of the tidal tail that in Fig.~\ref{fig:ha}
  is highlighted by the outermost \ha contours in the left of the map.
}
of the (southern) tidal tail. These could be related to the stretching of the
material by the tidal forces that form both tidal tails.  A string of bright
\hiiregs can be seen in the same regions.
In the outskirts, away from the edge of the tidal tail, east-west filaments can
be seen as well. A hint of another such feature is marginally visible in the data
to the west of the merger.

In the targeted region near the end of the southern tidal tail, several
\hiiregs are apparent (Fig.~\ref{fig:ha}b). Some of the brighter ones are
surrounded by diffuse emission, but unlike the central region there are large
spatial gaps between the multiple \ha detections. Filamentary \ha of the same
type as around the central region is not visible anywhere in this field.
The regions detected already by \citet[][denoted I, II, and III]{MDL92} can be
associated with some of the brightest regions detected in the MUSE data. We
show the positions as recovered by \citet{HvdH+01}\footnote{We assume that the
  vertical axis of Fig.~6 of \citet{HvdH+01} is supposed to have coordinates
  $-18\degr59\arcmin$, $-19\degr00\arcmin$, $-19\degr01\arcmin$, and
  $-19\degr02\arcmin$, since the 10\arcmin\ distances between the southern
  axis ticks do not make sense.}
in Fig.~\ref{fig:ha}b.
If Fig.~1b of Mirabel et al.\ and Fig.~6 of Hibbard et al.\ are correct, then
the slit of those observations was located just in between close
pairs of \hiiregs. Their 1\farcs5 slit was probably wide enough to integrate
light from both components of each region. However, with that single slit, they
missed the brightest regions, and with EFOSC on the ESO 3.6m telescope they were
not able to detect any of the fainter regions.
Of the compact \ha detections in the comparatively shallow Fabry-Perot data of
\citet[][their Table~A.1 and Fig.~A.8]{BDA+04} only three (3, 4, and 6) can be
matched to something in our data, after correcting their positions by an offset
of 2\farcs9. Two other detections (1 and 8) are outside the field of our data,
three more (2, 5, and 7) are not present in our data, suggesting that they were
spurious sources. None of the other, similarly bright, \hiiregs
(like the Mirabel sources II and III) were picked up in the FP data. The
velocities measured by \citeauthor{BDA+04} are higher than the estimate using our
Gaussian fit by $\sim180$\kms (objects 3 and 4) and $\sim160$\kms (object 6),
far outside the $1\sigma = 3\dots5$\kms measurement error of those bright
regions in our data.

In the following two sections we will characterize the diffuse emission and
the \hiiregs.

\section{Diffuse ionized gas}\label{sec:dig}
\subsection{Verification of the visual appearance}
\begin{figure}
\includegraphics[width=\linewidth]{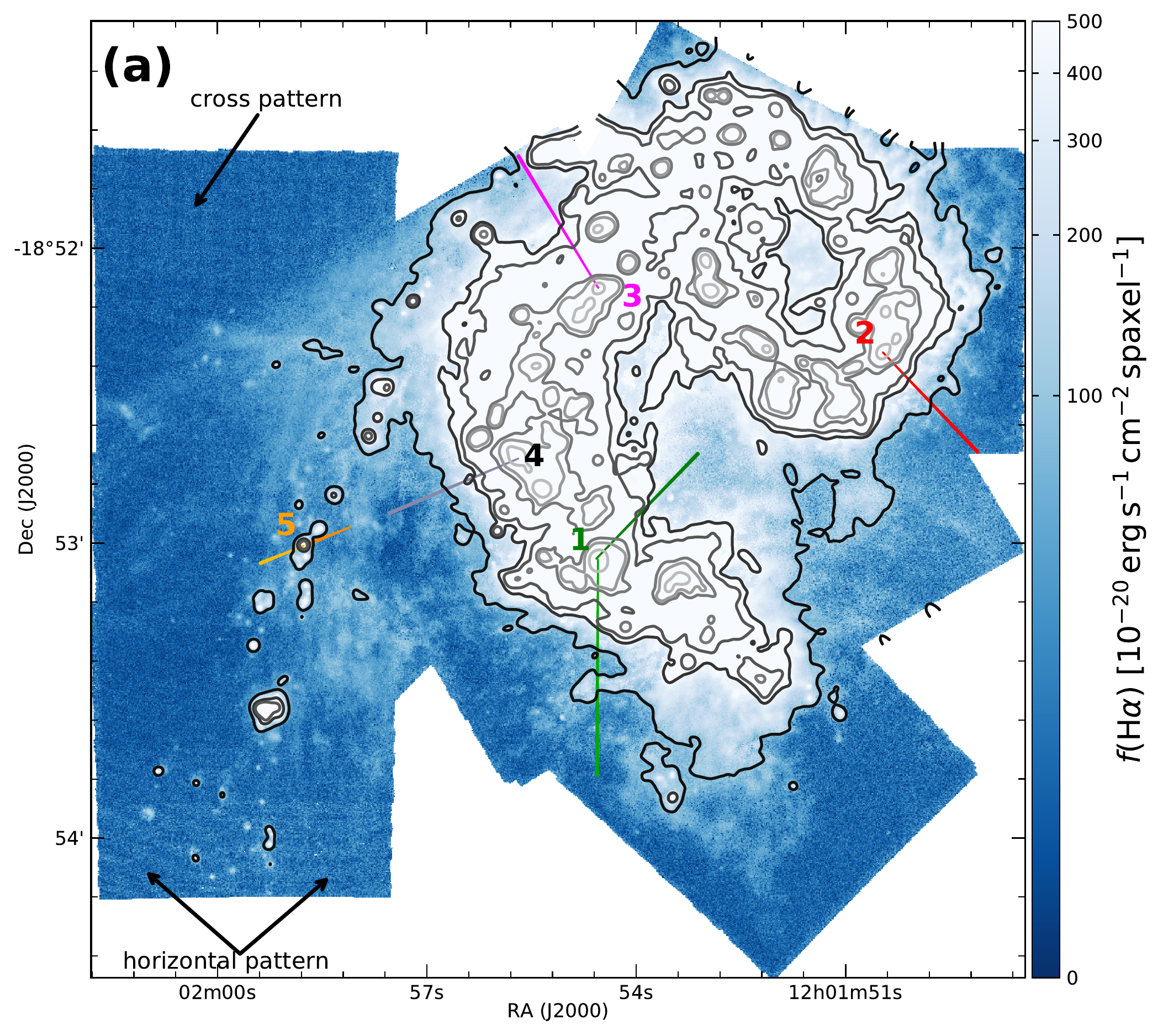}
\includegraphics[width=\linewidth]{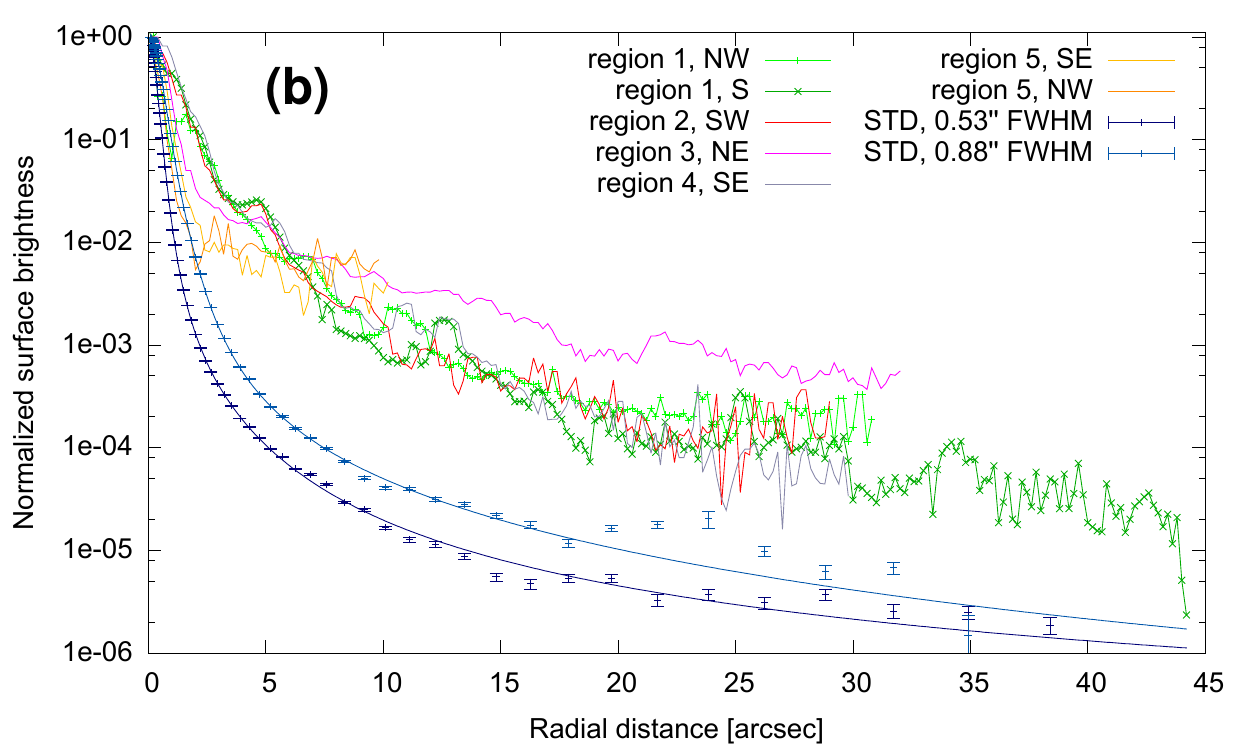}
\caption{Radial profiles of the \ha surface brightness compared to the PSF as
         derived from two typical standard star observations.
         {\bf (a)} \ha flux map from Gaussian line fits to the emission line,
         comparable to Fig.~\ref{fig:ha}a, but with radial extraction cuts
         marked and the corresponding \ha peaks numbered. Low-level patterns
         caused by the correlated noise in the MUSE cubes are marked with
         arrows.
         {\bf (b)} The radial flux distribution outwards from the \hiiregs.
         The markers in the map and the corresponding profiles are shown with
         the same color. The PSFs derived from the standard stars are shown
         as elliptical profiles with error bars and triple Moffat fits.}
\label{fig:ha_radial}
\end{figure}

To verify that the visual appearance in Fig.~\ref{fig:ha}a is correct and that
the diffuse outer \ha detection is not caused by the instrumental plus
atmospheric point spread function \citep[PSF, see discussion
by][]{2014A&A...567A..97S,2015A&A...577A.106S}, we compared the extended
emission to the PSF in two different ways, using a radial extraction of the
data and by convolving the high surface-brightness data with the PSF.

Since the stars in the Antennae MUSE data are too faint to construct a PSF over more
than a 1-2\arcsec\ radius, we used two typical standard stars (LTT\,3218,
observed on 2015-05-21 in 0\farcs88 seeing at $\sim$6600\,\AA, and LTT\,7987, of
2015-05-13, 0\farcs53 FWHM) observed in the same mode as the Antennae data.
These stars are bright and isolated enough to
construct a PSF, using ellipse fitting, out to a radius of at least 30\arcsec.
Both stars were observed in a 4-position dither pattern, reduced as a science frame
with the MUSE pipeline, and each combined into a deeper cube.  The PSF was then
determined on a 2D image averaged from the wavelength range 6570\dots6634\,\AA,
using the {\sc ellipse} task running in the IRAF/STSDAS environment, and
subsequently also fitted with a triple Moffat function. As
comparison, we extracted radial profiles, from the peak of a few of the
brightest \ha peaks toward surface brightness minima of the \ha map. The radial
cuts and the resulting radial profiles are shown in Fig.~\ref{fig:ha_radial}.
The PSF as measured from the standard stars is well constrained out to about
20\arcsec\ radius, \ie over 5 orders of magnitude. Beyond 25\arcsec\ the
variations are significant, since the standard stars are not bright
enough.\footnote{Since the exposure time of the standard stars is set to give
  optimal $S/N$ without saturation, it is not surprising that we did not find any
  brighter, isolated point sources that were observed with MUSE in a similar
  setup and with higher $S/N$ in the outer parts of the PSF. So we cannot
  currently derive any better PSF estimate.}
The extracted \ha profiles in this normalized view are consistently above the
PSF, up to 2 dex higher at 5\arcsec, 1.5 dex at $\sim15$\arcsec\, and still
1 dex higher than the mean PSF at radii approaching 30\arcsec\ and beyond.
This and the presence of small-scale structure indicates that scattered light
has only minor contributions to the extended emission.

As an alternative, we tested the \ha maps that would result when convolving
the bright parts of the \ha emission with a MUSE PSF. We therefore set all
pixels in the \ha map with a flux below $2.5\times10^{-16}$\fluxa to zero and
convolved the resulting image with both PSFs. Unsurprisingly, the resulting
images show a smooth outer appearance, and no structure in the outskirts
of the type that is visible in Fig.~\ref{fig:ha}a. After subtracting the
convolved images from the original \ha map, the features and especially the
filamentary structure in the outskirts are even more enhanced.
Tests with different cut-off levels ($6.25\times10^{-16}$ and
$2.5\times10^{-15}$\fluxa) show that it is not possible to explain the
observed faint features as wings of high surface brightness emission and a
typical MUSE PSF.

\begin{figure}
\includegraphics[width=\linewidth]{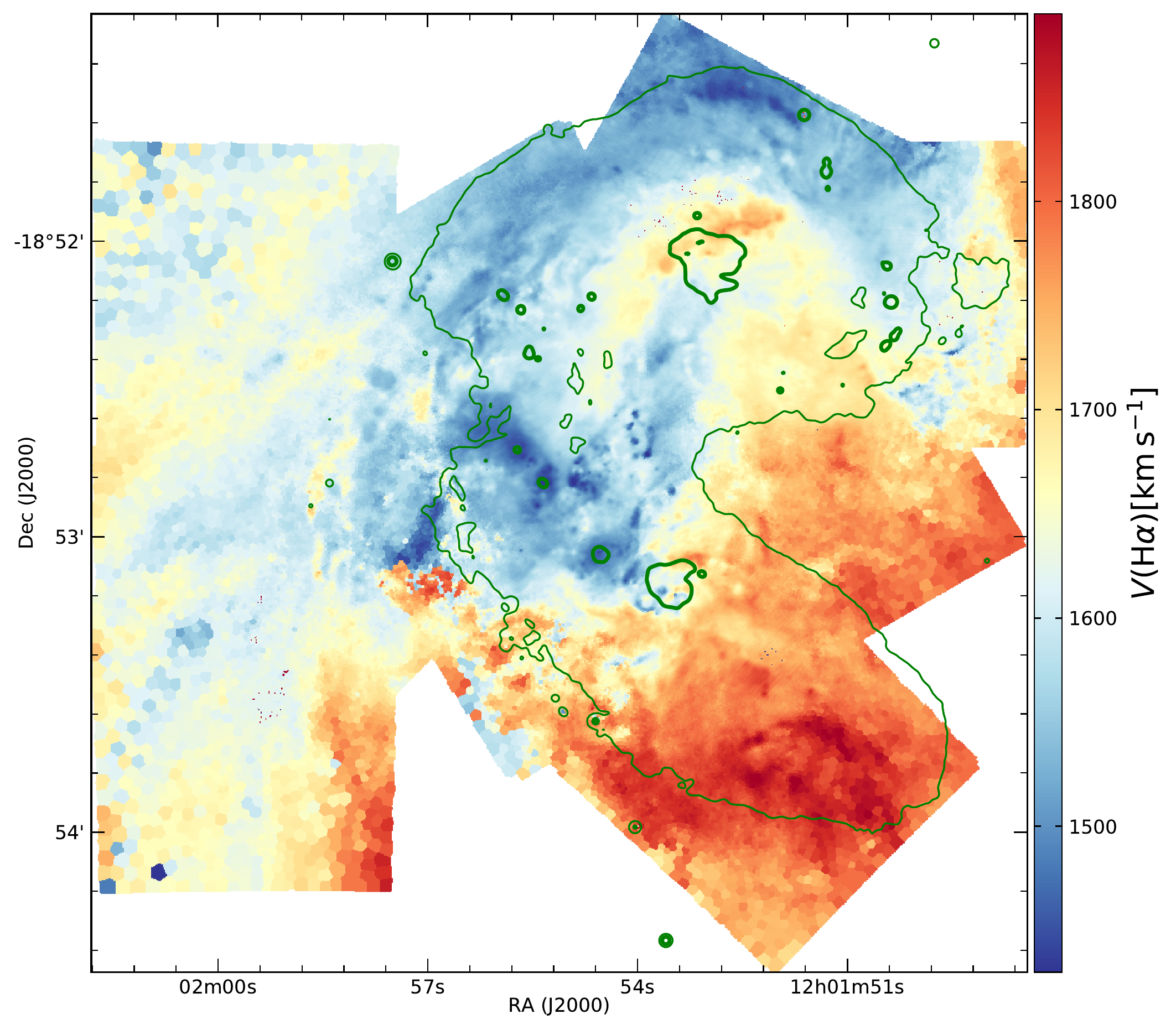}
\caption{Velocity derived from the \ha emission line in the central field of
         the Antennae. The velocities are corrected to the solar system
         barycenter and are computed over bins of $S/N\sim30$ (see text).
         The green lines are the same HST broad-band contours as in
         Fig.~\ref{fig:ant}.}
\label{fig:ha_velo}
\end{figure}

Finally, we looked at the velocities measured from the \ha line.  To derive
them, we employed the {\sc p3d} environment \citep{SBR+10,SWTV+12} to fit a
single Gaussian profile to the \ha line.\footnote{{\sc p3d} has the advantage
  of being able to fit line profiles on log-sampled spectra and can make use of
  Voronoi bins, while being extremely fast compared to other tools like pPXF.
  It is available from \url{http://p3d.sourceforge.net/}.}
We used the continuum-free cube (App.~\ref{sec:ppxf:cube}) binned to a \ha-$S/N$
of 30 as input. The resulting \ha velocity map is shown in
Fig.~\ref{fig:ha_velo}.  This map clearly shows variations of the measured
velocity, also in the outskirts where the faint \ha filaments are detected. If
they were due to scattered light, they would show a smooth distribution of
velocity in radial direction.

We conclude that the faint filamentary structures seen in the \ha emission
line are a real feature of the outskirts of the central Antennae field, and
that scattered light only plays a secondary role.

\subsection{Properties of the diffuse ionized gas}\label{sec:digprop}
Even at the depth of the MUSE spectroscopy, the diffuse ionized gas (DIG) is too
faint for us to derive physical properties with
good spatial resolution. We therefore start measuring spectra of large
integrated regions in the central merger. We divide the data into three surface
brightness levels:
{\it bright} ($\hamath \ge 10^{-16}$\fluxspx), {\it intermediate} ($10^{-17}
\leq \hamath < 10^{-16}$\fluxspx), and {\it faint} ($\hamath <
10^{-17}$\fluxspx). To integrate the spectra we exclude the regions determined
to be \hiiregs in Sect.~\ref{sec:hiiregs} below. We then follow the pPXF analysis
(App.~\ref{sec:ppxf} and \ref{sec:ppxf:hii}) to measure the average emission
line fluxes over these regions. We correct for extinction using the Balmer
decrement, and compute electron densities (using the the \sii6716,31 line
ratio) and temperatures (from the \nii6548,84 to \nii5755 ratio), using PyNeb
\citep[version 1.0.26;][]{PyNeb}.
\begin{table}
\caption{Properties of the diffuse ionized gas}
\label{tab:DIGprop}
\begin{tabular}{l | c cc}
name               & $c_{\hbmath}$ & $n_\mathrm{e}$ & $T_\mathrm{e}$ \\
                   & $^{(1)}$      & [cm$^{-3}$]    & [K]            \\
\hline
{\it bright}       & 0.043         & $53_{-32}^{+38}$ &  $7940_{-120}^{+130}$  \\
{\it intermediate} & 0.066         & $21_{-8}^{+9}$   &  $9290_{-60}^{+90}$    \\
{\it faint}        & 0.000         & $14_{-11}^{+14}$ & $11560_{-760}^{+710}$  \\
\hline
\end{tabular}\\
\footnotesize{
$^1$ Logarithmic extinction at the wavelength of \hb.
}
\end{table}
The results are shown in Table~\ref{tab:DIGprop}.
The errors quoted there are 68\% confidence limits, computed using 100
Monte-Carlo iterations, using the flux measurement errors propagated from the
lines involved in the cross-iteration of both quantities.
The values indicate very low extinction, and subsequently lower densities
and higher temperatures as we go to fainter \ha surface brightness. However,
for all three diffuse spectra, the \sii diagnostic ratio is close to the
low-density limit.

\begin{figure*}
\includegraphics[width=0.33\linewidth]{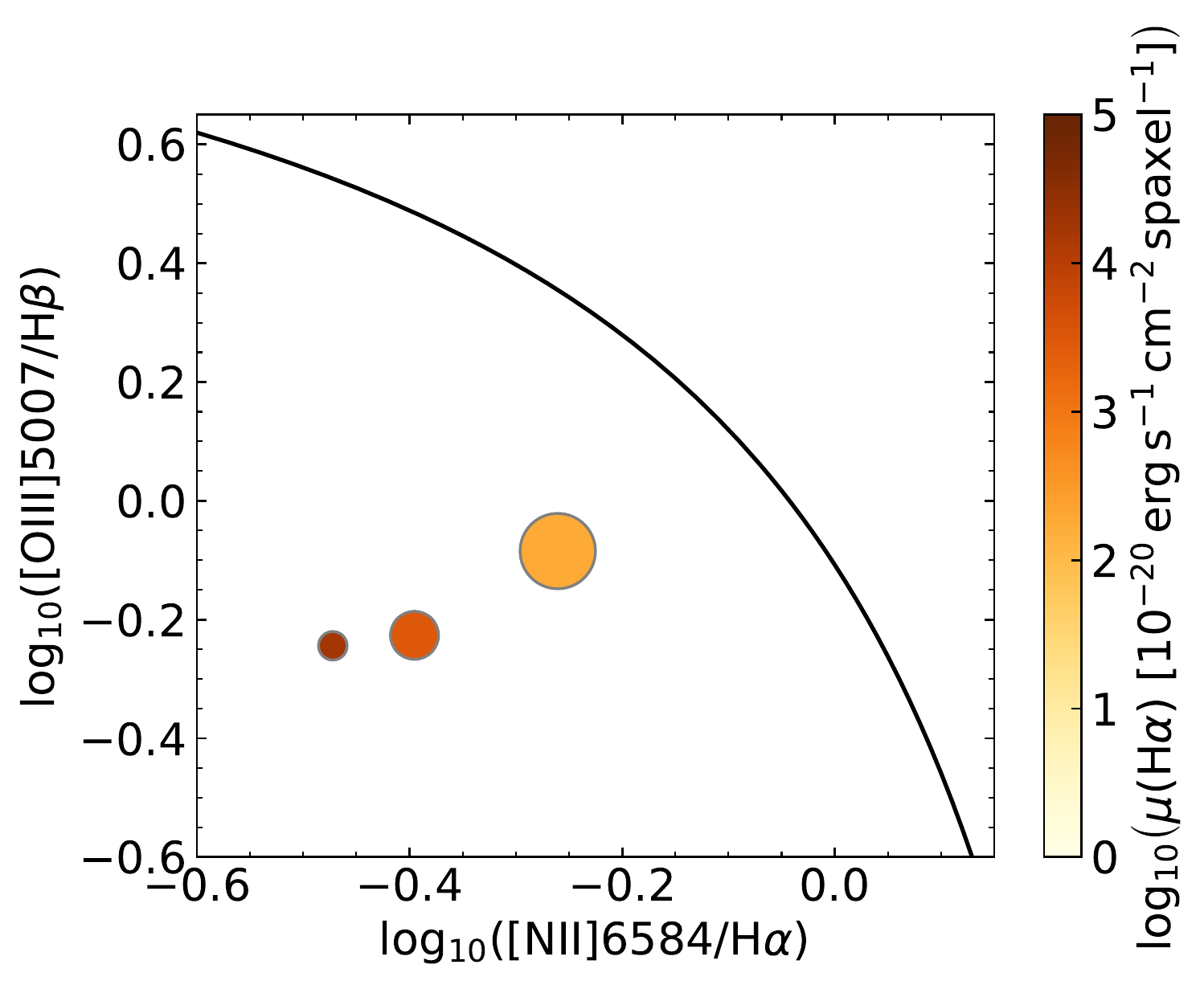}
\includegraphics[width=0.33\linewidth]{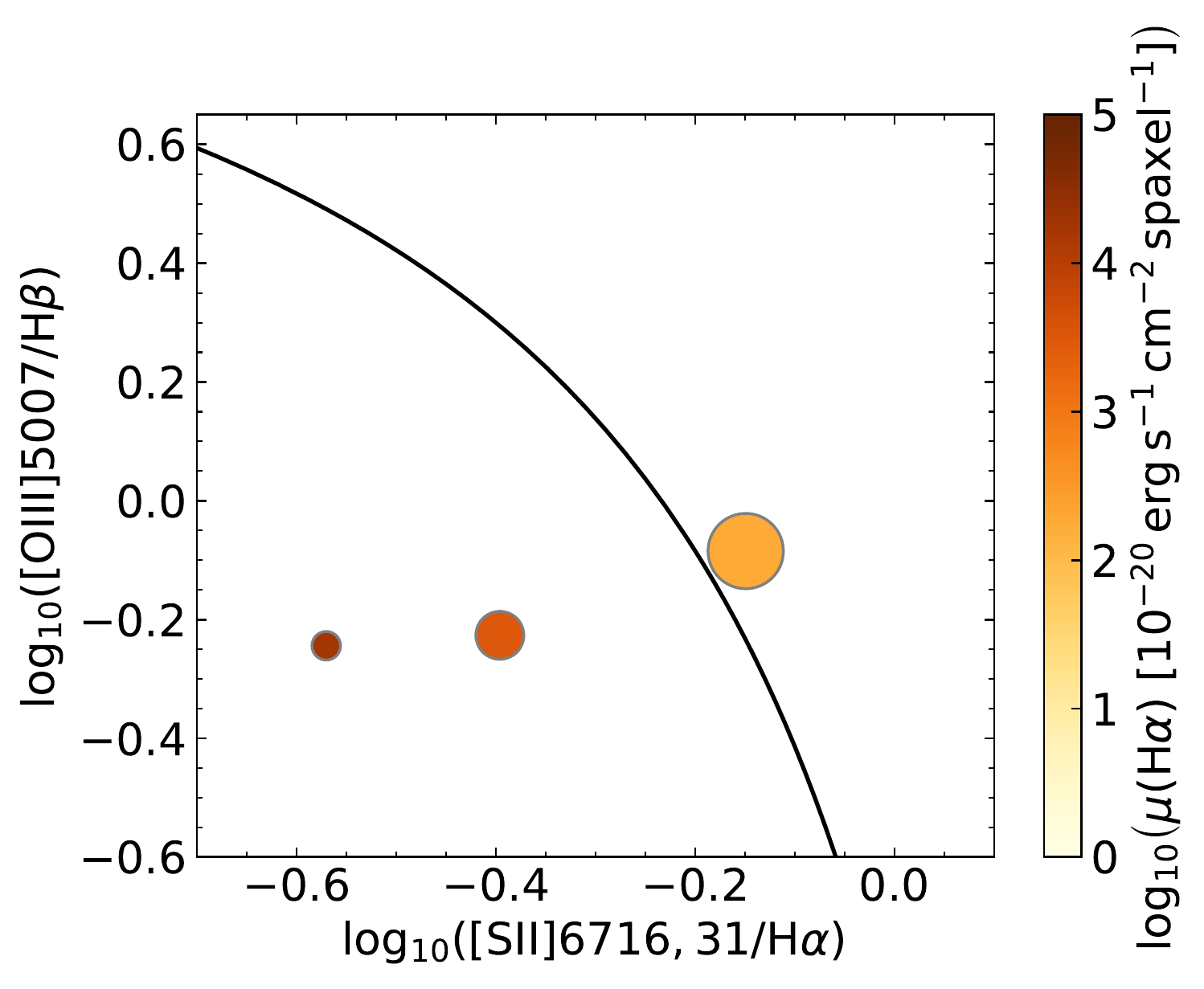}
\includegraphics[width=0.33\linewidth]{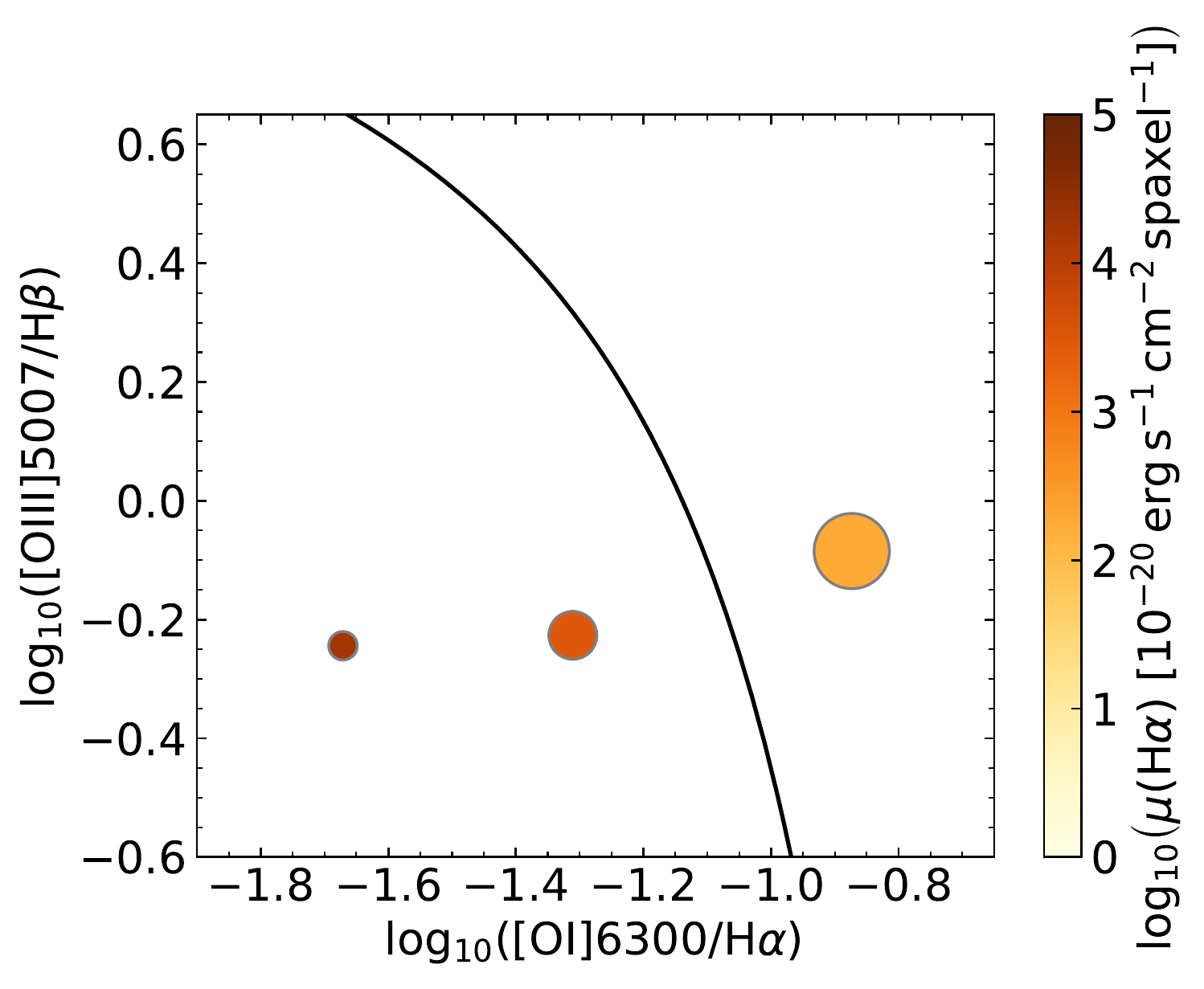}
\caption{Diagnostic diagrams showing the properties of the faint ionized gas.
         The radius of the three data points in each panel scales linearly
         to the square-root of the area, the error bars are smaller than the
         size of the points.
         The color is coded according to the average \ha surface brightness.
         The solid line indicates the nominal photoionization limit of
         \citet{2001ApJ...556..121K}.}
\label{fig:DIG_BPTs}
\end{figure*}

Using these measurements, we can also position the faint \ha emitting gas in
diagnostic diagrams, as presented in Fig.~\ref{fig:DIG_BPTs}. For reference,
we show the extreme photoionization line of \citet{2001ApJ...556..121K}, even
though the mechanism in the DIG may be different.
Except for the faintest level, the data lies within the
\citeauthor{2001ApJ...556..121K} limit, for all three diagnostics.

It is well known that DIG often shows \sii/\ha and \oi/\ha ratios in regions of
the diagnostic diagrams that usually indicate other types of ionization besides
photoionization \citep[\eg][]{2009RvMP...81..969H,2016ApJ...827..103K}.
Several explanations for this are discussed in the literature. Shocks are an
obvious candidate, but they cannot always explain all observed properties
\citep[see \eg][]{2016A&A...585A..79F} at the same time.
\citet{2017MNRAS.466.3217Z} find evolved stars as the most likely candidate
source of the ionizing photons.
\citet{2003ApJ...586..902H} and \citet{2006ApJ...644L..29V} suggest that this
can be caused by ionization of the DIG by leaking \hiiregs where the UV
spectrum is hardened by the absorption inside the ionized nebulae. This then
shifts the line ratio diagnostics beyond the usual photoionization limit.

In the case of the Antennae, the latter suggestion seems to fit the properties
discussed here as well.
Shocks have been reported in the Antennae \citep[\eg][]{1989AJ.....97..995C}
but they seem to be related to starburst activity in the denser regions
\citep{2005A&A...433L..17H}.  The low density of the gas in the outskirts, free
of any density gradients at our spatial resolution as determined by the
\sii6716,31 line ratio, makes it unlikely for the high \oi/\ha and \sii/\ha
ratios there to be caused by shocks.
Our three-zone measurements suggest that the ionizing spectrum is still close
to that of hot stars immediately surrounding the \hiiregs (the {\it bright}
regions). Only in the {\it faint} regions, furthest away from the \hiiregs, the
line ratios are beyond being photoionized, suggesting that the ionizing spectrum is
harder there. This fits well with the models of \citet{2003ApJ...586..902H}, if
we assume that to reach the gas to be ionized, the UV photons have to travel
through even higher column densities of gas, resulting in even harder UV
spectra.\footnote{That leaking LyC photons from \hiiregs contribute to the
  DIG in the Antennae was already mentioned by \citet{WGL+05}, but their
  argument seems to contradict what these photoionization models show.  However,
  they were only referring to DIG immediately surrounding the brightest star
  cluster complexes.}
Although most studies focus on the young stellar populations in the Antennae,
old stars with ages $>10$\,Gyr exist in the disks of both interacting galaxies
\citep{2003AJ....126.1276K}. So it may well be that evolved stars contribute to
the ionization of the DIG, however, as \citet{2017MNRAS.466.3217Z} point out,
they are unlikely to be the dominating source in starburst galaxies. We will
revisit the contribution of old stars in a future publication about the
properties of the stellar populations.
In the present paper, we focus on whether we can actually find evidence for
Lyman-continuum leakage from the star-forming regions (see
Sect.~\ref{sec:hii_leaks}).

\begin{figure*}
\includegraphics[width=0.499\linewidth]{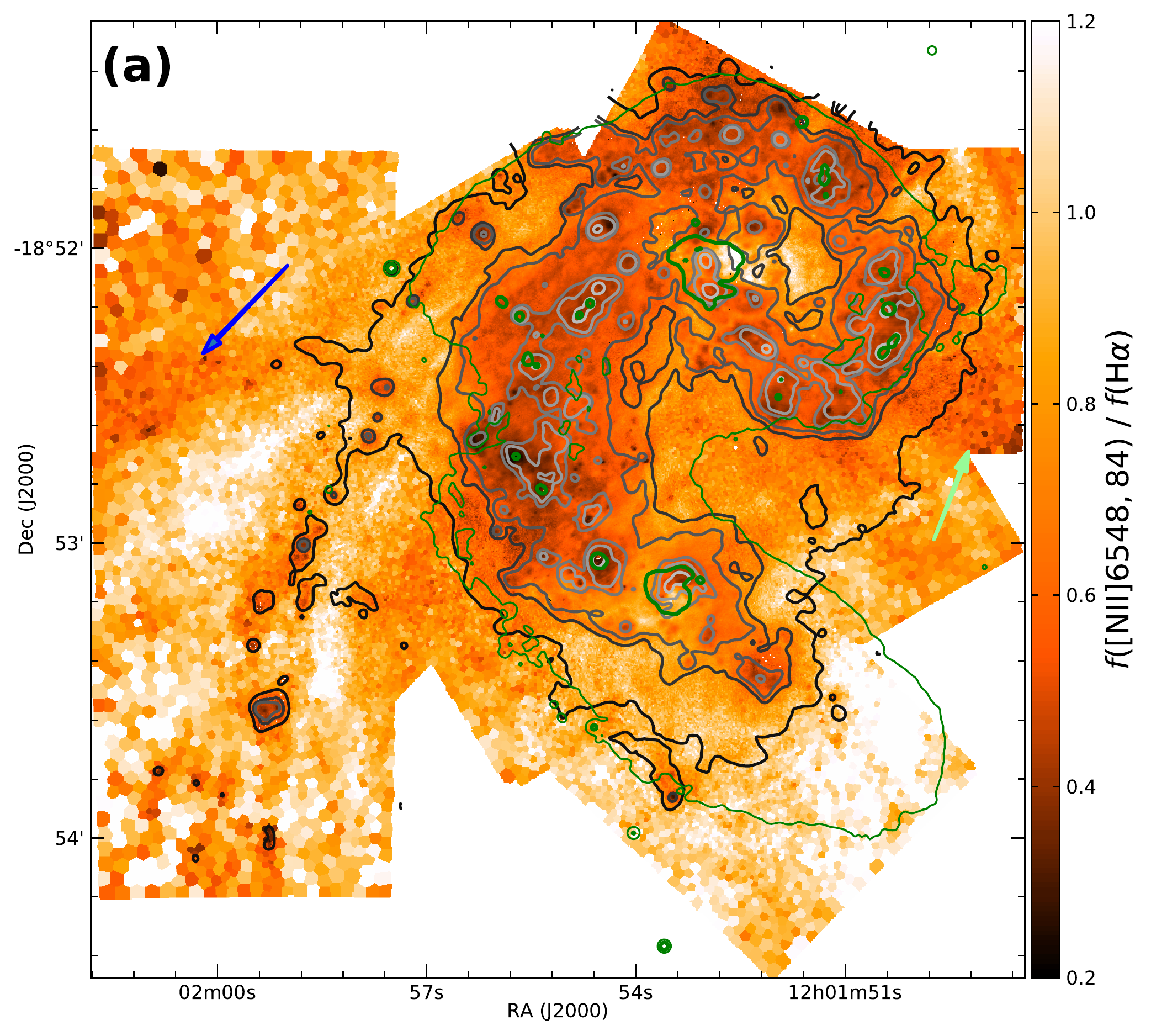}
\includegraphics[width=0.499\linewidth]{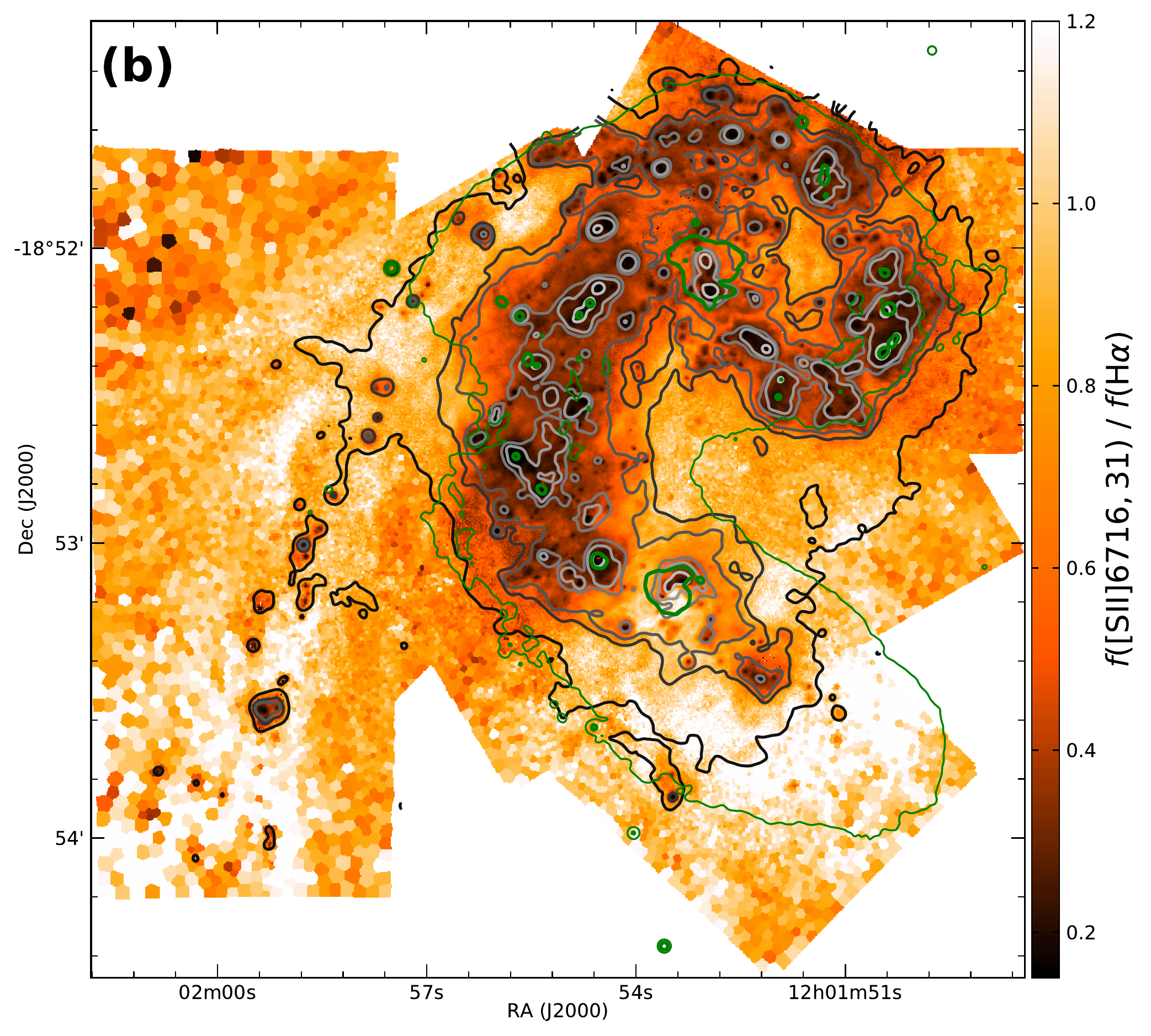}
\caption{Flux ratio maps, {\bf (a)} \nii6548,84/\ha, {\bf (b)} \sii6716,31/\ha.
         Both are Voronoi-binned to $S/N_{\mathrm{H}\alpha}\sim30$. Overplotted
         are the same \ha contours as in Fig.~\ref{fig:ha}a and broad-band
         levels as in Fig.~\ref{fig:ant}. Features discussed in the text are
         marked with arrows.}
\label{fig:ha_div_niisii}
\end{figure*}

In Fig.~\ref{fig:ha_div_niisii}a, we show \nii6548,84/\ha, the line ratio
with the highest $S/N$,
in a spatially resolved manner. This map was computed using
single Gaussians fit to \ha and both \nii lines, using the {\sc p3d} line
fitting tool, on the continuum-free cube discussed in
App.~\ref{sec:ppxf:cube}, after binning this cube to $S/N\approx30$ in
the \ha emission line. The faint ionized gas does not have uniform line ratios
at given surface brightness levels, but an overall trend is visible:
an increase of \nii with regard to \ha for fainter surface
brightness levels of the gas.
The most striking features of this map can be seen in the eastern part,
\ie the region of the tidal tail: along the ridge of the tail and around
the \hiiregs detected there, the \nii line is relatively weak (\nii/\ha$\
\lesssim 0.8$ or $\log_{10}(\nii/\hamath)\lesssim-0.08$). Next to this
ridgeline, however, we see strongly increased \nii emission (\nii/\ha reaches
values above 1.25, or $\log_{10}(\nii/\hamath)\gtrsim0.09$).
In the same way, the south-western edge of the field, in the outer disk of
\ngc39, we also see a strong increase of \nii with respect to \ha.
These features are visible in a similar way in the map of \sii6716,31/\ha
which we show in Fig.~\ref{fig:ha_div_niisii}b.
To the east, around a declination of -18\degr52\farcm5, we see a broad,
somewhat triangular, region with weak \nii (marked in
Fig.~\ref{fig:ha_div_niisii}a with a blue arrow), which lies slightly south of one of
the east-west filaments marked in Fig.~\ref{fig:ha}a. This region coincides
with a region of higher velocity gas as measured from the \ha line
(Fig.~\ref{fig:ha_velo}) but is not remarkable in any way in the \sii/\ha map.
A similar region lies near the border of our field at the western edge, around
a declination of -18\degr52\farcm7 (pointed to by the pale green arrow).
Here, the velocity field suggests the presence of
lower velocity gas with regard to surrounding regions. Both regions counter
the general trend of stronger \nii emission in fainter gas. Given the velocity
difference, it is tempting to think of these as outflows from more central
regions. Since the Antennae do not contain AGN \citep{BSdB+09} and no bright
ionizing source is located near the eastern region, a source of such outflows
remains unclear. The western \nii-strong region lies close to the strong
star-formation sites in the western spiral arm of \ngc38 \citep[regions R, S,
T, see][]{1970ApJ...160..801R} of which region S was as already found by
\citet{2007ApJ...668..168G} to be the source of a (local) outflow from
line-width measurements of the Br$\gamma$ line.

\section{The \hiiregs}\label{sec:hiiregs}
To probe a possible origin of the diffuse ionized gas, we turn to the \hiiregs.
Some of them are already directly visible in Fig.~\ref{fig:ha}.

\begin{figure*}
\includegraphics[width=0.486\linewidth]{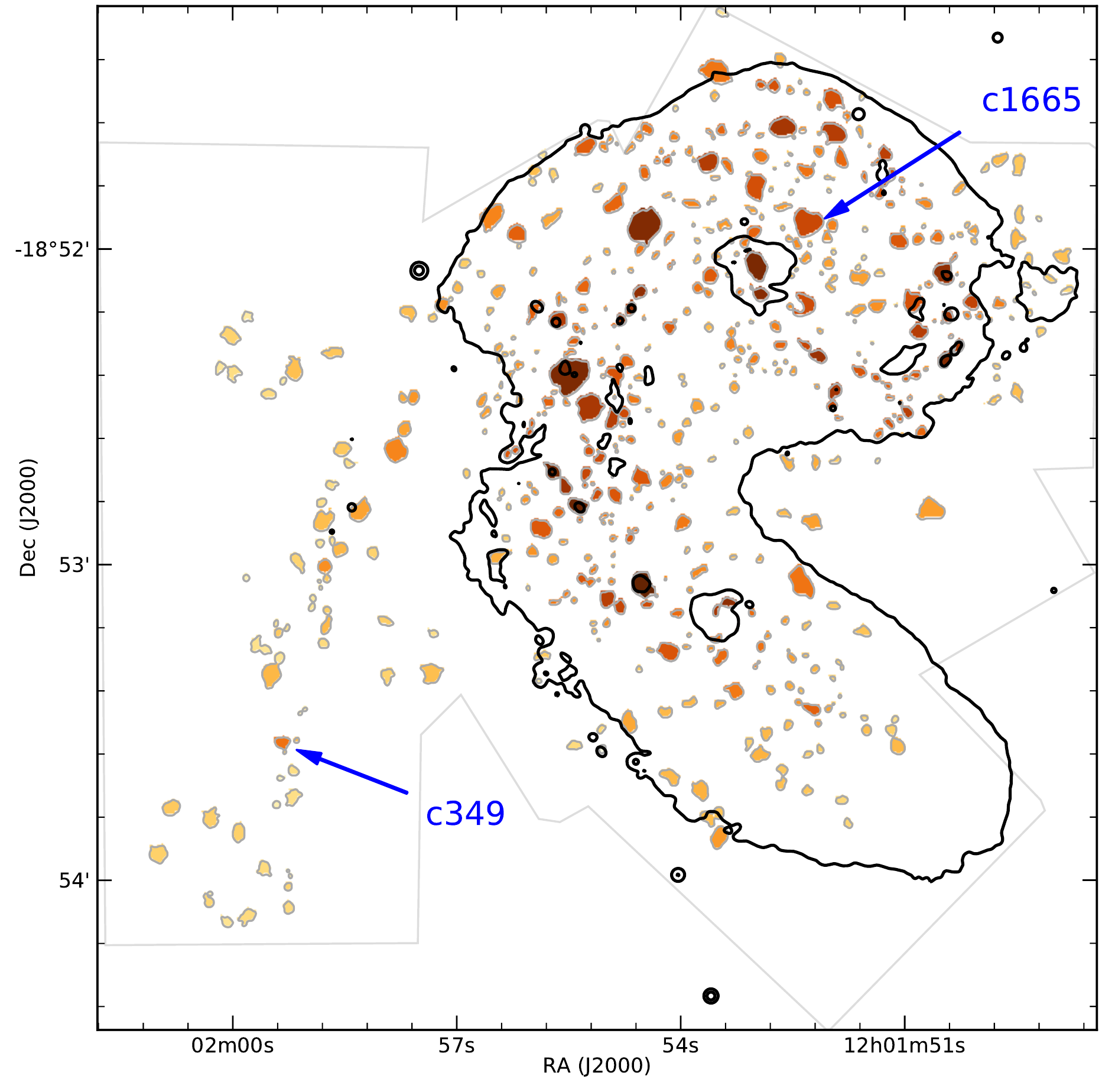}
\includegraphics[width=0.512\linewidth]{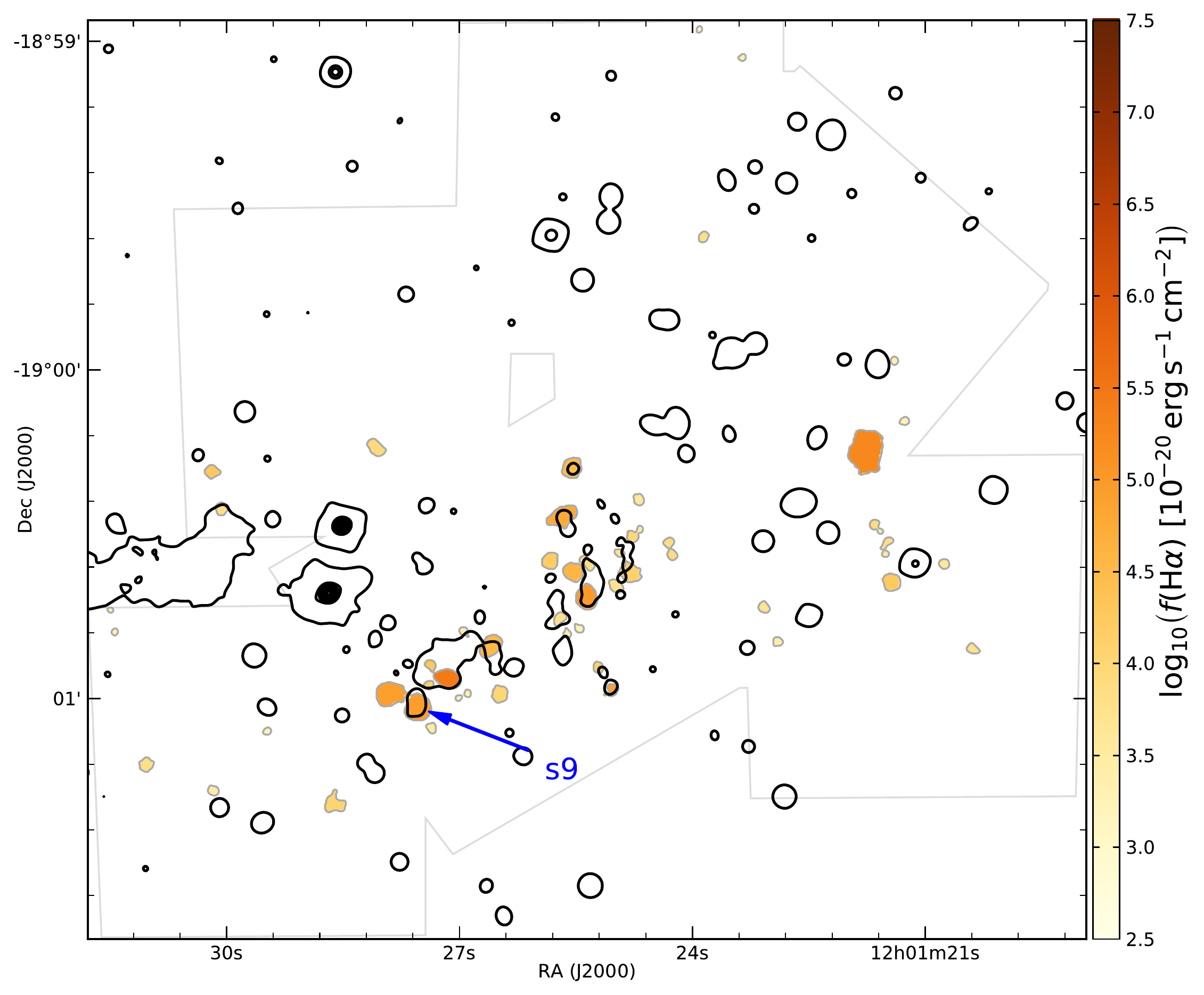}
\caption{\hiiregs detected in the Antennae using the {\sc astrodendro} package.
         The common color scale gives the \ha flux of each region.
         {\bf Left} we show the central and {\bf right} the southern region.
         The contours are the same broad-band levels as in
         Fig.~\ref{fig:ant}.
         \hiiregs whose spectra are displayed in Fig.~\ref{fig:hiispectra}
         are marked.}
\label{fig:hiireg_spatial_flux}
\end{figure*}

\subsection{Spectral extraction}
We extract peaks in the \ha flux maps using the {\sc dendrograms}
tool.\footnote{Available as the {\sc astrodendro} Python package from
  \url{http://dendrograms.org}.}
This tool detects local maxima and extracts them down to surface-brightness
levels where the corresponding contours join. This is repeated in a
hierarchical manner until a global lower limit is reached. The resulting tree
structure can be used to track hierarchical relations between the individual
detections. Here we only use the \emph{leaves} of the structure, i.e.\ the
individual local peaks, which we take as defining the size of the \hiiregs to
extract.
Since contours at surface brightness levels below the limits of the leaves
already encompass multiple peaks, other structures created by the dendrogram
are ignored here.

As input to compute the dendrograms, we use continuum-subtracted \ha images
computed directly from the MUSE cubes, see Sect.~\ref{sec:strct}. To prevent the
algorithm from identifying too many noise peaks, we process regions where only a
single or poor-quality exposure dominated the data -- these are the light-gray
regions visible in Fig.~\ref{fig:fields} -- with a
spatial 3x3 median filter.  We then filter the whole extent of the images with
a 2D Gaussian of 0\farcs6 FWHM to enhance compact sources, and configure {\sc
astrodendro} to find local maxima down to a level
of $2.625\times10^{-19}$\fluxspx and require that all \hiiregs have at least 7
spatially connected pixels.
These parameters result in 42901 dendrogram elements in the central and 1501 in
the southern field. From these, \hiiregs are selected as those leaves which
have a peak level over the background of at least $2.625\times10^{-19}$\fluxspx.
All parameters were found by trial and error, visually checking that both real
peaks in the bright central region and faint ones in the outskirts but no
diffuse regions or noise peaks were detected.
The resulting list comprises 556 for the central field and 63 for the southern
region.  The mask of each leaf is used to extract an average spectrum and
associated variance from the original MUSE cubes.

\subsection{Spectral analysis}\label{sec:hiianal}
To analyze the spectrum of each region, we use pPXF, with the setup described
in App.~\ref{sec:ppxf} and \ref{sec:ppxf:hii}. This gives us emission line
fluxes and error estimates.
Through the stellar population fit, the Balmer lines were corrected for
underlying absorption, with typical EW$(\hamath)$ in the range 1.7\dots2.7\,\AA.
The data for all emission lines are dereddened for further analysis using the
Balmer decrement of 2.86 and the parametrization of the starburst
attenuation curve of \citet{CAB+00}.  We again compute this using the
PyNeb tool. Regions with lower Balmer line ratios are not corrected, spectra
with \ha/\hb$<2$ are discarded (5 and 8 for the central and southern regions,
respectively), most of them are spectra with low $S/N$ or dominated by
foreground stars\footnote{For bright stars, the continuum subtraction in the
  detection image was imperfect, so some of them were false \hiireg detections
  in the dendrogram.}
where the emission line fit does not work well.

\begin{figure}
\centering
\includegraphics[width=0.9\linewidth]{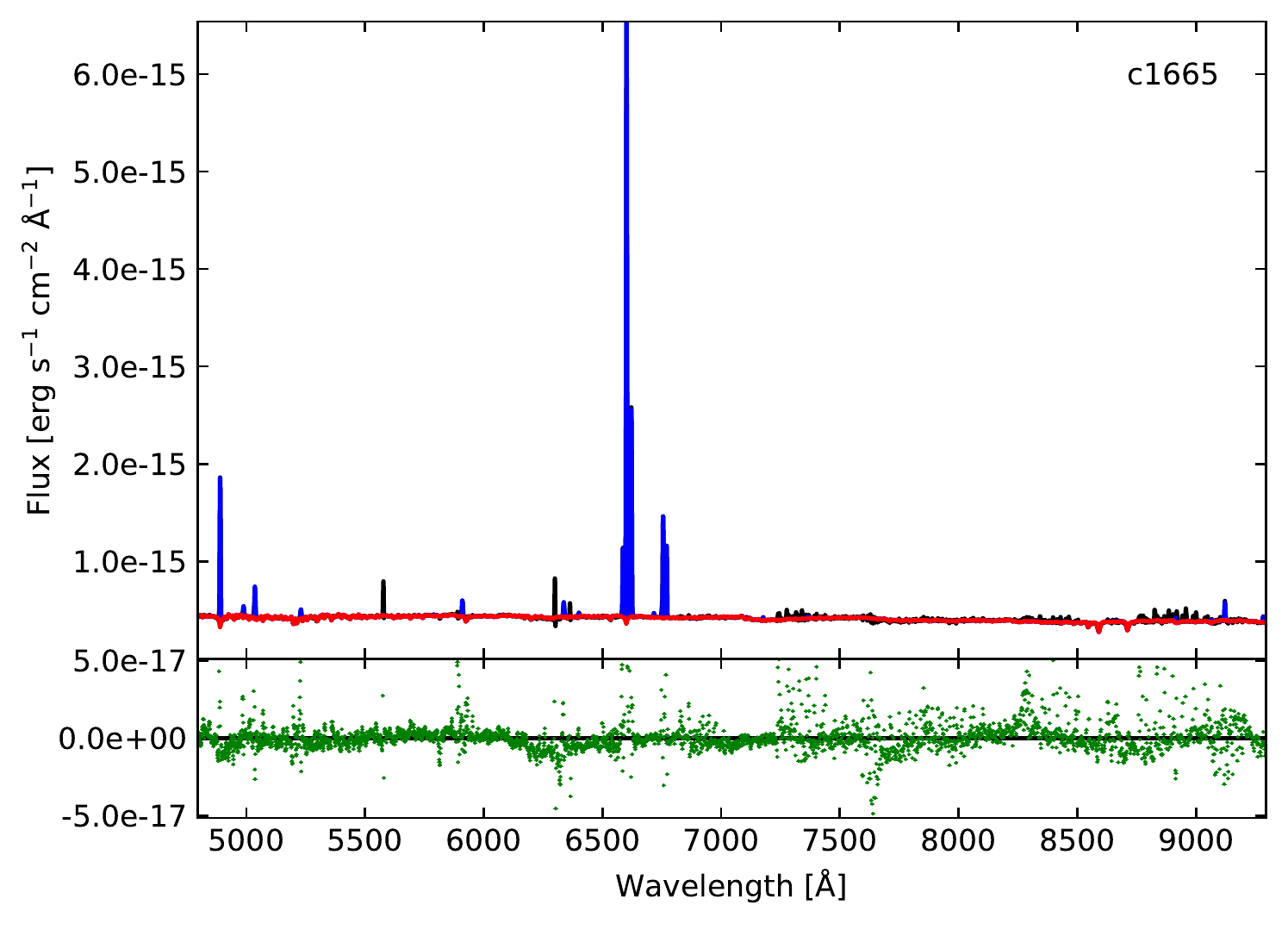}
\includegraphics[width=0.9\linewidth]{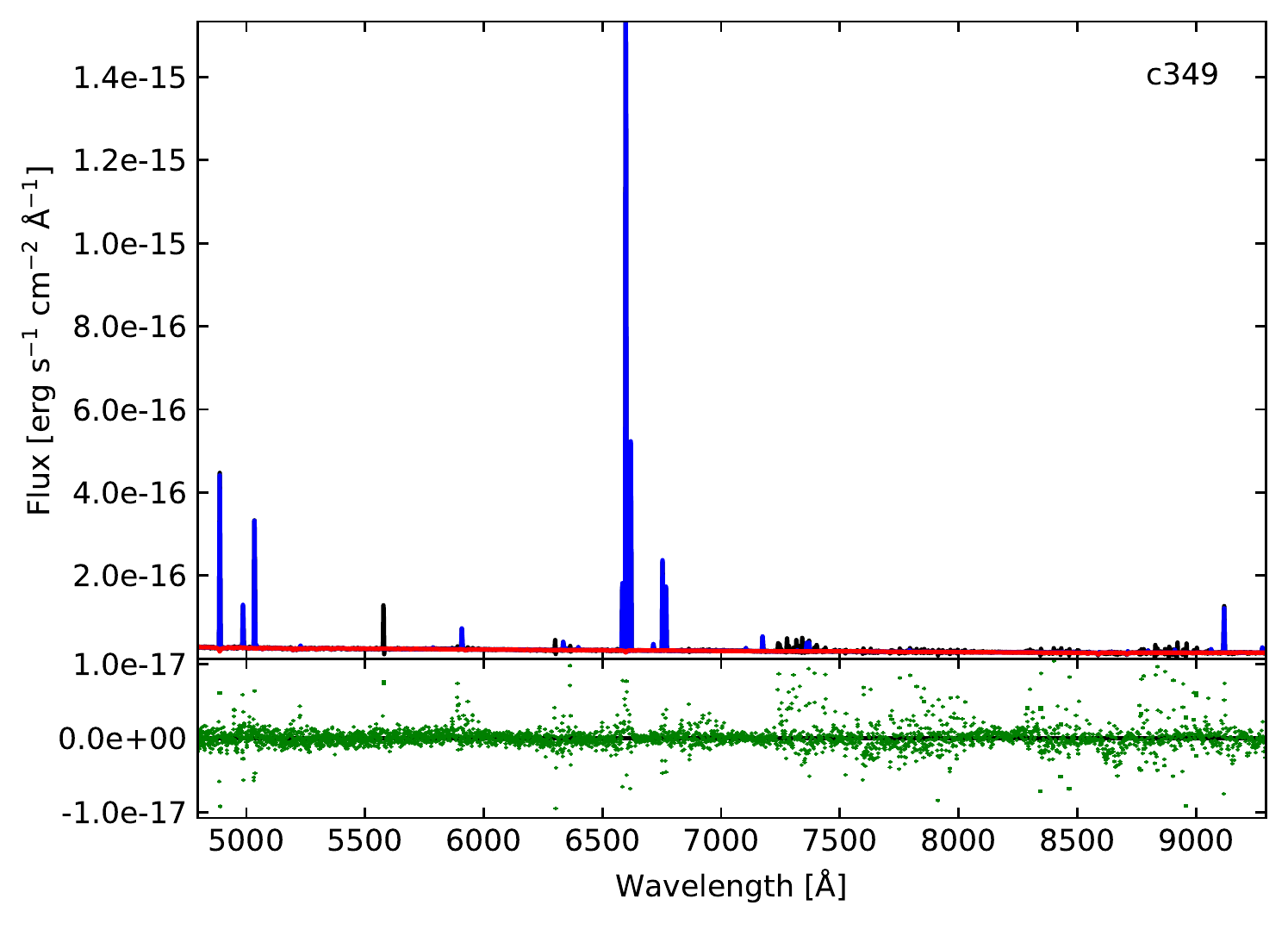}
\includegraphics[width=0.9\linewidth]{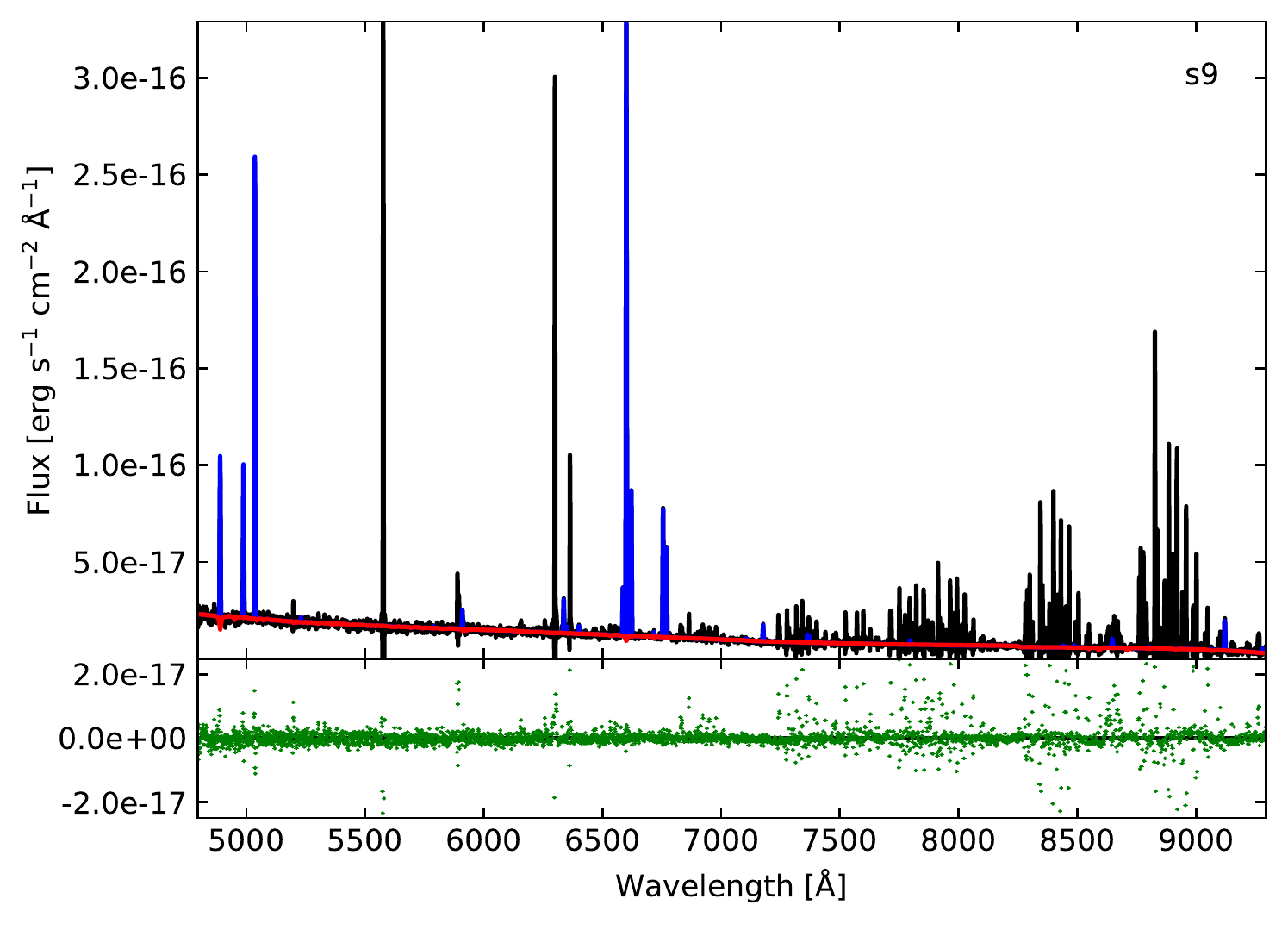}
\caption{Three example spectra of \hiiregs: c1665 is located near the \ngc38
         nucleus, c349 is in the tidal tail of the central field, s9 is in the
         southern field.  The corresponding regions are annotated in
         Fig.~\ref{fig:hiireg_spatial_flux}.
         In each upper panel, the black line shows the extracted data, the red
         line is the continuum fit and the blue lines represent the fit to gas
         emission.  The green points with error bars in the lower panels show
         the residuals of the pPXF fit.}
\label{fig:hiispectra}
\end{figure}

The final sample of \hiiregs therefore consists of 551 in the central and 55 in
the southern field.
In Fig.~\ref{fig:hiispectra} we present three typical spectra for the
\hiiregs that we detect and analyze using the MUSE data; they are also
annotated in Fig.~\ref{fig:hiireg_spatial_flux}. Of these, c1665 is one of the
few regions, which show enough continuum features to give a reliable continuum
fit. s9 is a fainter \hiireg where the sky line residuals become apparent in
the spectrum. However, the lines relevant to this study are located in spectral
regions without bright telluric emission lines, and hence the measurements are
unaffected by these artifacts.

\subsection{Basic properties of the \hiiregs}
\begin{figure}
\includegraphics[width=\linewidth]{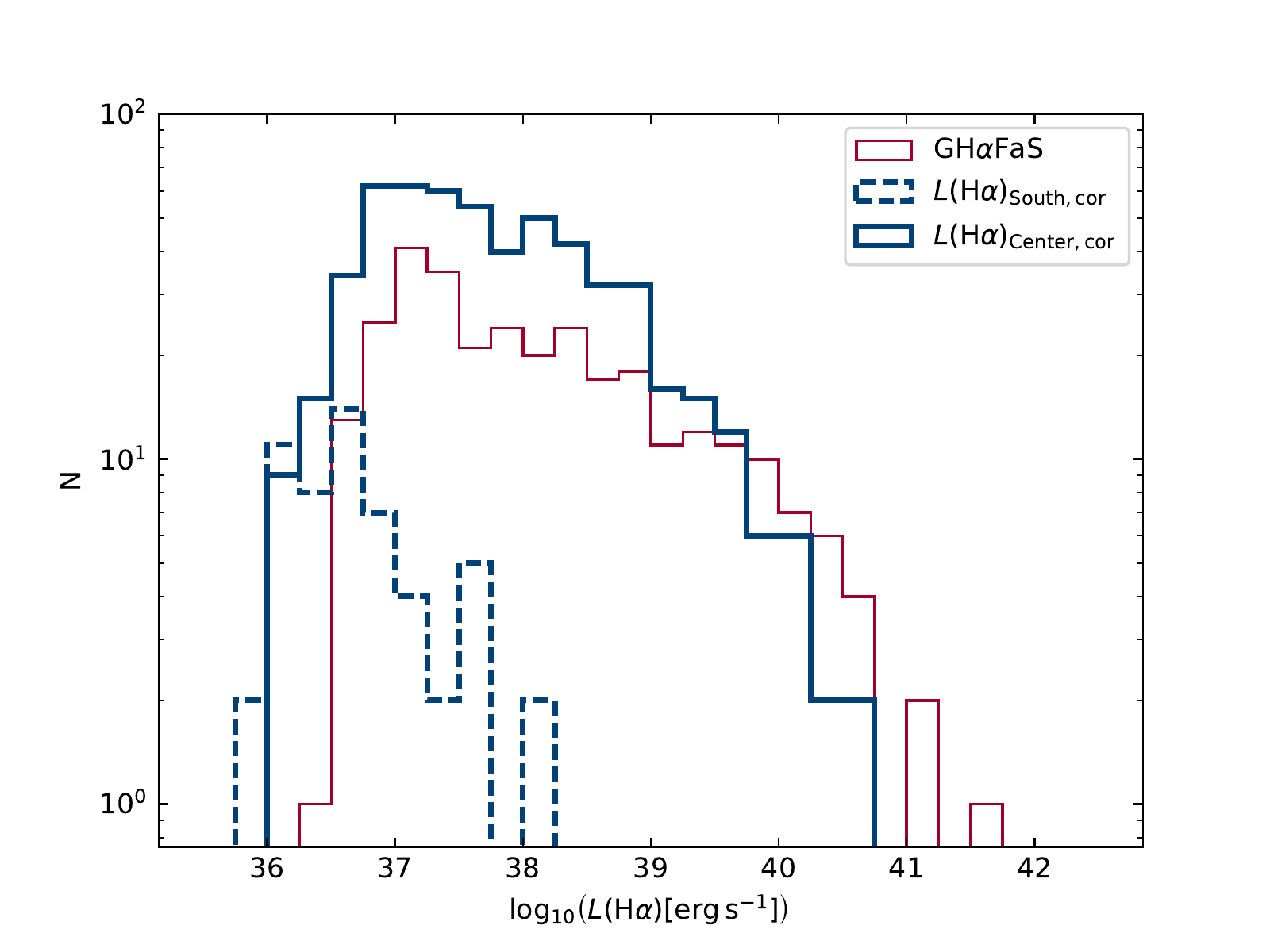}
\caption{The \ha luminosity function for the \hiiregs detected in the Antennae
         using the {\sc astrodendro} package. The bold blue lines (solid: central
         region, dashed: southern field) show the luminosity histogram in our
         MUSE data of the Antennae, after correction for internal extinction.
         The red solid line shows the luminosity function of
         \citet{2014MNRAS.445.1412Z}.}
\label{fig:hiireg_LF}
\end{figure}

A first result of this procedure is displayed in
Fig.~\ref{fig:hiireg_spatial_flux}, where the actual pixels of each extracted
\hiireg are color-coded with the total \ha flux of each region, before
extinction correction. It is apparent that the brighter \hiiregs are located
preferentially in the central part of the merger, and reach up to
$f(\hamath)=4.9\times10^{-13}$\flux, while the outskirts of the interacting center
and the region in the southern tail show only fainter regions,
up to $f(\hamath)=2.7\times10^{-15}$\flux.

We show the \ha luminosity function (LF) of the detected regions in
Fig.~\ref{fig:hiireg_LF}, for the central and southern fields. To derive the
luminosity, we used the reddening-corrected fluxes, computed using the
Balmer decrement, and assumed a distance of 22\,Mpc \citep{2008AJ....136.1482S}.
For comparison, we plot the LF derived from the
table of \hiiregs based on GH$\alpha$FaS Fabry-Perot observations, publicly
released by \citet{2014MNRAS.445.1412Z} who used the same distance. It is
apparent that we detect more regions in the luminosity range $\log_{10}L(\hamath)=36\dots39$ whereas
the GH$\alpha$FaS data shows regions with $\log_{10}L(\hamath)>39.5$.
Their most luminous regions are also detected in our data, but their flux determination is higher by up to one order of magnitude, owing to the extinction
correction (J.~Zaragoza Cardiel, priv.\ comm.). For the same regions we
infer only moderate reddening from
the Balmer decrement. Since our measurements are based on individual Balmer
lines instead of correction through narrow-band filters -- where the fluxes can
be affected by neighboring emission lines and the relative absorption under
each line can lead to an overestimate of the extinction\footnote{The
  equivalent width of the stellar absorption is larger for \hb than \ha. With
  narrow-band filters one cannot correct for this effect, hence underestimates
  the emission line fluxes, more so for \hb. In regions where this absorption is
  significant relative to the emission line, this can lead to an overestimate of
  the extinction.}
--, we believe that our estimates are more realistic.
The difference in the medium to low luminosity range can be explained by the
difference in atmospheric seeing \citep{PHF00,SPE+01}: in worse seeing
conditions, more regions blend with each other and hence form fewer, brighter
apparent regions, while more fainter regions remain undetected.  Since the
seeing in the GH$\alpha$FaS data was $\sim$0\farcs9 while our effective seeing
is around 0\farcs6 this difference is not unexpected. We also have deeper data
and can detect fainter regions, in the range below
$\log_{10}L(\hamath)\lesssim36.5$.  A more definitive \hiireg-LF for the
bright end would require a wide-field IFS with HST-like spatial resolution.
We also note that ongoing work on MUSE data from the nearby galaxy NGC\,300
(distance 1.87\,Mpc) is revealing even fainter compact \hiiregs (Roth et al.,
in prep.), below the detection threshold of our Antennae data.

We see again that the LF in the southern tidal tail is devoid of bright
\hiiregs. While the most luminous region in the central part reaches
$L(\hamath)_\mathrm{cor} = 3.9\times10^{40}$\lum, in the southern part we find only
$L(\hamath)_\mathrm{cor} = 1.6\times10^{38}$\lum.
To investigate, if this is just an effect of the different sizes of the two
samples, we randomly drew one million samples of 55 \hiiregs from the 551
regions in the central merger. All of these samples contained at least three
regions with $L(\hamath)_\mathrm{cor} > 1.6\times10^{38}$\lum.
Despite the small number of detections in the southern region, we also notice
that the slope of both LFs is different: while we detect a similar number of
regions in the histogram bin of $\log_{10}L(\hamath)_\mathrm{cor}=36.125$, the
central histogram shows a strong increase up to a turnover at
$\log_{10}L(\hamath)_\mathrm{cor}\sim37$. The numbers in the southern bins on
the other hand decrease almost monotonically to the maximum lumonosity of
$\log_{10}L(\hamath)_\mathrm{cor}=38.2$, without any turnover.
Using the same random sampling we find that the numbers of \hiiregs in the
southern field in the luminosity bins at 36.125, 36.375, and 36.625 are
$11.3\sigma$, $5.7\sigma$, and $6.3\sigma$ outside the expected range, if drawn
from the same population as the \hiiregs in the central field.
We conclude that the \hiireg samples in the central and the southern fields are
of intrinsically different luminosity distribution.

\subsection{Diffuse gas fraction}\label{sec:difffrac}
We can now compare the \ha flux measured in the \hiiregs with the flux
elsewhere to derive the fraction of the diffuse ionized gas in the Antennae.
The flux of all \hiiregs is the sum of the flux inside the masks of the
dendrogram leaves that are used as \hiiregs. By inverting the mask, we derive
the flux of the diffuse gas.

\begin{table*}
\caption{Integrated \ha fluxes in the different components (in units of \flux).}
\label{tab:fluxes}
\begin{tabular}{l | cccr | cccr}
estimate       & \multicolumn{4}{c|}{Center}              &  \multicolumn{4}{c}{South}       \\
               & total    & HII      & DIG      & \multicolumn{1}{c|}{$f_\mathrm{DIG}$}
                    &  total    & HII      & DIG      & \multicolumn{1}{c}{$f_\mathrm{DIG}$} \\
\hline
{\tt masking}  & 8.35\nexp{12} & 3.33\nexp{12} & 5.02\nexp{12} & 60.2\%  &  1.71\nexp{14} & 1.53\nexp{14} & 1.78\nexp{15} & 10.5\% \\[0.3em]
{\tt spectral} &9.36\nexp{12}& 3.79\nexp{12} &5.57\nexp{12} & 59.5\%  & [1.39\nexp{14}]& 1.22\nexp{14} & [{\tt masking}] &[12.8\%] \\
{\tt speccor}  &1.33\nexp{11}& 7.31\nexp{12} &6.02\nexp{12} & 45.2\%  & [1.42\nexp{14}]& 1.25\nexp{14} & [{\tt masking}] &[12.5\%] \\[0.3em]
{\tt specsub}  &[{\tt spectral}]&2.75\nexp{12} &[6.61\nexp{12}]&[70.7\%]&[1.39\nexp{14}]& 1.01\nexp{14} &[3.82\nexp{15}]&[27.4\%] \\
\hline
\end{tabular}
\footnotesize{
Values in square brackets denote values that could not be derived directly but
are extrapolated in a way. If a row name appears in brackets, then that value
was used to compute the fraction. See text for details.
}
\end{table*}

In the simplest and most consistent way, we apply both masks on the narrow-band
continuum-subtracted image of both fields as created at the beginning of this
investigation (Sect.~\ref{sec:strct}). This yields the fluxes presented in row {\tt
masking} in Table~\ref{tab:fluxes} from which we derive the most direct estimate
of the diffuse gas fraction of $\sim$60\% for the central merger and $\sim$10\%
for the southern field.\footnote{Note that
  since the dominant fraction of the southern field is blank sky and contains
  neither clumpy nor diffuse \ha line emission, the value given for this field
  depends strongly on how well the sky could be subtracted in the \ha wavelength
  region as well as on the level of artifacts left after continuum subtraction.
  The diffuse flux estimate and hence diffuse fraction therefore comes with
  a significant systematic error and should be understood as $10\pm5$\%.
}

For the {\em central field}, we can derive the integrated flux using
alternative approaches, using the spectra that we analyzed as detailed in
Sect.~\ref{sec:hiianal} and \ref{sec:digprop}.  Summing the flux over all
extracted spectra -- once for all \hiireg measurements, and once for the
three integrated spectra of the DIG, both before extinction correction --
gives the value in row {\tt spectral}. This gives a comparable value for the
central field, with about 60\% DIG fraction.

To relate that to the amount of Lyman-continuum photons available in the
\hiiregs, we also look at the \ha flux after extinction correction. Those
measurements are given in table row {\tt speccor}.  Since the extinction in
parts of the central merger is known to be high
\citep[\eg][]{2005A&A...443...41M,WCS+10} -- a property that we can confirm
with our measurements --, this strongly affects the estimate of the total \ha
flux. The corrected \hiireg flux is twice as high as without the correction. On
the other hand, the extinction in the diffuse gas is very low, so the
integrated flux in the DIG remains comparable. This results in a significantly
lower {\tt speccor} estimate of $f_\mathrm{DIG}\approx45\%$.

Since the \hiireg measurements are affected by the underlying diffuse component
\citep[see \eg][]{2016ApJ...827..103K}, we also performed a check using a spectral
extraction routine that aims at subtracting the surrounding background around
each region. For this, mask dilation is used to define a gap of approximately 2
pixels and then a background annulus with a width of about 3 pixels. This results in
background regions that typically have the same sizes as the areas of the
\hiiregs. To minimize the influence of other nearby \ha peaks, we subtract the
median spectrum over this background annulus. However, for many regions this
subtraction does not work very well, so that the resulting spectral properties
vary in unphysical ways. We therefore only use this spectral extraction as a
cross-check, for the sample as a whole.
The integrated \hiireg flux estimated from this extraction is given in table
row {\tt specsub}. It is about 30\% smaller than the {\tt spectral}
estimate. We then add this difference to the {\tt spectral} DIG estimate and
assume that the total \ha flux has not changed. This results in a moderate
change to a $f_\mathrm{DIG}\sim70\%$.

For the \emph{southern region}, the spectral estimates cannot be done in the same way.
Integrating the data over everything outside the \hiiregs cannot work, since
most of the data is filled either with residual noise, background galaxies, or
foreground stars. Such a procedure would therefore need extensive manual
editing of a spatial mask and rely on the information from the {\tt masking}
approach. To nevertheless give similar estimates for the DIG fraction, we correct
the total flux by the difference in \hiireg flux, and re-use the
DIG flux from the {\tt masking} technique.
For the southern field, we then estimate a {\tt spectral} DIG fraction of
about 13\%. Since the extinction is very low in most \hiiregs in that field,
the {\tt speccor} fraction is very similar, roughly 12\%. Both approximately
agree with the basic {\tt masking} estimate. For {\tt specsub} we assign the
\ha flux that the \hiiregs lost by subtracting the surrounding flux to the DIG
which then results in a significantly higher $f_\mathrm{DIG}\sim27\%$.

To summarize, the fraction of diffuse \ha emission in the Antennae
is about 60\% for our central field and about 10\% for the southern region.
After trying to correct for diffuse emission underlying the \hiiregs, we estimate
even higher DIG fractions of 70\% and 27\% for the central and southern fields,
respectively.

\subsection{Search for leaking \hiiregs}\label{sec:hii_leaks}
As the observations of \citet{HvdH+01} have shown, there is about
$5$\pexp{9}\,M$_\odot$ in atomic hydrogen available in the Antennae, that would
be ionized if a sufficient number of Lyman continuum (LyC) photons escaped
from the star-forming regions and
young star clusters. Indeed, \citeauthor{HvdH+01} suggested exactly this as
explanation for the gap at the base of the northern tidal tail that is visible
in the \hi data, but not covered by the MUSE data.
Using our sample of \hiiregs we therefore want to check which fraction of the
diffuse gas that is detected in our data can be due to leaking LyC photons from
those star-forming regions.

We have estimated that after correction for extinction, 45\% of the \ha flux is
detected as diffuse gas in the central field and 12\% in the southern field.
These values should not be compared to LyC escape fractions of {\em galaxies},
which are typically estimated to be on the order of a few
percent \citep{2013A&A...553A.106L,2016ApJ...823...64L}.
Instead, we need enough photons escaping from the {\em \hiiregs} into
the surrounding medium that is still bound to the interacting galaxies and
strongest in the high-surface brightness parts of the object, \ie close to the
star-forming regions themselves. And for \hiiregs in nearby galaxies, several
studies have found significant LyC escape fractions of 50\% and more for some
of the most luminous regions \citep[][]{2008AJ....135.1291V,2012ApJ...755...40P}.
Whether the galaxy as a whole is then a LyC leaking object depends on the
fraction of LyC radiation consumed to ionize the interstellar gas.

Our spectral fits using pPXF do not constrain the LyC
flux. These fits were only
created to approximate the absorption below the Balmer lines and hence take out the
continuum slope using polynomials (see App.~\ref{sec:ppxf}). We postpone a
full assessment of the stellar population using the MUSE data to a future
publication.
Instead, we compare to the analysis of the young star clusters by \citet{WCS+10}
who used broad and medium-band HST data to estimate masses and ages of all
compact sources that were detected with high enough $S/N$. This analysis is based
on several colors and does not rely on the identification of the spectral type
of single stars as in the NGC\,300 work of \citet{2016A&A...592A..47N}, who
find that when assessing the stellar sources inside \hiiregs just from stellar
broad-band data constraining the escape fraction has a high degree of
uncertainty. The \citeauthor{WCS+10} analysis fits stellar populations to
the multi-band ($U$ to $I$ and \ha) cluster data.
The broad-band filter selection agrees with the recommendations of
\citet{ABFA+03} and \citet{2010A&A...521A..22F} for the analysis of young
clusters and provides a solid base for the analysis.
Since the estimated masses of the young clusters with the strongest LyC flux
are above $10^4$\,M$_\odot$, and often beyond $10^6$\,M$_\odot$, the initial
mass function should be well-sampled, making the estimate of $Q(\mathrm{H}^0)$
relatively insensitive to effects of stochasticity \citep{2013ApJ...778..138A}.
Additionally, their use of the F658N filter effectively breaks the
age-extinction degeneracy \citep[see][]{2012ApJ...750...60F} and assigns
realistic ages, especially for young ages $<10$\,Myr which are important for
our analysis.
However, since it is a priori unclear which parts of the \hiiregs are
ionized by a given cluster and \citeauthor{WCS+10} only integrated the fraction
of \ha flux within the cluster aperture, the F658N filter cannot strictly help
to distinguish between different young ages, as long as \ha emission is
present. Another issue is that no UV data shortward of the F336W filter exists
with sufficient spatial resolution over the full field; this would otherwise
have better constrained the continuum slope and hence the age of the youngest
population.  The contribution of emission lines to broad-band filters is
another source of uncertainty \citep{AFA03}; these emission lines might mask
the appearance of the 4000\,\AA\ break in populations older than a few Myr and
cause the mass to be overestimated.
\citet{2012ApJ...750...60F}
thoroughly investigated the systematics involved in the determination of
cluster populations from $U,B,V\!,I$,\,\ha filters, and found a standard
deviation of 0.14\,dex for the age determination. They also note that clusters
with ages younger than $\sim$5\,Myr tend to be assigned older ages.
\citet{2016ApJ...824...71C}, who used the same analysis technique and filter
set as \citet{WCS+10}, estimated the $1\sigma$ error bar on the age and mass
estimates to be 0.3\,dex.
Since the LyC flux of a stellar population evolves nonlinearly, the
$\sim$0.2\,dex uncertainty in age can be translated into uncertainties in the
LyC rate between 1\% and one order of magnitude (depending on the actual age
estimate), but may be underestimated for individual star clusters.
Despite the fact that the uncertainties are significant, we assume that
masses and ages of the clusters as presented by \citeauthor{WCS+10} are robust
enough for our purpose, the assessment of LyC photon flux.

From the Whitmore catalog\footnote{While the \citet{WCS+10} publication comes
  with a selection of tables of most noteworthy objects, the full table of 60790
  objects was sent to us by B.~Whitmore by private mail. We corrected the
  coordinates in this table by $\Delta\alpha = -1\farcs345$ and
  $\Delta\delta = +1\farcs607$ to match the astrometric calibration of the MUSE
  data.}
we first select all clusters with valid measurements in at least the F336W,
F435W, and F814W filters. From these we further select those for which a valid
mass estimate exists and where the logarithmic age estimate is between 6.0 and 7.8.
These clusters should be responsible for the total LyC flux ionizing the
\hiiregs in the Antennae. From the total catalog, 2162 clusters meet these
criteria. We then match the position of the clusters to the extent of each
\hiireg on the sky.
We compute the LyC flux $Q(\mathrm{H}^0)$ expected from each cluster by comparison to a
solar-metallicity\footnote{We use solar metallicity models for comparison,
  since \citet{BEK+06} and \citet{2015ApJ...812..160L} estimate approximately
  solar metallicity for the central part of the Antennae. As regions in the
  outskirts are likely to have lower metal abundances, the solar estimate of the
  LyC photon flux represents a lower limit to the real LyC flux.}
GALEV model \citep{2009MNRAS.396..462K}, computed using the
Geneva isochrones. To be consistent with the \citeauthor{WCS+10} population
analysis and the origin of the LyC-to-\ha conversion factor given by
\citet{Ken98b} we used a \citet{Sal55} initial mass function with a mass range
0.1 to 100\,$M_\odot$. Then we compare the rate of LyC photons produced by the
population of young clusters inside the region with the rate of ionizing
photons computed from the \ha luminosity, using the conversion factor
$Q(\mathrm{H}^0) = 7.31 \times 10^{11} L(\hamath)$ (\citeauthor{Ken98b}),
resulting in a LyC escape fraction \fesc for each \hiireg.

\begin{figure}
\includegraphics[width=\linewidth]{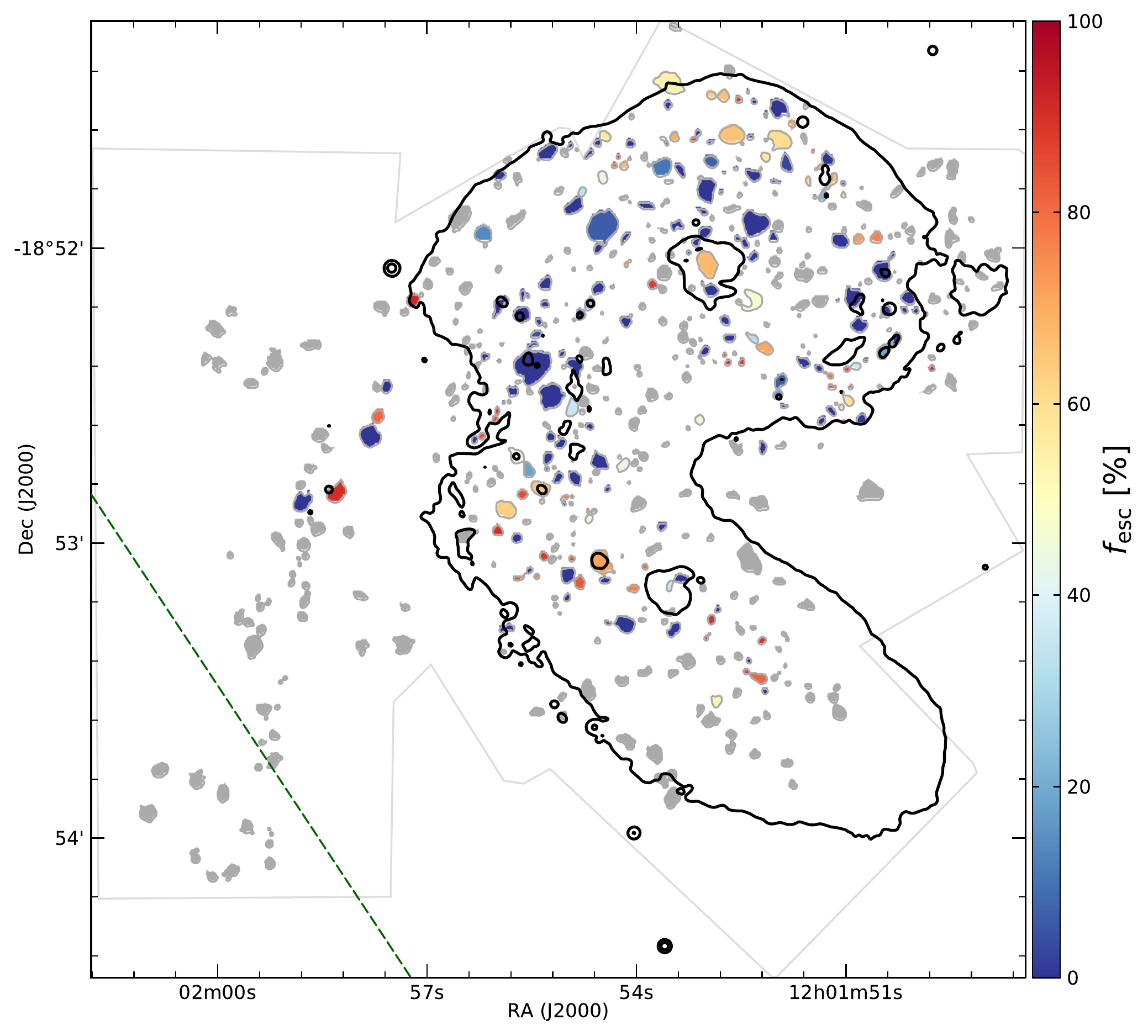}
\caption{\hiiregs detected in the central Antennae. Here, the color indicates
         the Lyman-continuum escape fraction \fesc, estimated by comparing \ha
         luminosity and illuminating young star cluster(s). All colored regions
         have a positive escape fraction, other regions are filled light gray.
         The contours are the same broad-band levels as in Fig.~\ref{fig:ant},
         the dashed dark green line shows the approximate limit of the HST
         imaging.}
\label{fig:hiireg_fesc}
\end{figure}

Since the GALEV code uses the table of \citet{SdK97} to compute the
Lyman-continuum flux, which was shown to overestimate the UV flux in some early
phases of young stellar evolution \citep{2002MNRAS.337.1309S}, we also
cross-checked the results using Starburst99
\citep[v7.0.1,][]{LSG+99,2010ApJS..189..309L,2012ApJ...751...67L} which
implements a more modern treatment of the UV spectrum and also makes use of
newer Geneva isochrones. In order to run Starburst99, we otherwise use inputs
comparable to the GALEV models, especially using the fixed-mass instaneous
star-formation history.

The results are as follows:
From comparison with the GALEV model, of the 551 \hiiregs, 108 leak LyC
photons at a rate of altogether $Q(\mathrm{H}^0)=2.7\times10^{53}$\,s$^{-1}$
while 112 regions are optically thick to UV radiation.
The results are presented in Fig.~\ref{fig:hiireg_fesc}. Most of
the \hiiregs with non-zero escape fraction are located in the inner parts of
the merger, \eg in the overlap region and in the center of \ngc38, the
northern nucleus. But there are also a few regions in its outskirts, especially
near the northern edge of our field of view, in the outer disk of \ngc38.
Note that there are 331 for which we cannot identify an ionizing source in form
of a young cluster in the sample of \citet{WCS+10} that spatially coincides.
These were assigned a light gray color for the presentation in
Fig.~\ref{fig:hiireg_fesc}. As marked in this figure, 14 of the \hiiregs
without ionizing source are located outside the limits of the HST multi-band
coverage. The others might be the result of the sensitivity limits of the HST
observations, possibly in conjunction with different extinction towards the
region of line emission and the stellar location \citep[see \eg][]{1999A&A...347..841H}.

If the estimates computed with SB99 are taken to be more realistic, then
127 of the 551 \hiiregs are partially optically thin to Lyman continuum
photons and leak an even higher number of
$Q(\mathrm{H}^0)=3.2\times10^{53}$\,s$^{-1}$ into the surrounding medium.

Since the \ha fluxes that we used in this section to estimate the LyC escape of
every \hiireg were computed in the apertures without correcting for the diffuse
background, the numbers given above might have been affected by the same
diffuse ionized gas for which we are trying to find the origin. To test this hypothesis,
we turn again to the background-subtracted \hiireg measurements that we already
used as cross-check in Sect.~\ref{sec:difffrac}. With this dataset and in
comparison to the GALEV estimate, we find a total of 115 optically thin nebulae
which appear to leak $2.9\times10^{53}$\,s$^{-1}$ LyC photons, with SB99 we
again find higher numbers of $3.4\times10^{53}$\,s$^{-1}$ leaked by 137
\hiiregs.

Another effect that might influence our estimate is the reddening by dust.
If the light of some of the \hiiregs are not only extincted along the line of
sight, but also into other directions, the LyC photons could be absorbed before
reaching the surrounding neutral medium. To model this, we assumed that the
extinction within the galaxies follows the same distribution of dust screens as
towards the observer. We therefore reduced the summed LyC photons inside each
region by the respective UV extinction, drawn randomly from the sample of
reddening values. We created 100 Monte-Carlo iterations with this method and
found a distribution around an average of
$Q(\mathrm{H}^0)_\mathrm{red}=(1.4\pm0.4)\times10^{53}$\,s$^{-1}$.
However, since star formation hidden from the observer but not necessarily from
the rest of the central merger is known to exist
\citep{1998A&A...333L...1M,2005A&A...443...41M}, the total number of LyC
photons that do not get absorbed by dust is likely higher than this estimate.
Since the dust reddening for each \hiireg within the galaxy cannot be made with
any certainty, we keep the above GALEV analysis for further discussion, with
the fiducial value of $Q(\mathrm{H}^0)=2.7\times10^{53}$\,s$^{-1}$.

\subsubsection{LyC leakage and emission line ratios}\label{sec:lyc_vs_ratios}
Several studies have recently used emission line ratios composed of lines of
different ionization energy to find \hiiregs (and whole galaxies) that are
likely to leak significant amounts of ionizing photons. The ratio \oiii/\oii is
frequently used, since it was found to be high (\oiii/\oii$\gtrsim5$)
in galaxies that were observed to leak LyC photons
\citep{2016Natur.529..178I,2016MNRAS.461.3683I}.
Similarly, high ratios of \oiii/\sii \citep{2012ApJ...755...40P} are indicative
of leakage of LyC photons.

Since MUSE is missing the blue wavelength coverage that would allow detection of the bright \oii3727,29 doublet for
such nearby galaxies, and the line ratio \oiii/\sii also depends on the
relative abundances of oxygen and sulfur, one may prefer to use
ionization-parameter sensitive line ratios of the same elements, namely
\siii/\sii \citep{2013ApJ...779...76Z} and \oiii/\oi
\citep{2015A&A...576A..83S}.  In normal, ionization-bounded \hiiregs one
expects a region of lower ionization gas on the outskirts \citep{OF05}.  So, if
the low-ionization line exhibits unusually low flux compared to a
high-ionization line in the same region, this would show a region to have an
unusually thin or non-existing transition region, as one might expect in
density-bounded, \ie LyC-leaking regions. However, since the size of the
transition region or the ionization front depends strongly on the ionizing
source \citep{OF05,2012ApJ...755...40P}, it is difficult to give an absolute
limit for any of these line ratios.
At the spatial resolution of about 60\,pc, we lack the ability to carry out
proper ionization parameter mapping \citep[cf.][]{2012ApJ...755...40P} of
the \hiiregs to find better evidence which of them is optically thin.

\begin{figure*}
\includegraphics[width=0.33\linewidth]{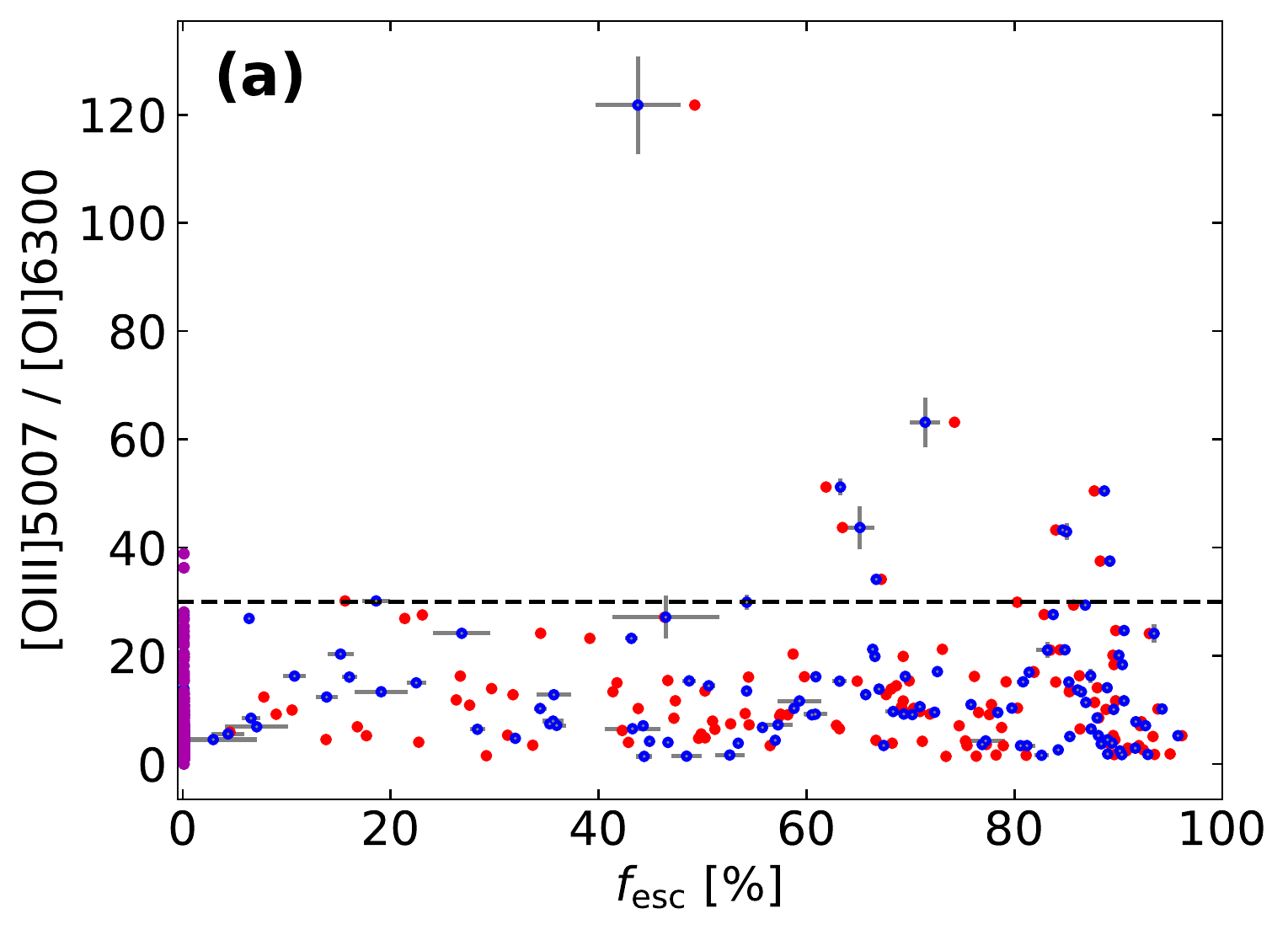}
\includegraphics[width=0.33\linewidth]{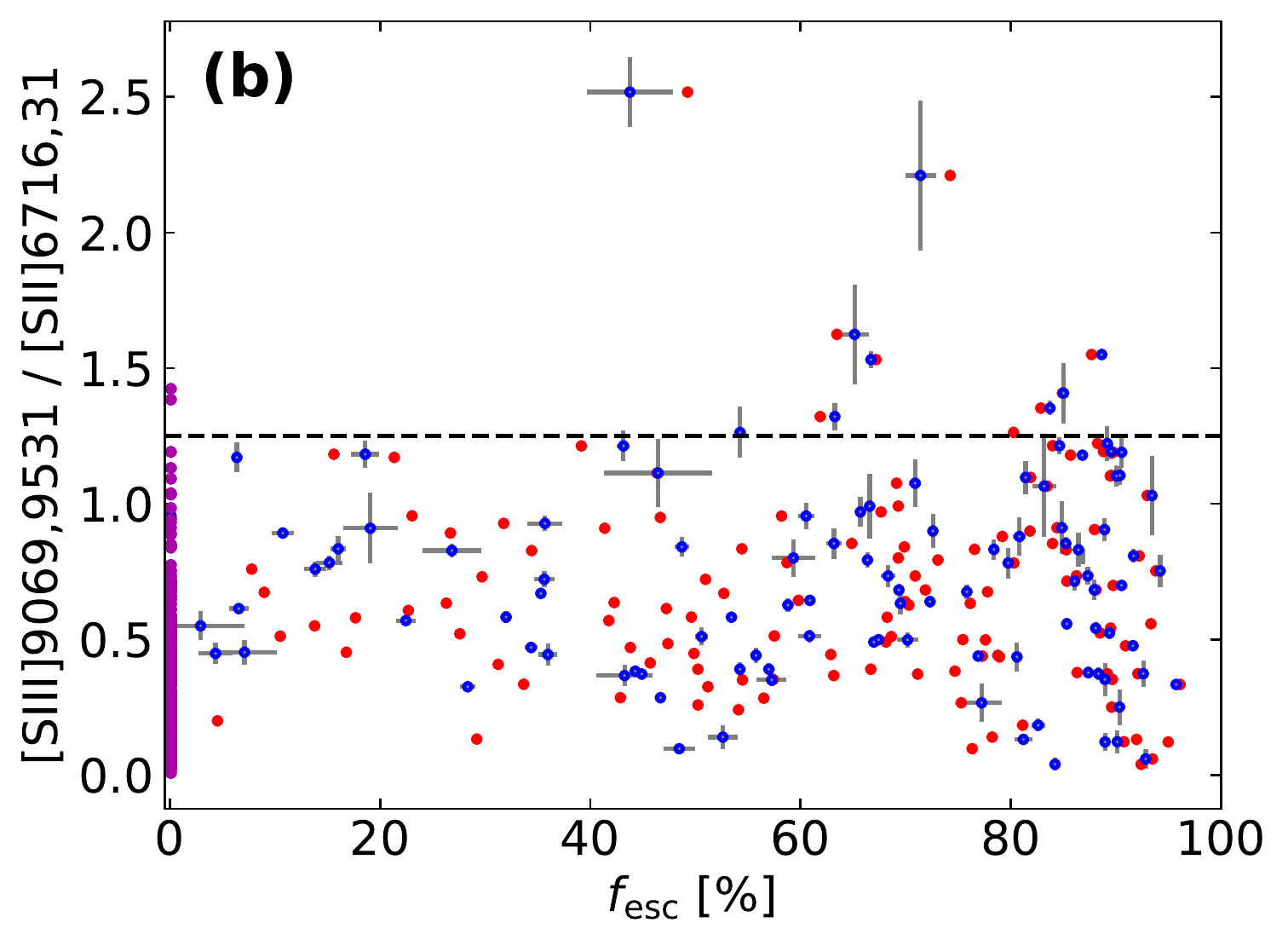}
\includegraphics[width=0.33\linewidth]{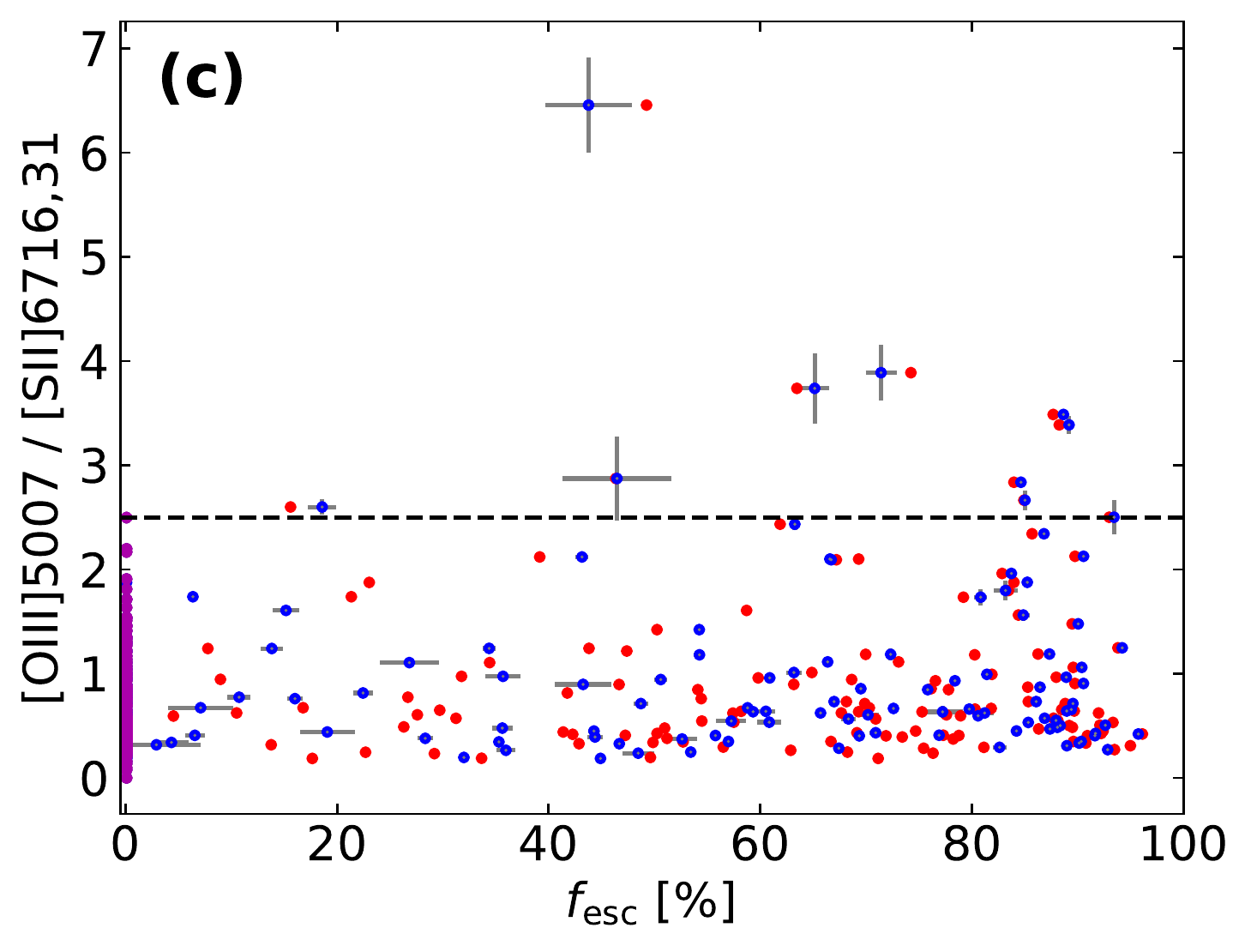}
\caption{Plots of ionization-parameter sensitive line ratios against
         Lyman-continuum escape fraction (\fesc) for our sample of \hiiregs.
         {\bf (a)} Shows the line ratio \oiii5007/\oi6300, {\bf (b)} presents
         \siii9068,9531/\sii6716,31, and {\bf (c)} \oiii5007//\sii6716,31.
         In all plots, the blue points are the results using the GALEV model
         while the red points shows the escape fraction computed using SB99. $1\sigma$
         ranges of the statistical error of both axes are plotted as gray
         lines for the GALEV results. Violet points at the left border are
         those regions which were found to be optically thick.
         Horizontal dashed lines denote lower limits of the line ratios with
         high escape fraction, as discussed in the text.}
\label{fig:fesc_corr}
\end{figure*}

However, assuming that our approach with the cluster population analysis
gives a good estimate of the real LyC escape fraction, we can check if there
is a correlation of the line ratios with \fesc. We show the results in
Fig.~\ref{fig:fesc_corr}, for all three line ratios that MUSE covers. We
discuss these results quoting numbers based on the more conservative GALEV
analysis. (SB99 values differ typically by only 10\%.)

The plot with the \oiii/\oi ratio (in Fig.~\ref{fig:fesc_corr}a) does not give
a strong visual impression of a correlation. However, there are 12 regions with
\oiii/\oi ratio above 30, with 10 of them showing positive LyC escape fractions of up
to 90\%, summing up to LyC rates of $1.3\times10^{53}$\,s$^{-1}$ that are
available to ionize \hi outside the star-forming regions.
The same picture emerges with the \siii/\sii line ratio
(Fig.~\ref{fig:fesc_corr}b). There is again a broad range of escape fractions
for low sulfur line ratios, without a strong correlation.
However, of the \hiiregs with the highest \siii/\sii ratio ($>1.25$), 9 of 11
are LyC leakers by the above criteria, producing $1.2\times10^{53}$\,s$^{-1}$
ionizing photons. There are 8 regions that are above the given limits for both
\oiii/\oi and \siii/\sii, and 7 of them are estimated to be at least partially
density-bounded.
For the line ratio \oiii/\sii as plotted in Fig.~\ref{fig:fesc_corr}c, all
10 of the \hiiregs with \oiii/\sii$>2.5$ appear to be optically thin, and leak
$5.7\times10^{52}$\,s$^{-1}$ LyC photons into their environment.

To conclude this section we can say that the Lyman-continuum escape
fraction is not directly correlated with the ionization-parameter sensitive
line ratios.  Very high values of all three of the studied line ratios, above
certain limits, on the other hand are a clear sign that an \hiireg is density
bounded, and seem to signify LyC leakage at a rate between about 20 and up to
90\%.

\subsubsection{Lyman-continuum photon budget}\label{sec:lyc_budget}
As we have shown above, MUSE data in conjunction with HST cluster analysis
suggest that many \hiiregs in the central part of the Antennae merger are
optically thin and leak LyC photons at a very high rate of
$Q(H^0)=2.7\times10^{53}$\,s$^{-1}$, as the most conservative estimate
without internal extinction.

If our computation of an \ha flux in the diffuse gas component of
$f(\hamath,\mathrm{DIG})\approx6.02\times10^{-12}$\flux (after extinction
correction, see Sect.~\ref{sec:difffrac}) is correct, a rate of
$2.55\times10^{53}$\,s$^{-1}$ LyC photons is needed to ionize the \hi in
the central Antennae. This now suggests that all photons necessary to
ionize the diffuse medium can be provided by leakage from the star-forming
regions.

Unfortunately, the catalog of \citet{WCS+10} only covers our central
field; we cannot use it to run the same analysis for the southern part of the
MUSE data. While some HST data exists over the MUSE pointings, the coverage is
incomplete and consists of only a subset of the filters that are
required to do a proper analysis of the stellar populations.
Therefore, as an alternative approach, we use the 386 \hiiregs in the central field
that have as low luminosities as the ones in the southern field
($\log_{10}L(\hamath)\leq38.25$, see Fig.~\ref{fig:hiireg_LF}).
Of these nebulae, 38 have a significant, positive \fesc with a mean of
72\%.\footnote{Here, we use the conservative estimate using the GALEV code
  without dust reddening, since the extinction in the southern tail is very low.
  The standard deviation of \fesc among these regions is 22\%, the median 83\%.}
This means that the {\em overall} LyC escape for all low-luminosity regions could be
extrapolated to be about 7\% on average.

Under the assumption that our estimate of the flux of the diffuse ionized gas
from Sect.~\ref{sec:difffrac} is correct, we need $7.55\times10^{49}$\,s$^{-1}$
LyC photons to ionize the neutral hydrogen. Converting the
extinction-corrected \ha flux to an estimate of the LyC rate, we find that all
\hiiregs in the southern part are ionized by
$Q(\mathrm{H}^0)=5.3\times10^{50}$\,s$^{-1}$ photons. (This corresponds to the
{\tt speccor} estimate in Table~\ref{tab:fluxes}). If indeed 7\% escape
as extrapolated above, this would leave roughly $3.7\times10^{49}$\,s$^{-1}$
available to ionize the surrounding medium, about half the number of ionizing
photons needed.
If we use instead the more conservative {\tt specsub} estimate, that we derived
in Sect.~\ref{sec:difffrac} by subtracting the background surrounding the
\hiiregs, we still find that about $3.0\times10^{49}$\,s$^{-1}$ LyC photons
would be available to create a diffuse ionized component around the \hiiregs,
only $2.5\times$ smaller than the required number.

If we take only the 5 regions where our line measurements show valid \oiii
and \oi flux values and a ratio \oiii/\oi$>30$, we find that their
combined \ha luminosity corresponds to $Q(\mathrm{H}^0)\approx
2.3$\pexp{50}\,s$^{-1}$ (in the case without background subtraction).
Hence, these regions only need to have a moderate escape fraction of 23.3\% to
fill the gap.

Like for the center of the interacting galaxies, this estimate for the
\hiiregs at the end of the southern tidal tail of the Antennae suggests that
there are enough LyC photons available through escape from star-forming regions
to explain the amount of diffuse ionized gas that we detect in the MUSE data.

Since these estimates show that enough Lyman-continuum photons are available
from the young star clusters in both the central and southern region to explain
the observed ionized gas, and since it is likely that other sources of UV
photons (such as shocks and hot evolved stars) are present in the Antennae, it
is likely that overall, the \ngc38/39 system is a Lyman-continuum leaker.

\section{Summary and conclusions}\label{sec:concl}
In this paper we presented a new set of data on the Antennae Galaxy
(\ngc38/39), observed with the integral field spectrograph MUSE at the VLT.  We
targeted two fields, a region arranged in an irregular pattern covering
$7.5\sq\arcmin$ at the location of the central merger and another irregular
region of $5.8\sq\arcmin$ near the tip of the southern tidal tail.

We show that these MUSE data are of unprecedented depth which enables us to detect
\ha to considerably lower levels than before: $\sim14$\% of this faint diffused ionized gas
was undetected in previous, less deep observations of the central region.
Since the detected faint gas shows a filamentary morphology and different
kinematics from the parts with high surface brightness, it represents a real
detection and not an instrumental artifact. Similarly, we detect more and
brighter \hiiregs in the southern field than were known before. These are also
surrounded by diffuse gas. The diffuse gas fractions are about 60\% in the
central field and 10\% in the southern region, but may be as high as 70\% and
30\% after accounting for diffuse emission underlying the \hiiregs.

We use a peak-detection algorithm on the continuum-subtracted \ha image to
search for \hiiregs. From those locations we extract spectra of about 550
\hiiregs in the central and 50 in the southern field. We compare our detections
with previous work and find reasonable agreement.

Using the existing HST catalog of young star clusters of \citet{WCS+10} we
assess the Lyman-continuum photon production of the stellar populations
inside each \hiireg in the central field. In comparison with our \ha luminosity
measurements, we estimate that about 100 of them leak high fractions of the UV
photons produced by the stars inside them. Summing up these escape fractions
results in ionizing photon rates that are enough to explain the amount of
diffuse ionized emission that we detect.

We compare three line ratios that are sensitive to the
ionization-parameter to this estimate of the Lyman-continuum escape
fractions and find that in the environment of the central Antennae,
\oiii/\oi$=30$, \siii/\sii$=1.25$, and \oiii/\sii$=2.5$ are limits above which
most \hiiregs appear to be optically thin.
However, no systematic trend between these line ratios and escape fraction is
found, so it appears difficult to estimate LyC leakage from galaxies at the
epoch of reionization from the measurement of these line ratios with JWST, as
recently proposed
\citep{2013ApJ...766...91J,2014MNRAS.442..900N,2016ApJ...829...99F}. At least,
more preparatory work (models and observations) needs to be carried out to
better understand the link between LyC escape from \hiiregs and its impact on
nebular line ratios.

By applying these results from the central region, we argue that the \hiiregs
in the southern field also leak enough UV photons to explain the diffuse gas
detected there.

The current paper only addresses one particular topic. Since the MUSE
mosaic of the Antennae represents a much richer dataset, we plan to
present further work on topics such as the interstellar medium, the stellar
populations, as well as the kinematics of the system very soon.
With this paper, we publicly release the \ha flux and velocity maps as FITS
format images and also publish the measurements of the \hiiregs in electronic
form.

\begin{acknowledgements}
We thank Brad Whitmore for sharing the full HST cluster catalog and are
grateful to Jeremy Walsh for useful discussions about \hiiregs and analysis
steps during visits to ESO Garching.
We also thank the other members of the MUSE collaboration, especially Eric
Emsellem and Anna Feltre, for help at various stages of this work.
PMW, SK, and SD received support through BMBF Verbundforschung (project MUSE-AO,
grants 05A14BAC and 05A14MGA).
AMI acknowledges support from the Spanish MINECO through project
AYA2015-68217-P.
AV is supported by a Marie Heim V\"ogtlin fellowship of the Swiss National
Foundation.
SD received additional support through DFG (project DR281/35-1).
Based on observations collected at the European Organisation for Astronomical
Research in the Southern Hemisphere under ESO programs 095.B-0042, 096.B-0017, and
097.B-0346.
We thank Bill Joye who keeps making DS9 better for use with MUSE data.
We also used AstroPy, APLpy, astrodendro, IRAF, STSDAS, MPDAF, and topcat, to
mention only the tip of the iceberg of packages used, and are grateful to the
communities producing such great software.
\end{acknowledgements}

\bibliographystyle{aa}
\bibliography{antennae.bib}

\begin{thebibliography}{115}
\expandafter\ifx\csname natexlab\endcsname\relax\def\natexlab#1{#1}\fi

\bibitem[{{Alonso-Herrero} {et~al.}(2010){Alonso-Herrero},
  {Garc{\'{\i}}a-Mar{\'{\i}}n}, {Rodr{\'{\i}}guez Zaur{\'{\i}}n},
  {Monreal-Ibero}, {Colina}, \& {Arribas}}]{2010A&A...522A...7A}
{Alonso-Herrero}, A., {Garc{\'{\i}}a-Mar{\'{\i}}n}, M., {Rodr{\'{\i}}guez
  Zaur{\'{\i}}n}, J., {et~al.} 2010, A\&A, 522, A7

\bibitem[{{Amram} {et~al.}(1992){Amram}, {Marcelin}, {Boulesteix}, \& {Le
  Coarer}}]{AMB+92}
{Amram}, P., {Marcelin}, M., {Boulesteix}, J., \& {Le Coarer}, E. 1992, A\&A,
  266, 106

\bibitem[{Anders {et~al.}(2004)Anders, Bissantz, Fritze-v. Alvensleben, \&
  de~Grijs}]{ABFA+03}
Anders, P., Bissantz, N., Fritze-v. Alvensleben, U., \& de~Grijs, R. 2004,
  MNRAS, 347, 196

\bibitem[{{Anders} \& {Fritze-v.~Alvensleben}(2003)}]{AFA03}
{Anders}, P. \& {Fritze-v.~Alvensleben}, U. 2003, A\&A, 401, 1063

\bibitem[{{Anders} {et~al.}(2013){Anders}, {Kotulla}, {de Grijs}, \&
  {Wicker}}]{2013ApJ...778..138A}
{Anders}, P., {Kotulla}, R., {de Grijs}, R., \& {Wicker}, J. 2013, ApJ, 778,
  138

\bibitem[{{Bacon} {et~al.}(2010){Bacon}, {Accardo}, {Adjali}, {Anwand},
  {Bauer}, {Biswas}, {Blaizot}, {Boudon}, {Brau-Nogue}, {Brinchmann},
  {Caillier}, {Capoani}, {Carollo}, {Contini}, {Couderc}, {Daguis{\'e}},
  {Deiries}, {Delabre}, {Dreizler}, {Dubois}, {Dupieux}, {Dupuy}, {Emsellem},
  {Fechner}, {Fleischmann}, {Fran{\c c}ois}, {Gallou}, {Gharsa}, {Glindemann},
  {Gojak}, {Guiderdoni}, {Hansali}, {Hahn}, {Jarno}, {Kelz}, {Koehler},
  {Kosmalski}, {Laurent}, {Le Floch}, {Lilly}, {Lizon}, {Loupias}, {Manescau},
  {Monstein}, {Nicklas}, {Olaya}, {Pares}, {Pasquini}, {P{\'e}contal-Rousset},
  {Pell{\'o}}, {Petit}, {Popow}, {Reiss}, {Remillieux}, {Renault}, {Roth},
  {Rupprecht}, {Serre}, {Schaye}, {Soucail}, {Steinmetz}, {Streicher}, {Stuik},
  {Valentin}, {Vernet}, {Weilbacher}, {Wisotzki}, \& {Yerle}}]{Bacon+10}
{Bacon}, R., {Accardo}, M., {Adjali}, L., {et~al.} 2010, in Proc.~{SPIE}, Vol.
  7735, {Ground-based and Airborne Instrumentation for Astronomy III}

\bibitem[{{Bacon} {et~al.}(2012){Bacon}, {Accardo}, {Adjali}, {Anwand},
  {Bauer}, {Blaizot}, {Boudon}, {Brinchmann}, {Brotons}, {Caillier}, {Capoani},
  {Carollo}, {Comin}, {Contini}, {Cumani}, {Daguis}, {Deiries}, {Delabre},
  {Dreizler}, {Dubois}, {Dupieux}, {Dupuy}, {Emsellem}, {Fleischmann}, {Fran{\c
  c}ois}, {Gallou}, {Gharsa}, {Girard}, {Glindemann}, {Guiderdoni}, {Hahn},
  {Hansali}, {Hofmann}, {Jarno}, {Kelz}, {Kiekebusch}, {Knudstrup}, {Koehler},
  {Kollatschny}, {Kosmalski}, {Laurent}, {Le Floch}, {Lilly}, {Lizon {\`a}
  L'Allemand}, {Loupias}, {Manescau}, {Monstein}, {Nicklas}, {Niemeyer},
  {Olaya}, {Palsa}, {Par{\`e}s}, {Pasquini}, {P{\'e}contal-Rousset}, {Pello},
  {Petit}, {Piqueras}, {Popow}, {Reiss}, {Remillieux}, {Renault}, {Rhode},
  {Richard}, {Roth}, {Rupprecht}, {Schaye}, {Slezak}, {Soucail}, {Steinmetz},
  {Streicher}, {Stuik}, {Valentin}, {Vernet}, {Weilbacher}, {Wisotzki},
  {Yerle}, \& {Zins}}]{BAA+12}
{Bacon}, R., {Accardo}, M., {Adjali}, L., {et~al.} 2012, The Messenger, 147, 4

\bibitem[{{Bastian} {et~al.}(2006){Bastian}, {Emsellem}, {Kissler-Patig}, \&
  {Maraston}}]{BEK+06}
{Bastian}, N., {Emsellem}, E., {Kissler-Patig}, M., \& {Maraston}, C. 2006,
  A\&A, 445, 471

\bibitem[{{Bastian} {et~al.}(2009){Bastian}, {Trancho}, {Konstantopoulos}, \&
  {Miller}}]{BTK+09}
{Bastian}, N., {Trancho}, G., {Konstantopoulos}, I.~S., \& {Miller}, B.~W.
  2009, ApJ, 701, 607

\bibitem[{{Bigiel} {et~al.}(2015){Bigiel}, {Leroy}, {Blitz}, {Bolatto}, {da
  Cunha}, {Rosolowsky}, {Sandstrom}, \& {Usero}}]{Bigiel15}
{Bigiel}, F., {Leroy}, A.~K., {Blitz}, L., {et~al.} 2015, ApJ, 815, 103

\bibitem[{Bournaud {et~al.}(2004)Bournaud, Duc, Amram, Combes, \&
  Gach}]{BDA+04}
Bournaud, F., Duc, P.-A., Amram, P., Combes, F., \& Gach, J.-L. 2004, A\&A, in
  press

\bibitem[{{Brandl} {et~al.}(2005){Brandl}, {Clark}, {Eikenberry}, {Wilson},
  {Henderson}, {Barry}, {Houck}, {Carson}, \& {Hayward}}]{Brandl05}
{Brandl}, B.~R., {Clark}, D.~M., {Eikenberry}, S.~S., {et~al.} 2005, ApJ, 635,
  280

\bibitem[{{Brandl} {et~al.}(2009){Brandl}, {Snijders}, {den Brok}, {Whelan},
  {Groves}, {van der Werf}, {Charmandaris}, {Smith}, {Armus}, {Kennicutt}, \&
  {Houck}}]{BSdB+09}
{Brandl}, B.~R., {Snijders}, L., {den Brok}, M., {et~al.} 2009, ApJ, 699, 1982

\bibitem[{{Calzetti} {et~al.}(2000){Calzetti}, {Armus}, {Bohlin}, {Kinney},
  {Koornneef}, \& {Storchi-Bergmann}}]{CAB+00}
{Calzetti}, D., {Armus}, L., {Bohlin}, R.~C., {et~al.} 2000, ApJ, 533, 682

\bibitem[{{Campbell} \& {Willner}(1989)}]{1989AJ.....97..995C}
{Campbell}, A. \& {Willner}, S.~P. 1989, AJ, 97, 995

\bibitem[{{Cappellari} \& {Copin}(2003)}]{2003MNRAS.342..345C}
{Cappellari}, M. \& {Copin}, Y. 2003, MNRAS, 342, 345

\bibitem[{{Cappellari} \& {Emsellem}(2004)}]{2004PASP..116..138C}
{Cappellari}, M. \& {Emsellem}, E. 2004, PASP, 116, 138

\bibitem[{{Chandar} {et~al.}(2016){Chandar}, {Whitmore}, {Dinino}, {Kennicutt},
  {Chien}, {Schinnerer}, \& {Meidt}}]{2016ApJ...824...71C}
{Chandar}, R., {Whitmore}, B.~C., {Dinino}, D., {et~al.} 2016, ApJ, 824, 71

\bibitem[{{Conselice}(2003)}]{2003ApJS..147....1C}
{Conselice}, C.~J. 2003, \apjs, 147, 1

\bibitem[{{Dopita} \& {Sutherland}(1995)}]{dop95}
{Dopita}, M.~A. \& {Sutherland}, R.~S. 1995, ApJ, 455, 468

\bibitem[{{Faisst}(2016)}]{2016ApJ...829...99F}
{Faisst}, A.~L. 2016, ApJ, 829, 99

\bibitem[{{Falc{\'o}n-Barroso} {et~al.}(2011){Falc{\'o}n-Barroso},
  {S{\'a}nchez-Bl{\'a}zquez}, {Vazdekis}, {Ricciardelli}, {Cardiel}, {Cenarro},
  {Gorgas}, \& {Peletier}}]{2011A&A...532A..95F}
{Falc{\'o}n-Barroso}, J., {S{\'a}nchez-Bl{\'a}zquez}, P., {Vazdekis}, A.,
  {et~al.} 2011, A\&A, 532, A95

\bibitem[{{Fensch} {et~al.}(2016){Fensch}, {Duc}, {Weilbacher}, {Boquien}, \&
  {Zackrisson}}]{2016A&A...585A..79F}
{Fensch}, J., {Duc}, P.-A., {Weilbacher}, P.~M., {Boquien}, M., \&
  {Zackrisson}, E. 2016, A\&A, 585, A79

\bibitem[{{Ferguson} {et~al.}(1996){Ferguson}, {Wyse}, \&
  {Gallagher}}]{1996AJ....112.2567F}
{Ferguson}, A.~M.~N., {Wyse}, R.~F.~G., \& {Gallagher}, J.~S. 1996, AJ, 112,
  2567

\bibitem[{{Fouesneau} \& {Lan{\c c}on}(2010)}]{2010A&A...521A..22F}
{Fouesneau}, M. \& {Lan{\c c}on}, A. 2010, A\&A, 521, A22

\bibitem[{{Fouesneau} {et~al.}(2012){Fouesneau}, {Lan{\c c}on}, {Chandar}, \&
  {Whitmore}}]{2012ApJ...750...60F}
{Fouesneau}, M., {Lan{\c c}on}, A., {Chandar}, R., \& {Whitmore}, B.~C. 2012,
  ApJ, 750, 60

\bibitem[{{Gaia Collaboration} {et~al.}(2016){Gaia Collaboration}, {Brown},
  {Vallenari}, {Prusti}, {de Bruijne}, {Mignard}, {Drimmel}, {Babusiaux},
  {Bailer-Jones}, {Bastian}, \& et~al.}]{GaiaDR1}
{Gaia Collaboration}, {Brown}, A.~G.~A., {Vallenari}, A., {et~al.} 2016, A\&A,
  595, A2

\bibitem[{{Gilbert} \& {Graham}(2007)}]{2007ApJ...668..168G}
{Gilbert}, A.~M. \& {Graham}, J.~R. 2007, ApJ, 668, 168

\bibitem[{{Haas} {et~al.}(2005){Haas}, {Chini}, \&
  {Klaas}}]{2005A&A...433L..17H}
{Haas}, M., {Chini}, R., \& {Klaas}, U. 2005, A\&A, 433, L17

\bibitem[{{Haffner} {et~al.}(2009){Haffner}, {Dettmar}, {Beckman}, {Wood},
  {Slavin}, {Giammanco}, {Madsen}, {Zurita}, \&
  {Reynolds}}]{2009RvMP...81..969H}
{Haffner}, L.~M., {Dettmar}, R.-J., {Beckman}, J.~E., {et~al.} 2009, Reviews of
  Modern Physics, 81, 969

\bibitem[{{Heckman} {et~al.}(1995){Heckman}, {Dahlem}, {Lehnert}, {Fabbiano},
  {Gilmore}, \& {Waller}}]{HDL+95}
{Heckman}, T.~M., {Dahlem}, M., {Lehnert}, M.~D., {et~al.} 1995, ApJ, 448, 98

\bibitem[{{Heidt} {et~al.}(2003){Heidt}, {Appenzeller}, {Gabasch}, {J{\"
  a}ger}, {Seitz}, {Bender}, {B{\" o}hm}, {Snigula}, {Fricke}, {Hopp}, {K{\"
  u}mmel}, {M{\" o}llenhoff}, {Szeifert}, {Ziegler}, {Drory}, {Mehlert},
  {Moorwood}, {Nicklas}, {Noll}, {Saglia}, {Seifert}, {Stahl}, {Sutorius}, \&
  {Wagner}}]{HAG+03}
{Heidt}, J., {Appenzeller}, I., {Gabasch}, A., {et~al.} 2003, A\&A, 398, 49

\bibitem[{{Heydari-Malayeri} {et~al.}(1999){Heydari-Malayeri}, {Charmandaris},
  {Deharveng}, {Rosa}, \& {Zinnecker}}]{1999A&A...347..841H}
{Heydari-Malayeri}, M., {Charmandaris}, V., {Deharveng}, L., {Rosa}, M.~R., \&
  {Zinnecker}, H. 1999, A\&A, 347, 841

\bibitem[{{Hibbard} {et~al.}(2001){Hibbard}, {van der Hulst}, {Barnes}, \&
  {Rich}}]{HvdH+01}
{Hibbard}, J., {van der Hulst}, J., {Barnes}, J., \& {Rich}, R. 2001, AJ, 122,
  2969

\bibitem[{{Hibbard} {et~al.}(2005){Hibbard}, {Bianchi}, {Thilker}, {Rich},
  {Schiminovich}, {Xu}, {Neff}, {Seibert}, {Lauger}, {Burgarella}, {Barlow},
  {Byun}, {Donas}, {Forster}, {Friedman}, {Heckman}, {Jelinsky}, {Lee},
  {Madore}, {Malina}, {Martin}, {Milliard}, {Morrissey}, {Siegmund}, {Small},
  {Szalay}, {Welsh}, \& {Wyder}}]{Hibbard05}
{Hibbard}, J.~E., {Bianchi}, L., {Thilker}, D.~A., {et~al.} 2005, ApJL, 619,
  L87

\bibitem[{{Hoopes} \& {Walterbos}(2003)}]{2003ApJ...586..902H}
{Hoopes}, C.~G. \& {Walterbos}, R.~A.~M. 2003, ApJ, 586, 902

\bibitem[{{Ivison} {et~al.}(2012){Ivison}, {Smail}, {Amblard}, {Arumugam}, {De
  Breuck}, {Emonts}, {Feain}, {Greve}, {Haas}, {Ibar}, {Jarvis}, {Kov{\'a}cs},
  {Lehnert}, {Nesvadba}, {R{\"o}ttgering}, {Seymour}, \&
  {Wylezalek}}]{Ivison12}
{Ivison}, R.~J., {Smail}, I., {Amblard}, A., {et~al.} 2012, MNRAS, 425, 1320

\bibitem[{{Izotov} {et~al.}(2016{\natexlab{a}}){Izotov}, {Orlitov{\'a}},
  {Schaerer}, {Thuan}, {Verhamme}, {Guseva}, \&
  {Worseck}}]{2016Natur.529..178I}
{Izotov}, Y.~I., {Orlitov{\'a}}, I., {Schaerer}, D., {et~al.}
  2016{\natexlab{a}}, Nature, 529, 178

\bibitem[{{Izotov} {et~al.}(2016{\natexlab{b}}){Izotov}, {Schaerer}, {Thuan},
  {Worseck}, {Guseva}, {Orlitov{\'a}}, \& {Verhamme}}]{2016MNRAS.461.3683I}
{Izotov}, Y.~I., {Schaerer}, D., {Thuan}, T.~X., {et~al.} 2016{\natexlab{b}},
  MNRAS, 461, 3683

\bibitem[{{Jaskot} \& {Oey}(2013)}]{2013ApJ...766...91J}
{Jaskot}, A.~E. \& {Oey}, M.~S. 2013, ApJ, 766, 91

\bibitem[{{Karl} {et~al.}(2010){Karl}, {Naab}, {Johansson}, {Kotarba}, {Boily},
  {Renaud}, \& {Theis}}]{KNJ+10}
{Karl}, S.~J., {Naab}, T., {Johansson}, P.~H., {et~al.} 2010, ApJL, 715, L88

\bibitem[{{Kassin} {et~al.}(2003){Kassin}, {Frogel}, {Pogge}, {Tiede}, \&
  {Sellgren}}]{2003AJ....126.1276K}
{Kassin}, S.~A., {Frogel}, J.~A., {Pogge}, R.~W., {Tiede}, G.~P., \&
  {Sellgren}, K. 2003, AJ, 126, 1276

\bibitem[{{Kehrig} {et~al.}(2012){Kehrig}, {Monreal-Ibero}, {Papaderos},
  {V{\'{\i}}lchez}, {Gomes}, {Masegosa}, {S{\'a}nchez}, {Lehnert}, {Cid
  Fernandes}, {Bland-Hawthorn}, {Bomans}, {Marquez}, {Mast}, {Aguerri},
  {L{\'o}pez-S{\'a}nchez}, {Marino}, {Pasquali}, {Perez}, {Roth},
  {S{\'a}nchez-Bl{\'a}zquez}, \& {Ziegler}}]{Kehrig12}
{Kehrig}, C., {Monreal-Ibero}, A., {Papaderos}, P., {et~al.} 2012, A\&A, 540,
  A11

\bibitem[{{Kennicutt}(1998)}]{Ken98b}
{Kennicutt}, R. 1998, ARA\&A, 36, 189

\bibitem[{{Kewley} {et~al.}(2001){Kewley}, {Dopita}, {Sutherland}, {Heisler},
  \& {Trevena}}]{2001ApJ...556..121K}
{Kewley}, L.~J., {Dopita}, M.~A., {Sutherland}, R.~S., {Heisler}, C.~A., \&
  {Trevena}, J. 2001, ApJ, 556, 121

\bibitem[{{Kotulla} {et~al.}(2009){Kotulla}, {Fritze}, {Weilbacher}, \&
  {Anders}}]{2009MNRAS.396..462K}
{Kotulla}, R., {Fritze}, U., {Weilbacher}, P., \& {Anders}, P. 2009, MNRAS,
  396, 462

\bibitem[{{Kreckel} {et~al.}(2016){Kreckel}, {Blanc}, {Schinnerer}, {Groves},
  {Adamo}, {Hughes}, \& {Meidt}}]{2016ApJ...827..103K}
{Kreckel}, K., {Blanc}, G.~A., {Schinnerer}, E., {et~al.} 2016, ApJ, 827, 103

\bibitem[{{Kroupa}(2002)}]{Kro02}
{Kroupa}, P. 2002, Science, 295, 82

\bibitem[{{Lacey} \& {Cole}(1993)}]{1993MNRAS.262..627L}
{Lacey}, C. \& {Cole}, S. 1993, MNRAS, 262, 627

\bibitem[{{Lardo} {et~al.}(2015){Lardo}, {Davies}, {Kudritzki}, {Gazak},
  {Evans}, {Patrick}, {Bergemann}, \& {Plez}}]{2015ApJ...812..160L}
{Lardo}, C., {Davies}, B., {Kudritzki}, R.-P., {et~al.} 2015, ApJ, 812, 160

\bibitem[{{Lee} {et~al.}(2016){Lee}, {Veilleux}, {McDonald}, \&
  {Hilbert}}]{2016ApJ...817..177L}
{Lee}, J.~C., {Veilleux}, S., {McDonald}, M., \& {Hilbert}, B. 2016, ApJ, 817,
  177

\bibitem[{{Lehnert} \& {Heckman}(1996)}]{1996ApJ...462..651L}
{Lehnert}, M.~D. \& {Heckman}, T.~M. 1996, ApJ, 462, 651

\bibitem[{{Leitet} {et~al.}(2013){Leitet}, {Bergvall}, {Hayes}, {Linn{\'e}}, \&
  {Zackrisson}}]{2013A&A...553A.106L}
{Leitet}, E., {Bergvall}, N., {Hayes}, M., {Linn{\'e}}, S., \& {Zackrisson}, E.
  2013, A\&A, 553, A106

\bibitem[{{Leitherer} {et~al.}(2016){Leitherer}, {Hernandez}, {Lee}, \&
  {Oey}}]{2016ApJ...823...64L}
{Leitherer}, C., {Hernandez}, S., {Lee}, J.~C., \& {Oey}, M.~S. 2016, ApJ, 823,
  64

\bibitem[{{Leitherer} {et~al.}(2010){Leitherer}, {Ortiz Ot{\'a}lvaro},
  {Bresolin}, {Kudritzki}, {Lo Faro}, {Pauldrach}, {Pettini}, \&
  {Rix}}]{2010ApJS..189..309L}
{Leitherer}, C., {Ortiz Ot{\'a}lvaro}, P.~A., {Bresolin}, F., {et~al.} 2010,
  ApJS, 189, 309

\bibitem[{{Leitherer} {et~al.}(1999){Leitherer}, {Schaerer}, {Goldader},
  {Delgado}, {Robert}, {Kune}, {de Mello}, {Devost}, \& {Heckman}}]{LSG+99}
{Leitherer}, C., {Schaerer}, D., {Goldader}, J., {et~al.} 1999, ApJS, 123, 3

\bibitem[{{Levesque} {et~al.}(2012){Levesque}, {Leitherer}, {Ekstrom},
  {Meynet}, \& {Schaerer}}]{2012ApJ...751...67L}
{Levesque}, E.~M., {Leitherer}, C., {Ekstrom}, S., {Meynet}, G., \& {Schaerer},
  D. 2012, ApJ, 751, 67

\bibitem[{{Luridiana} {et~al.}(2015){Luridiana}, {Morisset}, \& {Shaw}}]{PyNeb}
{Luridiana}, V., {Morisset}, C., \& {Shaw}, R.~A. 2015, A\&A, 573, A42

\bibitem[{{Madsen} {et~al.}(2006){Madsen}, {Reynolds}, \& {Haffner}}]{Madsen06}
{Madsen}, G.~J., {Reynolds}, R.~J., \& {Haffner}, L.~M. 2006, ApJ, 652, 401

\bibitem[{{Mathis}(2000)}]{2000ApJ...544..347M}
{Mathis}, J.~S. 2000, ApJ, 544, 347

\bibitem[{{Mengel} {et~al.}(2005){Mengel}, {Lehnert}, {Thatte}, \&
  {Genzel}}]{2005A&A...443...41M}
{Mengel}, S., {Lehnert}, M.~D., {Thatte}, N., \& {Genzel}, R. 2005, A\&A, 443,
  41

\bibitem[{Mirabel {et~al.}(1992)Mirabel, Dottori, \& Lutz}]{MDL92}
Mirabel, I., Dottori, H., \& Lutz, D. 1992, A\&A, 256, L19

\bibitem[{{Mirabel} {et~al.}(1998){Mirabel}, {Vigroux}, {Charmandaris},
  {Sauvage}, {Gallais}, {Tran}, {Cesarsky}, {Madden}, \&
  {Duc}}]{1998A&A...333L...1M}
{Mirabel}, I.~F., {Vigroux}, L., {Charmandaris}, V., {et~al.} 1998, A\&A, 333,
  L1

\bibitem[{{Monreal-Ibero} {et~al.}(2006){Monreal-Ibero}, {Arribas}, \&
  {Colina}}]{mon06}
{Monreal-Ibero}, A., {Arribas}, S., \& {Colina}, L. 2006, ApJ, 637, 138

\bibitem[{{Monreal-Ibero} {et~al.}(2010){Monreal-Ibero}, {Arribas}, {Colina},
  {Rodr{\'{\i}}guez-Zaur{\'{\i}}n}, {Alonso-Herrero}, \&
  {Garc{\'{\i}}a-Mar{\'{\i}}n}}]{MonrealIbero10}
{Monreal-Ibero}, A., {Arribas}, S., {Colina}, L., {et~al.} 2010, A\&A, 517, A28

\bibitem[{{Munari} {et~al.}(2005){Munari}, {Sordo}, {Castelli}, \&
  {Zwitter}}]{MSCZ05}
{Munari}, U., {Sordo}, R., {Castelli}, F., \& {Zwitter}, T. 2005, A\&A, 442,
  1127

\bibitem[{{Nakajima} \& {Ouchi}(2014)}]{2014MNRAS.442..900N}
{Nakajima}, K. \& {Ouchi}, M. 2014, MNRAS, 442, 900

\bibitem[{{Niederhofer} {et~al.}(2016){Niederhofer}, {Hilker}, {Bastian}, \&
  {Ercolano}}]{2016A&A...592A..47N}
{Niederhofer}, F., {Hilker}, M., {Bastian}, N., \& {Ercolano}, B. 2016, A\&A,
  592, A47

\bibitem[{{Osterbrock} \& {Ferland}(2005)}]{OF05}
{Osterbrock}, D.~E. \& {Ferland}, G.~J. 2005, {Astrophysics of Gaseous Nebulae
  and Active Galactic Nuclei, Second Edition} ({University Science Books})

\bibitem[{{Papaderos} {et~al.}(2013){Papaderos}, {Gomes}, {V{\'{\i}}lchez},
  {Kehrig}, {Lehnert}, {Ziegler}, {S{\'a}nchez}, {Husemann}, {Monreal-Ibero},
  {Garc{\'{\i}}a-Benito}, {Bland-Hawthorn}, {Cortijo-Ferrero}, {de
  Lorenzo-C{\'a}ceres}, {del Olmo}, {Falc{\'o}n-Barroso}, {Galbany},
  {Iglesias-P{\'a}ramo}, {L{\'o}pez-S{\'a}nchez}, {Marquez}, {Moll{\'a}},
  {Mast}, {van de Ven}, \& {Wisotzki}}]{Papaderos13}
{Papaderos}, P., {Gomes}, J.~M., {V{\'{\i}}lchez}, J.~M., {et~al.} 2013, A\&A,
  555, L1

\bibitem[{{Pellegrini} {et~al.}(2012){Pellegrini}, {Oey}, {Winkler}, {Points},
  {Smith}, {Jaskot}, \& {Zastrow}}]{2012ApJ...755...40P}
{Pellegrini}, E.~W., {Oey}, M.~S., {Winkler}, P.~F., {et~al.} 2012, ApJ, 755,
  40

\bibitem[{{Pleuss} {et~al.}(2000){Pleuss}, {Heller}, \& {Fricke}}]{PHF00}
{Pleuss}, P., {Heller}, C., \& {Fricke}, K. 2000, A\&A, 361, 913

\bibitem[{{Renaud} {et~al.}(2015){Renaud}, {Bournaud}, \&
  {Duc}}]{2015MNRAS.446.2038R}
{Renaud}, F., {Bournaud}, F., \& {Duc}, P.-A. 2015, MNRAS, 446, 2038

\bibitem[{{Rich} {et~al.}(2011){Rich}, {Kewley}, \& {Dopita}}]{RKD11}
{Rich}, J.~A., {Kewley}, L.~J., \& {Dopita}, M.~A. 2011, ApJ, 734, 87

\bibitem[{{Rossa} \& {Dettmar}(2003)}]{2003A&A...406..493R}
{Rossa}, J. \& {Dettmar}, R.-J. 2003, A\&A, 406, 493

\bibitem[{{Rossa} {et~al.}(2004){Rossa}, {Dettmar}, {Walterbos}, \&
  {Norman}}]{2004AJ....128..674R}
{Rossa}, J., {Dettmar}, R.-J., {Walterbos}, R.~A.~M., \& {Norman}, C.~A. 2004,
  AJ, 128, 674

\bibitem[{{Rubin} {et~al.}(1970){Rubin}, {Ford}, \&
  {D'Odorico}}]{1970ApJ...160..801R}
{Rubin}, V.~C., {Ford}, Jr., W.~K., \& {D'Odorico}, S. 1970, ApJ, 160, 801

\bibitem[{{Salpeter}(1955)}]{Sal55}
{Salpeter}, E. 1955, ApJ, 121, 161

\bibitem[{Sanders \& Mirabel(1996)}]{SM96}
Sanders, D. \& Mirabel, I. 1996, ARA\&A, 34, 749

\bibitem[{{Sandin}(2014)}]{2014A&A...567A..97S}
{Sandin}, C. 2014, A\&A, 567, A97

\bibitem[{{Sandin}(2015)}]{2015A&A...577A.106S}
{Sandin}, C. 2015, A\&A, 577, A106

\bibitem[{{Sandin} {et~al.}(2010){Sandin}, {Becker}, {Roth}, {Gerssen},
  {Monreal-Ibero}, {B{\"o}hm}, \& {Weilbacher}}]{SBR+10}
{Sandin}, C., {Becker}, T., {Roth}, M.~M., {et~al.} 2010, A\&A, 515, A35

\bibitem[{Sandin {et~al.}(2012)Sandin, Weilbacher, Tabataba-Vakili, Kamann, \&
  Streicher}]{SWTV+12}
Sandin, C., Weilbacher, P., Tabataba-Vakili, F., Kamann, S., \& Streicher, O.
  2012, in Proc.~{SPIE}, Vol. 8451, {Software and Cyberinfrastructure for
  Astronomy II}

\bibitem[{{Schaerer} \& {de Koter}(1997)}]{SdK97}
{Schaerer}, D. \& {de Koter}, A. 1997, A\&A, 322, 598

\bibitem[{{Schaye} {et~al.}(2015){Schaye}, {Crain}, {Bower}, {Furlong},
  {Schaller}, {Theuns}, {Dalla Vecchia}, {Frenk}, {McCarthy}, {Helly},
  {Jenkins}, {Rosas-Guevara}, {White}, {Baes}, {Booth}, {Camps}, {Navarro},
  {Qu}, {Rahmati}, {Sawala}, {Thomas}, \& {Trayford}}]{2015MNRAS.446..521S}
{Schaye}, J., {Crain}, R.~A., {Bower}, R.~G., {et~al.} 2015, MNRAS, 446, 521

\bibitem[{{Schirm} {et~al.}(2014){Schirm}, {Wilson}, {Parkin}, {Kamenetzky},
  {Glenn}, {Rangwala}, {Spinoglio}, {Pereira-Santaella}, {Baes}, {Barlow},
  {Clements}, {Cooray}, {De Looze}, {Karczewski}, {Madden}, {R{\'e}my-Ruyer},
  \& {Wu}}]{Schirm14}
{Schirm}, M.~R.~P., {Wilson}, C.~D., {Parkin}, T.~J., {et~al.} 2014, ApJ, 781,
  101

\bibitem[{{Schweizer} {et~al.}(2008){Schweizer}, {Burns}, {Madore}, {Mager},
  {Phillips}, {Freedman}, {Boldt}, {Contreras}, {Folatelli}, {Gonz{\'a}lez},
  {Hamuy}, {Krzeminski}, {Morrell}, {Persson}, {Roth}, \&
  {Stritzinger}}]{2008AJ....136.1482S}
{Schweizer}, F., {Burns}, C.~R., {Madore}, B.~F., {et~al.} 2008, AJ, 136, 1482

\bibitem[{{Scoville} {et~al.}(2001){Scoville}, {Polletta}, {Ewald}, {Stolovy},
  {Thompson}, \& {Rieke}}]{SPE+01}
{Scoville}, N.~Z., {Polletta}, M., {Ewald}, S., {et~al.} 2001, AJ, 122, 3017

\bibitem[{{Skrutskie} {et~al.}(2006){Skrutskie}, {Cutri}, {Stiening},
  {Weinberg}, {Schneider}, {Carpenter}, {Beichman}, {Capps}, {Chester},
  {Elias}, {Huchra}, {Liebert}, {Lonsdale}, {Monet}, {Price}, {Seitzer},
  {Jarrett}, {Kirkpatrick}, {Gizis}, {Howard}, {Evans}, {Fowler}, {Fullmer},
  {Hurt}, {Light}, {Kopan}, {Marsh}, {McCallon}, {Tam}, {Van Dyk}, \&
  {Wheelock}}]{2MASS}
{Skrutskie}, M.~F., {Cutri}, R.~M., {Stiening}, R., {et~al.} 2006, AJ, 131,
  1163

\bibitem[{{Smith} {et~al.}(2002){Smith}, {Norris}, \&
  {Crowther}}]{2002MNRAS.337.1309S}
{Smith}, L.~J., {Norris}, R.~P.~F., \& {Crowther}, P.~A. 2002, MNRAS, 337, 1309

\bibitem[{{Soto} {et~al.}(2012){Soto}, {Martin}, {Prescott}, \&
  {Armus}}]{Soto12}
{Soto}, K.~T., {Martin}, C.~L., {Prescott}, M.~K.~M., \& {Armus}, L. 2012, ApJ,
  757, 86

\bibitem[{{Stasi{\'n}ska} {et~al.}(2015){Stasi{\'n}ska}, {Izotov}, {Morisset},
  \& {Guseva}}]{2015A&A...576A..83S}
{Stasi{\'n}ska}, G., {Izotov}, Y., {Morisset}, C., \& {Guseva}, N. 2015, A\&A,
  576, A83

\bibitem[{{Steinmetz} \& {Navarro}(2002)}]{2002NewA....7..155S}
{Steinmetz}, M. \& {Navarro}, J.~F. 2002, New Astronomy, 7, 155

\bibitem[{{Tacconi} {et~al.}(2008){Tacconi}, {Genzel}, {Smail}, {Neri},
  {Chapman}, {Ivison}, {Blain}, {Cox}, {Omont}, {Bertoldi}, {Greve},
  {F{\"o}rster Schreiber}, {Genel}, {Lutz}, {Swinbank}, {Shapley}, {Erb},
  {Cimatti}, {Daddi}, \& {Baker}}]{Tacconi08}
{Tacconi}, L.~J., {Genzel}, R., {Smail}, I., {et~al.} 2008, ApJ, 680, 246

\bibitem[{Toomre \& Toomre(1972)}]{TT72}
Toomre, A. \& Toomre, J. 1972, ApJ, 178, 623

\bibitem[{{Vazdekis} {et~al.}(2012){Vazdekis}, {Ricciardelli}, {Cenarro},
  {Rivero-Gonz{\'a}lez}, {D{\'{\i}}az-Garc{\'{\i}}a}, \&
  {Falc{\'o}n-Barroso}}]{2012MNRAS.424..157V}
{Vazdekis}, A., {Ricciardelli}, E., {Cenarro}, A.~J., {et~al.} 2012, MNRAS,
  424, 157

\bibitem[{Ventou {et~al.}(2017)Ventou, Contini, Bouch\'e, Epinat, Brinchmann,
  Bacon, Inami, Lam, Drake, Garel, Michel-Dansac, Steinmetz, Weilbacher, \&
  Wisotzki}]{Ventou17}
Ventou, E., Contini, T., Bouch\'e, N., {et~al.} 2017, A\&A, submitted

\bibitem[{{Vogelsberger} {et~al.}(2014){Vogelsberger}, {Genel}, {Springel},
  {Torrey}, {Sijacki}, {Xu}, {Snyder}, {Nelson}, \&
  {Hernquist}}]{2014MNRAS.444.1518V}
{Vogelsberger}, M., {Genel}, S., {Springel}, V., {et~al.} 2014, MNRAS, 444,
  1518

\bibitem[{{Voges} {et~al.}(2008){Voges}, {Oey}, {Walterbos}, \&
  {Wilkinson}}]{2008AJ....135.1291V}
{Voges}, E.~S., {Oey}, M.~S., {Walterbos}, R.~A.~M., \& {Wilkinson}, T.~M.
  2008, AJ, 135, 1291

\bibitem[{{Voges} \& {Walterbos}(2006)}]{2006ApJ...644L..29V}
{Voges}, E.~S. \& {Walterbos}, R.~A.~M. 2006, ApJL, 644, L29

\bibitem[{{Weilbacher}(2015)}]{2015scop.confE..53W}
{Weilbacher}, P. 2015, in Science Operations 2015: Science Data Management - An
  ESO/ESA Workshop, ESO Garching.
  (https://www.eso.org/sci/meetings/2015/SciOps2015.html)

\bibitem[{Weilbacher {et~al.}(2003)Weilbacher, Duc, \& Fritze-von
  Alvensleben}]{WDF03}
Weilbacher, P.~M., Duc, P.-A., \& Fritze-von Alvensleben, U. 2003, A\&A, 397,
  545

\bibitem[{{Weilbacher} {et~al.}(2015{\natexlab{a}}){Weilbacher},
  {Monreal-Ibero}, {Kollatschny}, {Ginsburg}, {McLeod}, {Kamann}, {Sandin},
  {Palsa}, {Wisotzki}, {Bacon}, {Selman}, {Brinchmann}, {Caruana}, {Kelz},
  {Martinsson}, {P{\'e}contal-Rousset}, {Richard}, \& {Wendt}}]{WeilbacherM42}
{Weilbacher}, P.~M., {Monreal-Ibero}, A., {Kollatschny}, W., {et~al.}
  2015{\natexlab{a}}, A\&A, 582, A114

\bibitem[{{Weilbacher} {et~al.}(2015{\natexlab{b}}){Weilbacher},
  {Monreal-Ibero}, {Mc Leod}, {Ginsburg}, {Kollatschny}, {Sandin}, {Wendt},
  {Wisotzki}, \& {Bacon}}]{2015Msngr.162...37W}
{Weilbacher}, P.~M., {Monreal-Ibero}, A., {Mc Leod}, A.~F., {et~al.}
  2015{\natexlab{b}}, The Messenger, 162, 37

\bibitem[{{Weilbacher} {et~al.}(2012){Weilbacher}, {Streicher}, {Urrutia},
  {Jarno}, {P{\'e}contal-Rousset}, {Bacon}, \& {B{\"o}hm}}]{WSU+12}
{Weilbacher}, P.~M., {Streicher}, O., {Urrutia}, T., {et~al.} 2012, in
  Proc.~{SPIE}, Vol. 8451, {Software and Cyberinfrastructure for Astronomy II}

\bibitem[{{Weilbacher} {et~al.}(2014){Weilbacher}, {Streicher}, {Urrutia},
  {P{\'e}contal-Rousset}, {Jarno}, \& {Bacon}}]{2014ASPC..485..451W}
{Weilbacher}, P.~M., {Streicher}, O., {Urrutia}, T., {et~al.} 2014, in
  ASP~Conf.~Ser., Vol. 485, Astronomical Data Analysis Software and Systems
  XXIII, ed. N.~{Manset} \& P.~{Forshay}, 451

\bibitem[{{White} \& {Rees}(1978)}]{WR78}
{White}, S. \& {Rees}, M. 1978, MNRAS, 183, 341

\bibitem[{{Whitmore} {et~al.}(1999){Whitmore}, {Zhang}, {Leitherer}, {Fall},
  {Schweizer}, \& {Miller}}]{WZL99}
{Whitmore}, B., {Zhang}, Q., {Leitherer}, C., {et~al.} 1999, AJ, 118, 1551

\bibitem[{{Whitmore} {et~al.}(2014){Whitmore}, {Brogan}, {Chandar}, {Evans},
  {Hibbard}, {Johnson}, {Leroy}, {Privon}, {Remijan}, \& {Sheth}}]{Whitmore14}
{Whitmore}, B.~C., {Brogan}, C., {Chandar}, R., {et~al.} 2014, ApJ, 795, 156

\bibitem[{{Whitmore} {et~al.}(2010){Whitmore}, {Chandar}, {Schweizer},
  {Rothberg}, {Leitherer}, {Rieke}, {Rieke}, {Blair}, {Mengel}, \&
  {Alonso-Herrero}}]{WCS+10}
{Whitmore}, B.~C., {Chandar}, R., {Schweizer}, F., {et~al.} 2010, AJ, 140, 75

\bibitem[{{Whitmore} {et~al.}(2005){Whitmore}, {Gilmore}, {Leitherer}, {Fall},
  {Chandar}, {Blair}, {Schweizer}, {Zhang}, \& {Miller}}]{WGL+05}
{Whitmore}, B.~C., {Gilmore}, D., {Leitherer}, C., {et~al.} 2005, AJ, 130, 2104

\bibitem[{{Zaragoza-Cardiel} {et~al.}(2014){Zaragoza-Cardiel}, {Font},
  {Beckman}, {Garc{\'{\i}}a-Lorenzo}, {Erroz-Ferrer}, \&
  {Guti{\'e}rrez}}]{2014MNRAS.445.1412Z}
{Zaragoza-Cardiel}, J., {Font}, J., {Beckman}, J.~E., {et~al.} 2014, MNRAS,
  445, 1412

\bibitem[{{Zastrow} {et~al.}(2013){Zastrow}, {Oey}, {Veilleux}, \&
  {McDonald}}]{2013ApJ...779...76Z}
{Zastrow}, J., {Oey}, M.~S., {Veilleux}, S., \& {McDonald}, M. 2013, ApJ, 779,
  76

\bibitem[{{Zhang} {et~al.}(2017){Zhang}, {Yan}, {Bundy}, {Bershady}, {Haffner},
  {Walterbos}, {Maiolino}, {Tremonti}, {Thomas}, {Drory}, {Jones}, {Belfiore},
  {S{\'a}nchez}, {Diamond-Stanic}, {Bizyaev}, {Nitschelm}, {Andrews},
  {Brinkmann}, {Brownstein}, {Cheung}, {Li}, {Law}, {Roman Lopes}, {Oravetz},
  {Pan}, {Storchi Bergmann}, \& {Simmons}}]{2017MNRAS.466.3217Z}
{Zhang}, K., {Yan}, R., {Bundy}, K., {et~al.} 2017, MNRAS, 466, 3217

\bibitem[{{Zurita} {et~al.}(2002){Zurita}, {Beckman}, {Rozas}, \&
  {Ryder}}]{2002A&A...386..801Z}
{Zurita}, A., {Beckman}, J.~E., {Rozas}, M., \& {Ryder}, S. 2002, A\&A, 386,
  801

\end{thebibliography}

\begin{appendix}
\section{Spectral analysis using pPXF}\label{sec:ppxf}
As described in the main text, we use a number of different techniques to
derive properties from spectra, each adapted to the data from which we
want to derive a measurement.
One of the techniques involves stellar population fitting. This needs a
significant number of parameters and inputs. As this technique was finally
only used for a subset of the analysis, we describe this in detail in this
appendix and only refer to it from the main text, where needed.

We compute the stellar population fits with the pPXF tool \citep[Python version
6.0.3, dated 2016-12-01,][]{2004PASP..116..138C}.  This allows us to
simultaneously measure the emission line fluxes while modeling the stellar
continuum, mainly to correct for Balmer absorption below the emission lines.
We use two sets of templates. The first set is used to model the stellar
continuum and the second for the emission lines.

Our primary continuum library is made up of 75 simple stellar populations
(SSPs) from the GALEV code \citep{2009MNRAS.396..462K} which we couple to the
synthetic stellar library of \citet{MSCZ05} to get sufficient coverage of the Hertzsprung–Russell diagram at a
resolution of $R=20000$. Since it is unknown a priori where in the Antennae a
given stellar population dominates, we input a set of 15 ages (4\,Myr to
10\,Gyr) and 5 metallicities ([Fe/H]=-1.7 to 0.4) into pPXF, so that the
minimization routine can select from a broad set of possible spectra. We
create the SSPs using a \citet{Kro02} initial mass function.
The advantages of this library are that its spectral resolution is
high enough to be applied to MUSE data and that its wavelength range covers the
full MUSE range.  It also provides stellar populations including young ages
(starting at 4\,Myr). However, since it is based on a theoretical library
without broad-band contributions from predicted lines, the continuum shape may
not be real. Hence, we have to use polynomials to model the medium-scale
continuum shape.

We cross-check the continuum analysis with SSP templates from the MIUSCAT
library \citep[][version 10.0]{2012MNRAS.424..157V}. These in general give less
good fits with slightly larger $\chi^2$ values. The inferred absorption below
the Balmer lines does not change significantly. But they lack the young stellar
populations (the youngest SSP is 63.1\,Myr old) and hence the GALEV models
should be a better representation of large sections of the Antennae. We
therefore select the GALEV SSPs for the continuum for our pPXF runs.

The other set of templates is made of the 36 brightest emission lines in the
MUSE wavelength range; these are modeled as single Gaussian peaks.  We use a
multiplicative polynomial of 7th degree to take out medium-scale flux
calibration residuals. To keep the {\em relative} variance of all pixels along
the wavelength direction unaffected by an additional resampling step, we input
observed spectra to pPXF that are taken from the log-sampled cubes, so that we
only log-rebin and convolve the templates to the MUSE resolution.\footnote{We use the
  wavelength dependent instrumental profile derived by the pipeline, average it
  over all spatial elements, and convolve it with a logarithmic sampling
  corresponding to a velocity scale of 53.5\kms. The wavelength dependence of
  the FWHM of the LSF can then be modeled by the polynomial
  \[\mathrm{LSF}(\lambda) = 4.3459 -3.6704\times10^{-4}\lambda -6.0942\times10^{-9}\lambda^2 + 2.8458\times10^{-12}\lambda^3\]
  where $\lambda$ is the wavelength in \AA.  This LSF has a minimum of 2.45\,\AA\
  FWHM, narrower than the internal width of the MILES
  \citep{2011A&A...532A..95F} and MIUSCAT spectral libraries.
}

As further input parameters we set starting values for the kinematics to have
the central velocity value of the Antennae and a $\sigma_\mathrm{cont}=75$\kms for the
stars and $\sigma_\mathrm{gas}=45$\kms for the emission lines.

\subsection{Full datacube analysis}\label{sec:ppxf:cube}
When dealing with the full datacube of the central Antennae, we first
improved the $S/N$ to get reliable fits, and produced data with \emph{continuum}
$S/N$ of 30, 50, 100, and 250 with the Voronoi binning technique
\citep{2003MNRAS.342..345C}. We use the two high-$S/N$ sets to compute initial
velocities, and combined and spatially smooth the results, separately for stars
and gas. Then we run pPXF on the lower $S/N$ data, using the maps with
initial velocities as input.
In this case, the emission line templates are mainly used to improve the
continuum fit, instead of masking the spectral regions around the emission lines. The stellar fit is
then subtracted from the data to produce a cube that contains only the emission
of ionized gas, and where the Balmer lines should be corrected for underlying
absorption.  We can then estimate $S/N$ of the \ha emission line in this
continuum-free cube and Voronoi-bin that to a $S/N$ adapted to the strength of
the gas emission. Note that this binning will have a different spatial layout
than the continuum binning to the same target $S/N$.

As this sequence of analysis steps is complex and the resulting
continuum-free cube is not free of artifacts, we aim to do as many analysis
steps on the original cubes as possible, and only use the continuum-free cube where technically
necessary, and to cross-check the results.

Since the southern region consists largely of sky background and
contains several bright foreground stars as well as a number of background
galaxies, and since the object itself is non-contiguous, the Voronoi scheme does not
create a sensible two-dimensional grid. We therefore did not attempt to create
a similar continuum-free cube for the southern field.

\subsection{Individual spectra}\label{sec:ppxf:hii}
When analyzing integrated spectra (\eg the \hiiregs) instead, no extra spatial
binning needs to be done. The pPXF result then gives us the kinematical data of
each spectrum, and the continuum-corrected best-fit emission line fluxes. To also
derive corresponding error estimates, we use the fit residuals of the spectrum.
To get a reasonable estimate of the total noise for each emission line, we
compute the sliding standard deviation of the residuals over a width of 6
wavelength pixels (\ie about 2$\times$FWHM).\footnote{The MUSE data
  carries a variance estimate which provides a good estimate of the {\em per
  voxel} noise in the presence of significant background. However, this
  propagated variance is correlated to neighboring voxels due to the cube
  reconstruction, therefore one cannot trust its absolute value when binning
  spectra over a region. Hence, the effective noise that we use here should
  provide a better error estimate.}
This noise array is then used to iterate 100 Monte-Carlo fits with pPXF,
leaving the best-fit continuum and the kinematics constant but computing new
emission line fluxes for every iteration.
\end{appendix}

\end{document}